\documentclass[aps,prd,onecolumn,showpacs,showkeys,amsmath,amssymb]{revtex4}
\usepackage{amsfonts}
\usepackage{amsmath}
\usepackage{graphicx}
\usepackage{subfigure}
\usepackage{dcolumn}
\usepackage{bm}
\usepackage{booktabs}
\usepackage[dvips]{color}
\makeatletter
\newcommand{\figcaption}{\def\@captype{figure}\caption}
\newcommand{\tabcaption}{\def\@captype{table}\caption}

\newcommand{\Rmnum}[1]{\expandafter\@slowromancap\romannumeral #1@}
\def\hlinewd#1{%
  \noalign{\ifnum0=`}\fi\hrule \@height #1 \futurelet
   \reserved@a\@xhline}
\makeatother
\def\dab{\int^{\alpha_{max}}_{\alpha_{min}}d\alpha\int^{\beta_{max}}_{\beta_{min}}d\beta}
\def\qq{\langle\bar qq\rangle}
\def\ss{\langle\bar ss\rangle}
\def\GGa{\langle GG\rangle}
\def\GGb{\langle g_s^2GG\rangle}
\def\qGqa{\langle\bar qGq\rangle}
\def\qGqb{\langle\bar q\sigma\cdot Gq\rangle}
\def\sGs{\langle\bar s\sigma\cdot Gs\rangle}
\def\efun{e^{-\frac{m^2}{\alpha(1-\alpha)M_B^2}}}
\def\f(s){[(\alpha+\beta)m^2-\alpha\beta s]}
\def\non{\\ \nonumber}

\begin{document}
%
%
\title{Exotic $QQ\bar{q}\bar{q}$, $QQ\bar{q}\bar{s}$, and $QQ\bar{s}\bar{s}$ states}
\author{Meng-Lin Du}
\email{duml@pku.edu.cn}
\author{Wei Chen}
\email{boya@pku.edu.cn}

\author{Xiao-Lin Chen}
\email{chenxl@pku.edu.cn}\affiliation{Department of Physics
and State Key Laboratory of Nuclear Physics and Technology\\
Peking University, Beijing 100871, China  }
\author{Shi-Lin Zhu}
\email{zhusl@pku.edu.cn} \affiliation{Department of Physics
and State Key Laboratory of Nuclear Physics and Technology\\
and Center of High Energy Physics, Peking University, Beijing
100871, China }
\begin{abstract}

After constructing the possible $J^P=0^-, 0^+, 1^-$, and $1^+$
$QQ\bar{q}\bar{q}$ tetraquark interpolating currents in a
systematic way, we investigate the two-point correlation functions
and extract the corresponding masses with the QCD sum rule
approach. We study the $QQ\bar{q}\bar{q}$, $QQ\bar{q}\bar{s}$, and
$QQ\bar{s}\bar{s}$ systems with various isospins $I=0, 1/2, 1$.
Our numerical analysis indicates that the masses of
doubly bottomed tetraquark states are below the threshold of the
two-bottom mesons, two-bottom baryons, and one doubly bottomed
baryon plus one antinucleon. Very probably these doubly bottomed
tetraquark states are stable.

\end{abstract}
\maketitle
\pagenumbering{arabic}
%
%
\section{Introduction}\label{sec:intro}
%

In the past decade, many charmonium or charmoniumlike states were
observed in the B factories, some of which do not fit in the
conventional quark model and are considered as the candidates of
the exotic states such as molecular states, tetraquark states,
hybrid mesons, baryonium states, etc. For experimental reviews of
these new states, see Refs.~\cite{2006-Swanson-p243-305,
2008-Zhu-p283-322, 2009-Bracko-p-, 2009-Yuan-p-,
2007-Rosner-p12002-12002}.

Hadronic molecular states are loosely bound states composed of a
pair of mesons. They are probably bound by the long-range,
color-singlet pion exchange. These interesting states generally
lie very close to the open-charm/bottom threshold. Some
near-threshold charmoniumlike states such as X(3872) and Z(4430) are
very good candidates of the molecular states composed of a pair of
charmed and anticharmed mesons. In fact, such a possibility was
studied extensively in Refs.~\cite{2009-Liu-p411-428,
2008-Liu-p63-73, 2004-Swanson-p197-202, 2004-Swanson-p189-195,
2004-Close-p119-123, 2008-Thomas-p34007-34007,
2009-Fernandez-Carames-p222001-222001}.

Tetraquarks are composed of four quarks. They are bound by
colored force between quarks. They decay through rearrangement.
Some are charged. Some carry strangeness. There are many states
within the same multiplet. The low-lying scalar mesons below 1 GeV
have been considered good candidates of the tetraquark states.
Some of the recently observed charmoniumlike states were also suggested
to be candidates of the hidden-charm tetraquark states
~\cite{2007-Matheus-p14005-14005, 2007-Maiani-p182003-182003,
2006-Ebert-p214-219,2011-Chen-p34010-34010,2010-Chen-p105018-105018,2012-Du-p-,2009-Jiao-p114034-114034}.
We have studied the possible pseudoscalar, scalar, vector, and
axial-vector hidden-charm/bottom tetraquark states systematically
in the framework of the QCD sum rule
\cite{2011-Chen-p34010-34010,2010-Chen-p105018-105018,2012-Du-p-,2009-Jiao-p114034-114034}.

Besides the hidden-charm/bottom $Q{\bar Q}q\bar{q}$-type
tetraquark states, the doubly-charmed/bottomed tetraquark states
$QQ\bar q\bar{q}$ are also very interesting, where Q denotes a
heavy quark (beauty or charm) and q denotes one
light quark (up, down, strange). Such a
four-quark configuration is allowed in QCD. In QED we have the
hydrogen atom. One light electron circles around the proton. When
two electrons are shared by two protons, a hydrogen molecule is
formed. In QCD we have the heavy meson. One light antiquark
circles around the heavy quark. If the two light antiquarks were
shared by the two heavy quarks in the doubly charmed/bottomed
tetraquark system, we would have the QCD analogue of the hydrogen
molecule!

On the other hand, if the heavy quark QQ pair is spatially close,
it would act as a pointlike antiheavy quark color source $\bar
Q$ and pick up two light quarks $\bar q\bar{q}$ to form the bound
state $QQ\bar q\bar{q}$. The existence and stability of such
systems have been studied in many different models, such as the MIT
bag model~\cite{1988-Carlson-p744-744}, chiral quark
model~\cite{2008-Zhang-p437-440, 1997-Pepin-p119-123}, constituent
quark model~\cite{2006-Vijande-p54018-54018,2006-Vijande-p074010-074010,
1998-Brink-p6778-6787,1993-Silvestre-Brac-p457-470,1986-Zouzou-p457-457},
and chiral perturbation theory~\cite{1993-Manohar-p17-33}. Some other useful
references
are~\cite{Navarra:2007yw,2007-Cui-p7-13,2003-Gelman-p296-304,1996-Moinester-p349-362,
1994-Bander-p5478-5480,1982-Ader-p2370-2370,1991-Richard-p254-257,
1986-Lipkin-p242-242,1973-Lipkin-p267-271,Carames:2011zz}. However, their
existence and stability are model dependent up to now. More
theoretical investigations will be helpful in the clarification of
the situation.

In this work, we will discuss the $QQ\bar{q}\bar{q}$ systems using
the method of QCD sum rule. We first construct the $QQ\bar{q}\bar{q}$
currents with $J^P=0^-$, $0^+$, $1^-$, and $1^+$ in a systematic way.
These currents have no definite C parity due to their special flavor
structures. The isospins of these currents can be $I=0, 1/2, 1$ with
specific quark contents. With the independent currents, we investigate
the two-point correlation functions and spectral densities. After
performing the QCD sum rule analysis, we extract the masses of the
possible $0^-$, $0^+$, $1^-$, $1^+$ $QQ\bar{q}\bar{q}$, $QQ\bar{q}\bar{s}$,
and $QQ\bar{s}\bar{s}$ states.

The paper is organized as follows. In Sec.~\ref{sec:current},
we construct the $QQ\bar{q}\bar{q}$ currents with $J^P=0^-$,
$0^+$, $1^-$, and $1^+$. In Sec.~\ref{sec:ope}, we calculate
the correlation functions with the operator product expansion(OPE)
method and extract the spectral densities. The results are
collected in Appendix~\ref{sec:rhos}. In Sec.~\ref{sec:num},
we perform the numerical analysis and extract the masses of these
tetraquark states. We
discuss the possible decay patterns of these doubly charmed/bottomed
tetraquark states in Sec.~\ref{sec:decay}. The last section is
a brief summary.

%
%
\section{tetraquark interpolating currents}\label{sec:current}
%
In this section, we construct the diquark-antidiquark currents
with $J^P=0^-, 0^+, 1^-$, and $1^+$ using the same technique in our
previous works
\cite{2011-Chen-p34010-34010,2010-Chen-p105018-105018}. One can
construct the tetraquark current with the $qq$ basis or $\bar{q}q$
basis, $(qq)(\bar{q}\bar{q})$ or $(\bar{q}q)(\bar{q}q)$. However,
they can be related by the Fierz transformation. In this work, we
consider the first set. Considering the Lorentz structures, there
are five independent diquark fields without derivatives: $q^T_a
Cq_b$, $q^T_a C\gamma_5q_b$, $q^T_aC\gamma_\mu q_b$,
 $q^T_aC\gamma_\mu\gamma_5q_b$, and $q^T_a C\sigma_{\mu\nu}q_b$, where a, b are the
color indices. Since $q^T_aC\sigma_{\mu\nu}\gamma_5q_b$ and
$q^T_aC\sigma_{\mu\nu}q_b$ carry different parity, we consider
both operators although they are equivalent. These diquark
bilinear can be in the symmetric $\mathbf 6$ representation or
antisymmetric $\mathbf{\bar 3}$ representation in the color and
flavor \emph{SU}(3) space. Considering the Lorentz structures, we
list the properties of these diquark operators in
Table~\ref{table00}.

\begin{center}
\renewcommand{\arraystretch}{1.5}
\begin{tabular*}{11cm}{l|c|c|r}
\hlinewd{.8pt}
$q\Gamma q$                         & $J^P$             & ~~  States   ~~   & (Flavor, Color) \\
\hline $q_a^TC\gamma_5q_b$                 & $0^+$             &
$^1S_0$     &$(\mathbf{6_f},\mathbf{6_c}),
                                                               (\mathbf{\bar 3_f},\mathbf{\bar 3_c})$ \\
$q_a^TCq_b$                         & $0^-$             &
$^3P_0$     &$(\mathbf{6_f},\mathbf{6_c}),
                                                               (\mathbf{\bar 3_f},\mathbf{\bar 3_c})$ \\
$q_a^TC\gamma_{\mu}\gamma_5q_b$     & $1^-$             &
$^3P_1$     &$(\mathbf{6_f},\mathbf{6_c}),
                                                               (\mathbf{\bar 3_f},\mathbf{\bar 3_c})$ \\
$q_a^TC\gamma_{\mu}q_b$             & $1^+$             &
$^3S_1$     &$(\mathbf{6_f},\mathbf{\bar 3_c}),
                                                               (\mathbf{\bar 3_f},\mathbf{6_c})$ \\
$q_a^TC\sigma_{\mu\nu}q_b$          &$\begin{cases}
                                     1^-,&\mbox{for}~\mu,\nu=1,2,3\\
                                     1^+,&\mbox{for}~\mu=0,\nu=1,2,3
                                     \end{cases}$       &$\begin{matrix}
                                                         ^1P_1\\^3S_1
                                                         \end{matrix}$   &$(\mathbf{6_f},\mathbf{\bar 3_c}),
                                                               (\mathbf{\bar 3_f},\mathbf{6_c})$ \\
$q_a^TC\sigma_{\mu\nu}\gamma_5q_b$  &$\begin{cases}
                                     1^+,&\mbox{for}~\mu,\nu=1,2,3\\
                                     1^-,&\mbox{for}~\mu=0,\nu=1,2,3
                                     \end{cases}$         &$\begin{matrix}
                                                         ^3S_1\\^1P_1
                                                         \end{matrix}$   &$(\mathbf{6_f},\mathbf{\bar 3_c}),
                                                               (\mathbf{\bar 3_f},\mathbf{6_c})$ \\
\hlinewd{.8pt}
\end{tabular*}
\tabcaption{Properties of the diquark operators.} \label{table00}
\end{center}
Being composed of two quark (antiquark) fields, the diquark
(antidiquark) fields should satisfy Fermi statistics. As shown in
Table~\ref{table00}, the flavor and color structures are entangled
for every diquark operator. For example, the flavor and color
structures of the scalar diquark operator $q_a^TC\gamma_5q_b$ are
either $(\mathbf{6_f},\mathbf{6_c})$ or $(\mathbf{\bar
3_f},\mathbf{\bar 3_c})$. For $QQ\bar{q}\bar{q}$ systems, the
heavy quark pair has the symmetric flavor structure
$\mathbf{6_f}$. Its flavor and color structures are then fixed as
$(\mathbf{6_f},\mathbf{6_c})$. To construct the color singlet
$QQ\bar{q}\bar{q}$ currents, the heavy quark pair $QQ$ and the
light antiquark pair $\bar{q}\bar{q}$ should have the same color
structures.

According to Ref.~\cite{2011-Chen-p34010-34010}, there are ten
color singlet $QQ\bar{q}\bar{q}$ currents with $J^P=0^-$:
\begin{equation}
\begin{split}
S_{\pm}&=Q^T_{a}CQ_{b}(\bar{q}_{a}\gamma_5 C\bar{q}^T_b\pm \bar{q}_{b}\gamma_5 C\bar{q}^T_a),\\
V_{\pm}&=Q^T_aC\gamma_5Q_b(\bar{q}_aC\bar{q}^T_b\pm \bar{q}_bC\bar{q}^T_a),\\
T_{\pm}&=Q^T_aC\sigma_{\mu\nu}Q_b(\bar{q}_a\sigma^{\mu\nu}\gamma_5C\bar{q}^T_b\pm
\bar{q}_b\sigma^{\mu\nu}\gamma_5C\bar{q}^T_a),\\
A_{\pm}&=Q^T_aC\gamma_\mu Q_b(\bar{q}_a\gamma^\mu\gamma_5
C\bar{q}^T_b\pm
\bar{q}_b\gamma^\mu\gamma_5 C\bar{q}^T_a),\\
P_{\pm}&=Q^T_aC\gamma_\mu\gamma_5Q_b(\bar{q}_a\gamma^\mu
C\bar{q}^T_b\pm\bar{q}_b\gamma^\mu C\bar{q}^T_a). \label{current0}
\end{split}
\end{equation}
where ``+" denotes the symmetric color structure
$[\mathbf{6_c}]_{QQ} \otimes [\mathbf{ \bar 6_c}]_{\bar{q}\bar q}$
and ``-" denotes the antisymmetric color structure $[\mathbf{\bar
3_c}]_{QQ} \otimes [\mathbf{3_c}]_{\bar{q}\bar q}$. Due to the
symmetry constraint, it's enough to keep one light diquark piece
only in the bracket of Eq. (\ref{current0}) within the
calculation. We keep two terms in Eq. (\ref{current0}) to
illustrate the color symmetry explicitly.

By considering the symmetric flavor structure for heavy quark pair
$QQ$ , only the currents which satisfy the Pauli principle
survive. For the pseudoscalar currents in Eq.~(\ref{current0}),
$S_+, V_+, T_-, A_-$, and $P_+$ survive and all the other currents
vanish. According to Table~\ref{table00}, the $QQ\bar{q}\bar{q}$
($q=u, d$) operators $S_+, V_+, T_-$ are isovector currents and
$A_-, P_+$ are isoscalar currents. Finally, we obtain the following
$QQ\bar{q}\bar{q}$ interpolating currents with $J^P=0^-$, $0^+$,
$1^-$, and $1^+$:
\begin{itemize}
\item The tetraquark interpolating currents with $J^P=0^-$ are
\begin{equation}
\begin{split}
\eta_1&=Q^T_aCQ_b(\bar{q}_a\gamma_5C\bar{q}_b^T+\bar{q}_b\gamma_5C\bar{q}^T_a),\\
\eta_2&=Q^T_aC\gamma_5Q_b(\bar{q}_aC\bar{q}_b^T+\bar{q}_bC\bar{q}^T_a),\\
\eta_3&=Q^T_aC\sigma_{\mu\nu}Q_b(\bar{q}_a
\sigma^{\mu\nu}\gamma_5C\bar{q}^T_b-\bar{q}_b\sigma^{\mu\nu}\gamma_5C\bar{q}^T_a),\\
\eta_4&=Q^T_aC\gamma_\mu Q_b(\bar{q}_a\gamma^\mu\gamma_5
C\bar{q}^T_b-
\bar{q}_b\gamma^\mu\gamma_5 C\bar{q}^T_a),\\
\eta_5&=Q^T_aC\gamma_\mu\gamma_5Q_b(\bar{q}_a\gamma^\mu
C\bar{q}^T_b+\bar{q}_b\gamma^\mu C\bar{q}^T_a).\label{current1}
\end{split}
\end{equation}
in which $\eta_1, \eta_2, \eta_3$ are isovector currents with $[\mathbf{\bar 6_f}]_{\bar{q}\bar q}$
and $I=1$, $\eta_4, \eta_5$ are isoscalar currents with $[\mathbf{3_f}]_{\bar{q}\bar q}$ and $I=0$.
\item The tetraquark interpolating currents with $J^P=0^+$ are
\begin{equation}
\begin{split}
\eta_1&=Q^T_aCQ_b(\bar{q}_aC\bar{q}_b^T+\bar{q}_bC\bar{q}^T_b),\\
\eta_2&=Q^T_aC\gamma_5Q_b(\bar{q}_a\gamma_5C\bar{q}_b^T+\bar{q}_b\gamma_5C\bar{q}^T_b),\\
\eta_3&=Q^T_aC\gamma_\mu Q_b(\bar{q}_a\gamma^\mu C\bar{q}_b^T-\bar{q}_b \gamma^\mu C\bar{q}^T_b),\\
\eta_4&=Q^T_aC\gamma_\mu\gamma_5Q_b(\bar{q}_a\gamma^\mu \gamma_5C\bar{q}_b^T+\bar{q}_b\gamma^\mu \gamma_5C\bar{q}^T_b),\\
\eta_5&=Q^T_aC\sigma_{\mu\nu}Q_b(\bar{q}_a\sigma^{\mu\nu}C\bar{q}^T_b-\bar{q}_b\sigma^{\mu\nu}C\bar{q}^T_a).\label{current2}
\end{split}
\end{equation}
and all the scalar interpolating currents are isovector currents with $[\mathbf{\bar 6_f}]_{\bar{q}\bar q}$
and $I=1$.
\item The tetraquark interpolating currents with $J^P=1^-$ are
\begin{equation}
\begin{split}
\eta_1&=Q^T_aC\gamma_\mu\gamma_5 Q_b(\bar{q}_a\gamma_5C\bar{q}_b^T+\bar{q}_b\gamma_5C\bar{q}^T_a),\\
\eta_2&=Q^T_aC\gamma_5Q_b(\bar{q}_a\gamma_\mu\gamma_5 C\bar{q}_b^T+\bar{q}_b\gamma_\mu\gamma_5 C\bar{q}_a^T),\\
\eta_3&=Q^T_aC\sigma_{\mu\nu}Q_b(\bar{q}_a\gamma^\nu C\bar{q}^T_b-\bar{q}_b\gamma^\nu C\bar{q}^T_a),\\
\eta_4&=Q^T_aC\gamma^\nu Q_b(\bar{q}_a\sigma_{\mu\nu}
C\bar{q}^T_b-\bar{q}_b\sigma_{\mu\nu}C\bar{q}^T_a),\\
\eta_5&=Q^T_aC\gamma_\mu Q_b(\bar{q}_aC\bar{q}_b^T-\bar{q}_bC\bar{q}^T_a),\\
\eta_6&=Q^T_aCQ_b(\bar{q}_a\gamma_\mu C\bar{q}_b^T+\bar{q}_b\gamma_\mu C\bar{q}_a^T),\\
\eta_7&=Q^T_aC\sigma_{\mu\nu}\gamma_5Q_b(\bar{q}_a\gamma^\nu\gamma_5 C\bar{q}^T_b-\bar{q}_b\gamma^\nu\gamma_5 C\bar{q}^T_a),\\
\eta_8&=Q^T_aC\gamma^\nu\gamma_5 Q_b(\bar{q}_a\sigma_{\mu\nu}\gamma_5
C\bar{q}^T_b+\bar{q}_b\sigma_{\mu\nu}\gamma_5C\bar{q}^T_a).\label{current3}
\end{split}
\end{equation}
in which $\eta_1, \eta_2, \eta_3, \eta_4$ are isovector currents with $[\mathbf{\bar 6_f}]_{\bar{q}\bar q}$
and $I=1$,
$\eta_5, \eta_6, \eta_7, \eta_8$ are isoscalar currents with $[\mathbf{3_f}]_{\bar{q}\bar q}$ and $I=0$.
\item The tetraquark interpolating currents with $J^P=1^+$ are
\begin{equation}
\begin{split}
\eta_1&=Q^T_aC\gamma_\mu\gamma_5 Q_b(\bar{q}_aC\bar{q}_b^T+\bar{q}_bC\bar{q}^T_a),\\
\eta_2&=Q^T_aCQ_b(\bar{q}_a\gamma_\mu\gamma_5 C\bar{q}_b^T+\bar{q}_b\gamma_\mu\gamma_5 C\bar{q}_a^T),\\
\eta_3&=Q^T_aC\sigma_{\mu\nu}\gamma_5 Q_b(\bar{q}_a\gamma^\nu C\bar{q}^T_b-\bar{q}_b\gamma^\nu C\bar{q}^T_a),\\
\eta_4&=Q^T_aC\gamma^\nu Q_b(\bar{q}_a\sigma_{\mu\nu}\gamma_5
C\bar{q}^T_b-\bar{q}_b\sigma_{\mu\nu}\gamma_5
C\bar{q}^T_a),\\
\eta_5&=Q^T_aC\gamma_\mu Q_b(\bar{q}_a\gamma_5C\bar{q}_b^T-\bar{q}_b\gamma_5C\bar{q}^T_a),\\
\eta_6&=Q^T_aC\gamma_5Q_b(\bar{q}_a\gamma_\mu C\bar{q}_b^T+\bar{q}_b\gamma_\mu C\bar{q}_a^T),\\
\eta_7&=Q^T_aC\sigma_{\mu\nu} Q_b(\bar{q}_a\gamma^\nu\gamma_5 C\bar{q}^T_b-\bar{q}_b\gamma^\nu\gamma_5 C\bar{q}^T_a),\\
\eta_8&=Q^T_aC\gamma^\nu\gamma_5 Q_b(\bar{q}_a\sigma_{\mu\nu}C\bar{q}^T_b+\bar{q}_b\sigma_{\mu\nu}
C\bar{q}^T_a). \label{current4}
\end{split}
\end{equation}
in which $\eta_1, \eta_2, \eta_3, \eta_4$ are isovector currents with $[\mathbf{\bar 6_f}]_{\bar{q}\bar q}$
and $I=1$,
$\eta_5, \eta_6, \eta_7, \eta_8$ are isoscalar currents with $[\mathbf{3_f}]_{\bar{q}\bar q}$ and $I=0$.
\end{itemize}

For the isovector ($I=1$) currents, we do not differentiate the up and down quarks in our
analysis and denote them by $q$. However, they should be differentiated for the isoscalar ($I=0$)
currents because the flavor structures of the light anti-diquark $\bar q\bar q$ are antisymmetric.
The quark contents are $QQ\bar u\bar d$ for these currents. For the $QQ\bar s\bar s$ systems
, only the currents with $[\mathbf{\bar 6_f}]_{\bar{q}\bar q}$ flavor structures survive. The isospins
for these systems are $I=0$. We will also discuss the $QQ\bar q\bar s$ systems ($I=1/2$) by using all
the currents in Eq.~(\ref{current1})$-$(\ref{current4}). We pick up the interpolating currents with
different quark contents in Table.~\ref{table01}. To calculate the two-point correlation functions,
the Wick contractions of the currents for $QQ\bar u\bar d$ and $QQ\bar q\bar s$ systems are different
from those for the $QQ\bar q\bar q$ and $QQ\bar s\bar s$ systems.

\begin{center}
\renewcommand{\arraystretch}{1.3}
\begin{tabular*}{13cm}{c|c|c|c|c|c|c}
\hlinewd{.8pt}
Quark Content     & ~$[\bar{q}\bar q]_\mathbf{f}$  ~ &  ~~  I  ~~   &   $J^P=0^-$   &   $J^P=0^+$   &  $J^P=1^-$   &  $J^P=1^+$ \\
\hline
$QQ\bar q\bar q$  & $\mathbf{\bar 6_f}$ &    1        &  $\eta_1,\eta_2,\eta_3$  &  $\eta_1,\eta_2,\eta_3,\eta_4,\eta_5$
                                 &  $\eta_1,\eta_2,\eta_3,\eta_4$  &  $\eta_1,\eta_2,\eta_3,\eta_4$ \\
$QQ\bar s\bar s$  & $\mathbf{\bar 6_f}$ &    0        &  $\eta_1,\eta_2,\eta_3$  &  $\eta_1,\eta_2,\eta_3,\eta_4,\eta_5$
                                 &  $\eta_1,\eta_2,\eta_3,\eta_4$  &  $\eta_1,\eta_2,\eta_3,\eta_4$ \\
$QQ\bar q\bar s$  & $\mathbf{\bar 6_f}$ &  $1/2$      &  $\eta_1,\eta_2,\eta_3,$
                                 &  $\eta_1,\eta_2,\eta_3,\eta_4,\eta_5$
                                 &  $\eta_1,\eta_2,\eta_3,\eta_4,$ &  $\eta_1,\eta_2,\eta_3,\eta_4,$\\
                  & $\mathbf{3_f}$      &&  $\eta_4,\eta_5$&&$\eta_5,\eta_6,\eta_7,\eta_8$&$\eta_5,\eta_6,\eta_7,\eta_8$ \\
$QQ\bar u\bar d$  & $\mathbf{3_f}$      &    0        &  $\eta_4,\eta_5$          &  $-$
                                 &  $\eta_5,\eta_6,\eta_7,\eta_8$  &  $\eta_5,\eta_6,\eta_7,\eta_8$ \\
\hlinewd{.8pt}
\end{tabular*}
\tabcaption{The quark contents and isospins of the tetraquark currents.} \label{table01}
\end{center}

%
\section{QCD sum rule}\label{sec:ope}
%
%
In QCD sum rule
\cite{1979-Shifman-p385-447,1985-Reinders-p1-1,2000-Colangelo-p-},
we consider the two-point correlation functions of the
interpolation currents. For the scalar and pseudoscalar currents,
the two-point correlation functions read
\begin{equation}
\Pi(q^{2})\equiv i \int d^4
xe^{iqx}\langle0|T\eta(x)\eta^\dag(0)|0\rangle, \label{equ:po}
\end{equation}
where $\eta(x)$ is the corresponding interpolating current. The
two-point correlation functions of the vector and axial-vector
currents are
\begin{equation}
\begin{split}
\Pi_{\mu\nu}(q^2)&=i\int d^4x e^{iqx}\langle 0|T \eta_\mu(x)\eta_\nu^\dagger(0)|0\rangle\\
&=\Pi(q^2)(\frac{q_\mu
q_\nu}{q^2}-g_{\mu\nu})+\Pi_0(q^2)\frac{q_\mu q_\nu}{q^2}.
\end{split}
\end{equation}
There are two parts of $\Pi_{\mu\nu}$ with different Lorentz
structures because $\eta_\mu$ is not a conserved current.
$\Pi(q^2)$ is related to the vector and axial-vector meson, while
$\Pi_0$ is the scalar and pseudoscalar current polarization
function. At the hadron level, we express the correlation function
in the
 form of the dispersion relation with spectral function,
\begin{equation}
\Pi(q^2)=(q^2)^N\int_{4(m_q+m_Q)^2}^{\infty}\frac{\rho(s)}{s^N(s-q^2-i\varepsilon)}ds+
\sum^{N-1}_{n=0}b_n(q^2)^n,\label{equ:pq}
\end{equation}
where
\begin{equation}
\begin{split}
\rho(s)\equiv&\sum_{n}\delta(s-m^{2}_{n})\langle0|\eta|n\rangle\langle
n|\eta^{\dag}|0\rangle\\
=&f^{2}_{X}\delta(s-m^{2}_{X})+\mbox{continuum} \; ,
\label{equ:Pi}
\end{split}
\end{equation}
where $m_X$ is the mass of the resonance $X$ and $f_X$ is the
decay constant of the meson,
\begin{equation}
\begin{split}
&\langle0|\eta|X\rangle=f_X,\\
&\langle0|\eta_\mu|X\rangle=f_X\epsilon_\mu^X,
\end{split}
\end{equation}
where $\epsilon_\mu^X$ is the polarization vector of X
($\epsilon^X\cdot q=0$).

One can calculate the correlation functions at the quark-gluon
level via the operator product expansion(OPE) method. Using the
same technique as in
Refs.~\cite{2011-Chen-p34010-34010,2010-Chen-p105018-105018,2012-Du-p-,2009-Albuquerque-p53-66,2009-Bracco-p240-244,
2007-Matheus-p14005-14005}, we calculate the Wilson coefficients
while the light quark propagator and heavy quark propagator are
adopted as
\begin{equation}
\begin{split}
iS_q^{ab}&=\frac{i\delta^{ab}}{2\pi^2x^4}\hat{x}+\frac{i}{32\pi^2}\frac{\lambda^n_{ab}}{2}g_sG^n_{\mu\nu}\frac{1}{x^2}(\sigma^{\mu\nu}\hat{x}
+\hat{x}\sigma^{\mu\nu})-\frac{\delta^{ab}}{12}\langle\bar{q}q\rangle \\
&+\frac{\delta^{ab}x^2}{192}\langle g_s\bar{q}\sigma \cdot Gq\rangle -\frac{m_q\delta^{ab}}{4\pi^2x^2}+\frac{i\delta^{ab}m_q\langle \bar{q}q\rangle}{48}\hat{x},\\
iS_Q^{ab}&=\frac{i\delta^{ab}}{\hat{p}-m_Q}+\frac{i}{4}g_s\frac{\lambda^n_{ab}}{2}G^n_{\mu\nu}\frac{\sigma^{\mu\nu}(\hat{p}+m_Q)+
(\hat{p}+m_Q)\sigma^{\mu\nu}}{(p^2-m_Q^2)^2}\\
&+\frac{i\delta^{ab}}{12}\langle g^2_s GG\rangle
m_Q\frac{p^2+m_Q\hat{p}}{(p^2-m_Q^2)^4},
\end{split}
\end{equation}
where $\hat{x}=\gamma_\mu x^\mu$, $\hat{p}=\gamma_\mu p^\mu$,
$\langle \bar{q}g_s\sigma\cdot Gq\rangle= \langle
g_s\bar{q}\sigma^{\mu\nu} G_{\mu\nu}q\rangle$, $\langle
g_s^2GG\rangle=\langle g_s^2 G_{\mu\nu}G^{\mu\nu}\rangle$. The
spectral density is obtained with:
$\rho(s)=\frac{1}{\pi}\mbox{Im}\Pi(q^2)$.

In order to suppress the higher-state contributions and remove the
subtraction terms in Eq. (\ref{equ:pq}), we perform the Borel
transformation to the correlation function,
\begin{eqnarray}
L_{M_B}\Pi(p^2)=\lim_{{-p^2,n\to \infty\atop -p^2/n\equiv
M_B^2}}\frac{1}{n!}(-p^2)^{n+1}(\frac{d}{dp^2})^n\Pi(p^2).
\end{eqnarray}
After performing the Borel transformation and equating the two
representations of the correlation function with the quark-hadron
duality, we obtain
\begin{equation}
\Pi(M_B^2)=f_X^2e^{-m_X^2/ M_B^2}=\int_{4(m_q+m_Q)^2}^{s_0} ds
e^{-s/M_B^2}\rho(s),
\end{equation}
where $s_0$ is the threshold parameter, and $M_B$ is the Borel
parameter. We can extract the meson mass $m_X$,
\begin{equation}
m_X^2=\frac{\int_{4(m_q+m_Q)^2}^{s_0} ds e^{-s/M_B^2} s
\rho^{OPE}(s)}{\int_{4(m_q+m_Q)^2}^{s_0} ds e^{-s/M_B^2}
\rho^{OPE}(s)}.
\end{equation}
For all the tetraquark currents in
Eq.~(\ref{current1})$-$(\ref{current4}), we collect the spectral
densities $\rho(s)$ in Appendix~\ref{sec:rhos}, respectively.
We neglect the three-gluon condensate $\langle g_s^3 fGGG\rangle$
because their contribution is negligible.

%
\section{Numerical Analysis}\label{sec:num}
%
In the QCD sum rule analysis, we use the following values of the
parameters \cite{1979-Shifman-p385-447,
2010-Nakamura-p75021-75021, 2001-Eidemuller-p203-210,
1999-Jamin-p300-303, 2002-Jamin-p237-243} in the chiral limit
($m_u=m_d=0$):
\begin{equation}
\begin{split}
&m_s(1 \mbox{GeV})=125\pm 20 \mbox{MeV},\\
&m_c(m_c)=(1.23\pm0.09)~\mbox{GeV},\\
&m_b(m_b)=(4.2\pm0.07)~\mbox{GeV},\\
&\langle \bar{q}q\rangle=-(0.23\pm0.03)^3~\mbox{GeV}^3,\\
&\langle \bar{s}s\rangle=(0.8\pm0.1)\langle \bar{q}q\rangle,\\
&\langle \bar{q}g_s \sigma\cdot G q\rangle=-M_0^2\langle \bar{q}q\rangle,\\
&M_0^2=(0.8\pm0.2)~\mbox{GeV},\\
&\langle g_s^2 GG\rangle=(0.48\pm0.14)~\mbox{GeV}^4.
\label{parameter}
\end{split}
\end{equation}
The Borel mass $M_B$ and the threshold value $s_0$ are two pivotal
parameters. Requiring the convergence of the OPE leads to the
lower
 bound  $M^2_{Bmin}$ of the Borel parameter. In the present work, we
require that the most important condensate contribution be less
than one fourth of the perturbative term. We require that the pole
contribution be larger than $30\%$, which determines the upper
bound $M^2_{Bmax}$ of the Borel parameter. The pole contribution
(PC) is defined as
\begin{equation}
\mbox{PC}=\frac{\int_{0}^{s_0} ds
e^{-s/M_B^2}\rho(s)}{\int_{0}^{\infty} ds e^{-s/M_B^2}\rho(s)},
\end{equation}
which depends on both the Borel mass $M^2_B$ and the threshold
value $s_0$. $s_0$ is chosen around the region where the variation
of $m_X$ with $M_B$ is minimum in the Borel working region. For a
genuine hadron state, the extracted mass from the sum rule
analysis is expected to be stable with the reasonable variation of
the Borel parameter $M_B^2$ and threshold $s_0$.

For all the isovector $cc\bar{q}\bar{q}$ and isoscalar
$cc\bar{u}\bar{d}$ systems, the most important nonperturbative
corrections come from the four-quark condensate
$\langle\bar{q}q\rangle^2$. Both the quark condensate
$\langle\bar{q}q\rangle$ and the quark-gluon mixed condensate
$\langle \bar{q}g_s \sigma\cdot G q\rangle$ vanish when we let
$m_u=m_d=m_q=0$. For the $(I, J^P)=(1, 0^-)$ $cc\bar{q}\bar{q}$
system, only the interpolating currents $\eta_1$ and $\eta_3$
lead to a stable mass sum rule after performing the QCD sum rule
analysis. In Fig.~\ref{fig1}, we show the
mass curves of the extracted hadron mass $m_X$ with $M_B^2$ and
$s_0$ for the current $\eta_3$ with $(I, J^P)=(1, 0^-)$.
The variation of $m_X$ with the Borel mass $M_B$ is very weak
around $s_0\sim 23$ GeV$^2$. For $\eta_2$, the stability of the
mass curves is much worse and $m_X$ grows monotonically with $s_0$
and $M_B$. The situation is very similar for the $(I, J^P)=(0, 0^-)$
$cc\bar{s}\bar{s}$ systems. Now we keep the $m_s$ related terms in
the spectral densities. These terms are very important corrections
for the OPE series. The dominant nonperturbative contribution is
the quark condensate $\langle\bar{s}s\rangle$ for $\eta_2^s$ and
$\eta_3^s$. We show the variations of $m_{X^s}$ with the Borel
mass $M_B^2$ and threshold parameter $s_0$ for the current
$\eta_3^s$ in Fig.~\ref{fig2}.

\begin{center}
\begin{tabular}{lr}
\scalebox{0.67}{\includegraphics{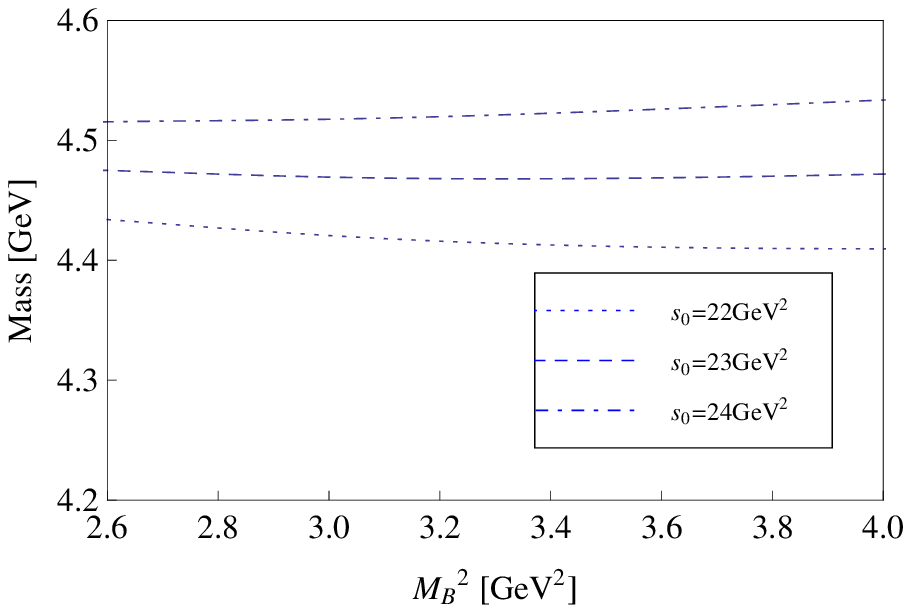}}&
\scalebox{0.67}{\includegraphics{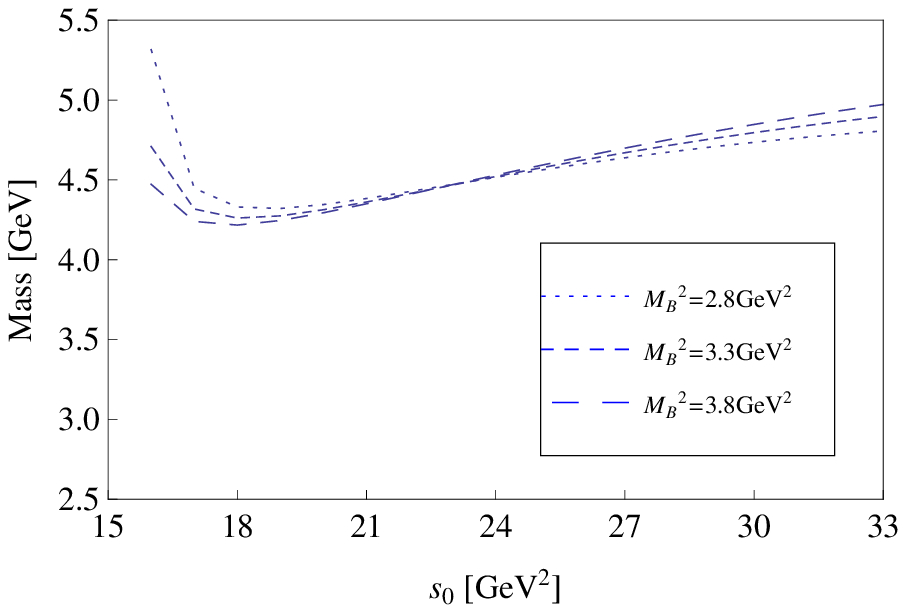}}
\end{tabular}
\figcaption{The variation of $m_X$ with $M_B$ and $s_0$ for the
current $\eta_3$ with $(I, J^P)=(1, 0^-)$ for $cc\bar{q}\bar{q}$ system.}
\label{fig1}
\end{center}
\begin{center}
\begin{tabular}{lr}
\scalebox{0.54}{\includegraphics{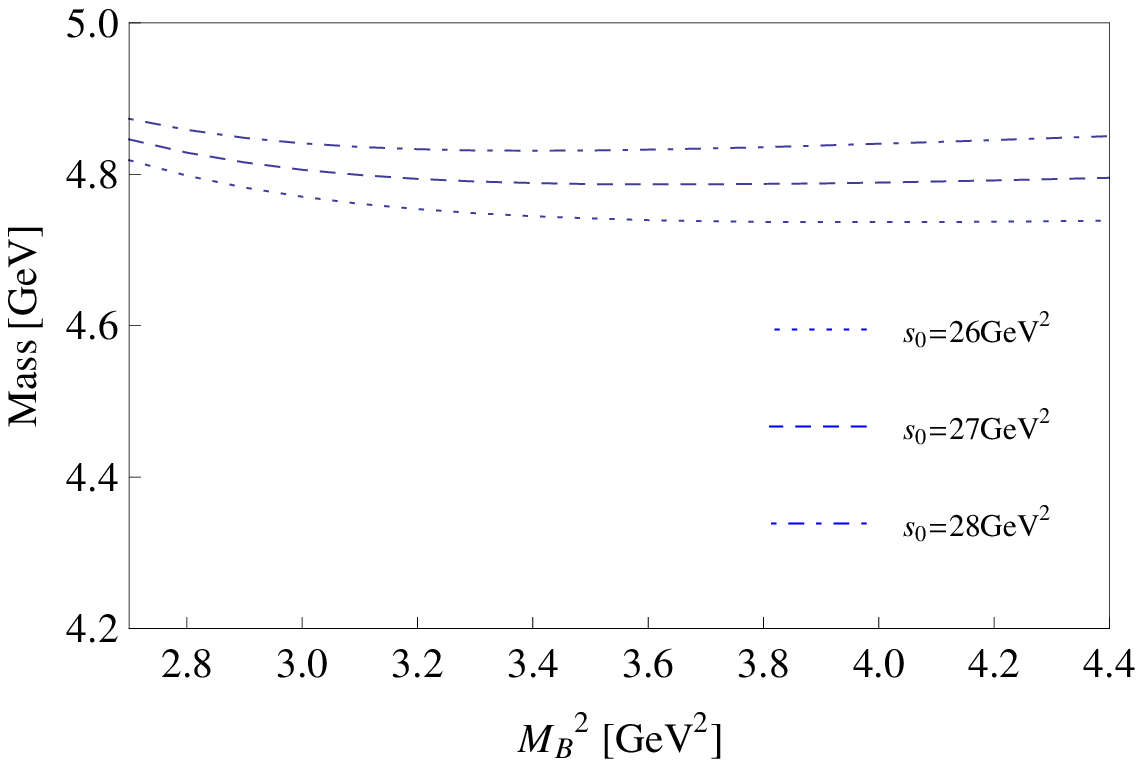}}&
\scalebox{0.54}{\includegraphics{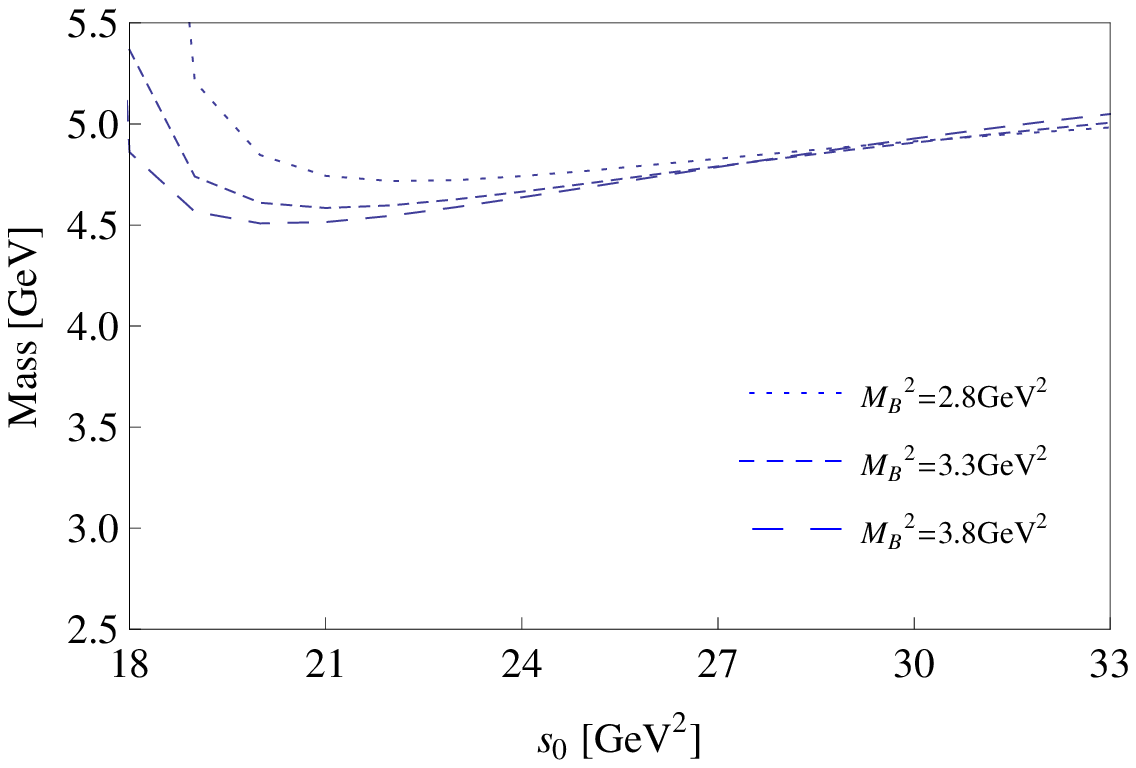}}
\end{tabular}
\figcaption{The variation of $m_{X^s}$ with $M_B$ and $s_0$ for
the current $\eta_3^s$ with $(I, J^P)=(0, 0^-)$ for $cc\bar{s}\bar{s}$
system.} \label{fig2}
\end{center}

With the parameters in Eq.~(\ref{parameter}), we list the working
region of the Borel parameters, threshold value $s_0$, the
extracted masses for the currents with $J^P=0^-$ in
Table~\ref{table1}. The pole contribution and the masses are
extracted using the corresponding threshold values $s_0$ and Borel
parameters $M_B^2$ listed in the table. For the isovector
currents $\eta_1, \eta_2$, and $\eta_3$, one can also investigate
the $QQ\bar u\bar d$ systems besides the $QQ\bar u\bar u$ and
$QQ\bar d\bar d$ systems. As mentioned in Sec.~\ref{sec:current},
the Wick contractions of the currents for the $QQ\bar u\bar d$
systems are different from those for the $QQ\bar d\bar d$ and
$QQ\bar u\bar u$ systems. However, the correlation functions are
the same in the chiral limit ($m_u=m_d=m_q=0$) except for a
constant coefficient. We denote them as $QQ\bar q\bar q$ when we
discuss the isovector systems. We take into account the
uncertainty of the values of the threshold parameter and variation
of the Borel mass to obtain the errors. The other possible error
sources are the truncation of the OPE series and the uncertainty
of the quark masses. The uncertainty of the condensate values are
not included.

Replacing $m_c$ with $m_b$ in the correlation functions and repeating
the same analysis procedures, we obtain the results of the doubly bottomed
systems. We also collect the numerical results of the $bb\bar{q}\bar{q}$,
$bb\bar{q}\bar{s}$, and $bb\bar{s}\bar{s}$ systems with $J^P=0^-$ in Table~\ref{table1}.

The derivation and analysis of the QCD sum rules of the $J^P=0^+, 1^-, 1^+$
$QQ\bar q\bar q$, $QQ\bar{q}\bar{s}$, and $QQ\bar s\bar s$ systems are very
similar to the $J^P=0^-$ case. We omit details and collect the numerical
results in Table~\ref{table2}, Table~\ref{table3}, and Table~\ref{table4}
for the $J^P=0^+, 1^-$ and $1^+$ systems, respectively.

There are no stable sum rules for the $0^+$ and $1^+$ $cc\bar q\bar q$ and
$cc\bar u\bar d$ systems. These tetraquarks probably do not exist. The
doubly bottomed systems are more stable as the heavy quark mass increases.

In the $J^P=1^+$ $cc\bar s\bar s$ system, the extracted mass is
about $5.03\sim 5.12$ GeV from the currents $\eta_1^s$ and
$\eta_2^s$. In contrast, the mass extracted from currents
$\eta_3^s$ and $\eta_4^s$ is around $4.17$ GeV, which is much
lower than that from $\eta_1^s$ and $\eta_2^s$. The same situations
occur in the $J^P=1^+$ $cc\bar q\bar s$, $bb\bar q\bar s$ and $bb\bar s\bar s$
cases. According to Table.~\ref{table00}, the diquark fields
$Q_a^TC\gamma_{\mu}\gamma_5Q_b$ and $Q_a^TCQ_b$ are P-wave operators while
$Q_a^TC\gamma_{\mu}Q_b$ and $Q_a^TC\sigma_{\mu\nu}\gamma_5Q_b$ are S-wave
operators. So the interpolating currents $\eta_1^s$ and $\eta_2^s$ contain
two P-wave operators, whereas $\eta_3^s$ and $\eta_4^s$ contain two
S-wave operators. In other words, the extracted masses from the
currents $\eta_3^s$ and $\eta_4^s$ correspond to the ground state
mass of the $J^P=1^+$ $cc\bar s\bar s$ system while the masses of
$\eta_1^s$ and $\eta_2^s$ correspond to the orbitally excited
state. That is the underlying mechanism which renders the
extracted mass from $\eta_1^s$ and $\eta_2^s$ is much higher than
that from $\eta_3^s$ and $\eta_4^s$. The similar situation occurs
in the current $\eta_1^s$ with $J^P=0^+$ for the $cc\bar s\bar s$
and $bb\bar s\bar s$ systems.

There is another intuitive way to understand the difference of the
various interpolating currents. According to textbook knowledge
about the quark model, the interaction between the quark pair for
the symmetric color structure $[\mathbf{6_c}]_{QQ} \otimes
[\mathbf{\bar 6_c}]_{\bar{q}\bar q}$ is repulsive, whereas the
interaction for the antisymmetric color structure $[\mathbf{\bar
3_c}]_{QQ} \otimes [\mathbf{3_c}]_{\bar{q}\bar q}$ is attractive.
Thus, the currents with the symmetric color structure will result
in higher mass than those with the antisymmetric color structure.
There were similar observations of the effect of the color
structure on the tetraquark spectrum in Refs.
\cite{1997-Pepin-p119-123, 1986-Zouzou-p457-457}.

\begin{center}
\begin{tabular}{l|c|c|c|c|c|c|c|c}
\hlinewd{.8pt}
  &Current & $s_0$  & [$M^2_{Bmin},M^2_{Bmax}$] & $M^2_B$ & $m_X $ & PC & ~~$f_X$~~ & open charm/bottom \\
                    &  & ($\mbox{GeV}^2$) & ($\mbox{GeV}^2$) &($\mbox{GeV}^2$)& $(\mbox{GeV})$ &(\%)&($\mbox{GeV}^5$) &threshold $(\mbox{GeV})$\\
                    \hline
$cc\bar{q}\bar{q}$  &$\eta_1$ & 24 & $3.0\sim 3.9$ & 3.4 & $4.43\pm0.12$  & 41.2  & 0.0674 & \\
                    &$\eta_3$ & 23 & $2.6\sim 3.6$ & 3.1 & $4.47\pm0.12$  & 42.6  & 0.312  & 3.872\\
$cc\bar{u}\bar{d}$  &$\eta_4$ & 22 & $2.7\sim 3.4$ & 3.0 & $4.43\pm0.13$  & 38.4  & 0.0870 & \\
                    &$\eta_5$ & 23 & $2.5\sim 3.7$ & 3.2 & $4.41\pm0.14$  & 41.5  & 0.106  & \\
                    \hline
                    &$\eta_1$ & 24 & $2.9\sim 3.8$ & 3.4 & $4.45\pm0.16$  & 40.1 & 0.0489 & \\
                    &$\eta_2$ & 24 & $2.7\sim 3.7$ & 3.4 & $4.68\pm0.12$  & 43.1 & 0.106 & \\
$cc\bar{q}\bar{s}$  &$\eta_3$ & 26 & $3.0\sim 4.4$ & 3.8 & $4.71\pm0.14$  & 40.6 & 0.245 & 3.975\\
                    &$\eta_4$ & 25 & $2.6\sim 4.1$ & 3.4 & $4.64\pm0.13$  & 44.9 & 0.136 & \\
                    &$\eta_5$ & 24 & $2.6\sim 4.1$ & 3.4 & $4.50\pm0.16$  & 45.9 & 0.124 & \\
                    \hline
$cc\bar{s}\bar{s}$  &$\eta_1$ & 25 & $2.8\sim 4.0$ & 3.4 & $4.46\pm0.13$  & 44.3 & 0.0731 & 4.081 \\
                    &$\eta_3$ & 27 & $2.7\sim 4.3$ & 3.4 & $4.79\pm0.17$  & 47.7 & 0.558  & \\
                    \hline
$bb\bar{q}\bar{q}$  &$\eta_1$ & 125 & $7.0\sim 9.6$ & 8.0 & $10.6\pm0.3$  & 48.6 & 0.207 &\\
                    &$\eta_3$ & 120 & $6.8\sim 9.4$ & 8.0 & $10.5\pm0.3$  & 43.7 & 1.60  & 10.60\\
$bb\bar{u}\bar{d}$  &$\eta_5$ & 115 & $7.0\sim 8.1$ & 7.5 & $10.3\pm0.2$  & 36.5 & 0.367 & \\
                    \hline
                    &$\eta_1$ & 124 & $7.2\sim 9.6$ & 8.5 & $10.6\pm0.2$  & 40.8 & 0.188 &\\
$bb\bar{q}\bar{s}$  &$\eta_3$ & 120 & $7.2\sim 9.1$ & 8.0 & $10.6\pm0.2$  & 40.2 & 0.853 & 10.69\\
                    &$\eta_5$ & 115 & $7.0\sim 7.9$ & 7.5 & $10.4\pm0.2$  & 33.8 & 0.378 & \\
                    \hline
$bb\bar{s}\bar{s}$  &$\eta_1$ & 125 & $6.7\sim 9.6$ & 8.0 & $10.6\pm0.3$  & 47.9 & 0.286 &10.78\\
                    &$\eta_3$ & 120 & $6.6\sim 8.9$ & 8.0 & $10.6\pm0.3$  & 38.0 & 1.74  & \\
\hlinewd{.8pt}
\end{tabular}
\tabcaption{The numerical results for the doubly-charmed/bottomed $QQ\bar q\bar q$,
$QQ\bar q\bar s$ and $QQ\bar s\bar s$ systems with $J^P=0^-$.} \label{table1}
\end{center}

\begin{center}
\begin{tabular}{l|c|c|c|c|c|c|c|c}
\hlinewd{.8pt}
  &Current & $s_0$  & [$M^2_{Bmin},M^2_{Bmax}$] & $M^2_B$ & $m_X $ & PC & ~~$f_X$~~ & open charm/bottom \\
                    &  & ($\mbox{GeV}^2$) & ($\mbox{GeV}^2$) &($\mbox{GeV}^2$)& $(\mbox{GeV})$ &(\%)&($\mbox{GeV}^5$) &threshold $(\mbox{GeV})$\\
                    \hline
$cc\bar{q}\bar{s}$  &$\eta_2$ & 22 & $2.8\sim 3.6$ & 3.2 & $4.16\pm0.14$  & 39.0  & 0.0548 & 3.833\\
                    &$\eta_3$ & 20 & $2.6\sim 3.4$ & 3.0 & $4.02\pm0.18$   & 39.3 & 0.0561 & \\
                    \hline
$cc\bar{s}\bar{s}$  & $\eta_1$ & 28 & $3.2\sim4.1$ & 3.4 & $5.05\pm0.15$  & 43.3 & 0.136 & 3.937\\
                    &$\eta_2$ & 22 & $2.6\sim3.8$ & 3.2 & $4.27\pm0.11$ & 43.2  & 0.0933 & \\
                    \hline
                    & $\eta_2$ & 120 & $7.0\sim 9.8$ & 8.2 & $10.3\pm0.3$  & 48.2 & 0.590 & \\
$bb\bar{q}\bar{q}$  & $\eta_3$ & 115 & $6.9\sim 9.0$ & 8.0 & $10.2\pm0.3$  & 40.3 & 0.539 & 10.56\\
                    & $\eta_5$ & 115 & $6.7\sim 8.8$ & 8.0 & $10.2\pm0.3$  & 39.4& 1.10   & \\
                    \hline
                    & $\eta_3$ & 115 & $6.5\sim 8.8$ & 8.0 & $10.2\pm0.3$  & 40.3 & 0.398 & \\
$bb\bar{q}\bar{s}$  & $\eta_4$ & 115 & $5.8\sim 8.6$ & 7.2 & $10.2\pm0.3$  & 45.6 & 0.337 & 10.65\\
                    & $\eta_5$ & 120 & $6.2\sim 9.8$ & 8.0 & $10.3\pm0.3$  & 49.3 & 0.806 & \\
                    \hline
                    & $\eta_1$ & 130 & $7.5\sim 9.8$ & 8.5 & $11.0\pm0.2$ & 41.4  & 0.391 & \\
                    &$\eta_2$  & 120 & $6.4\sim 9.8$ & 8.0 & $10.4\pm0.3$ & 49.7  & 0.632 & \\
$bb\bar{s}\bar{s}$  &$\eta_3$ & 115 & $6.3\sim 9.0$ & 8.0 & $10.2\pm0.3$  & 40.5  & 0.560 & 10.73\\
                    & $\eta_4$ & 120 & $6.2\sim 8.4$ & 8.0 & $10.4\pm0.3$  & 41.9 & 0.486 & \\
                    & $\eta_5$ & 115 & $6.2\sim 8.8$ & 8.0 & $10.2\pm0.3$  & 38.9 & 1.14  & \\
\hlinewd{.8pt}
\end{tabular}
\tabcaption{The numerical results for the doubly-charmed/bottomed $QQ\bar q\bar q$,
$QQ\bar q\bar s$ and $QQ\bar s\bar s$ systems with $J^P=0^+$.} \label{table2}
\end{center}

\begin{center}
\begin{tabular}{l|c|c|c|c|c|c|c|c}
\hlinewd{.8pt}
  &Current & $s_0$  & [$M^2_{Bmin},M^2_{Bmax}$] & $M^2_B$ & $m_X $ & PC & ~~$f_X$~~ & open charm/bottom \\
                    &  & ($\mbox{GeV}^2$) & ($\mbox{GeV}^2$) &($\mbox{GeV}^2$)& $(\mbox{GeV})$ &(\%)&($\mbox{GeV}^5$) &threshold $(\mbox{GeV})$\\
                    \hline
$cc\bar{q}\bar{q}$  &$\eta_1$ & 23 & $3.0\sim3.6$ & 3.3 & $4.35\pm0.14$ & 38.6 & 0.0490  & \\
$cc\bar{u}\bar{d}$  &$\eta_6$ & 23 & $3.1\sim3.7$ & 3.4 & $4.34\pm0.16$ & 37.9 & 0.0395  & 3.730\\
                    &$\eta_7$ & 22 & $2.6\sim3.4$ & 3.0 & $4.41\pm0.12$ & 39.4 & 0.0690  & \\
                    &$\eta_8$ & 23 & $2.6\sim3.5$ & 3.0 & $4.42\pm0.14$ & 41.1 & 0.0940  & \\
                    \hline
                    &$\eta_1$ & 23 & $2.7\sim3.6$ & 3.2 & $4.37\pm0.17$ & 39.1 & 0.0357 &\\
                    &$\eta_2$ & 24 & $2.9\sim3.8$ & 3.2 & $4.59\pm0.13$ & 43.0 & 0.0838 & \\
                    &$\eta_6$ & 23 & $2.9\sim3.7$ & 3.4 & $4.35\pm0.16$ & 37.7 & 0.0353 & 3.833\\
$cc\bar{q}\bar{s}$  &$\eta_7$ & 24 & $2.4\sim3.9$ & 3.4 & $4.58\pm0.14$ & 39.8 & 0.105  & \\
                    &$\eta_8$ & 24 & $2.4\sim3.9$ & 3.4 & $4.52\pm0.13$ & 41.1 & 0.114  & \\
                    \hline
                    &$\eta_1$ & 24 & $2.8\sim3.7$ & 3.3 & $4.47\pm0.13$  & 40.7 & 0.0603 & \\
$cc\bar{s}\bar{s}$  &$\eta_3$ & 23 & $2.5\sim3.5$ & 3.0 & $4.47\pm0.14$  & 40.9 & 0.101  & 3.937\\
                    &$\eta_4$ & 26 & $2.8\sim4.2$ & 3.3 & $4.74\pm0.17$ & 49.1  & 0.196  & \\
                    \hline
$bb\bar{q}\bar{q}$  &$\eta_1$ & 125 & $7.0\sim 9.6$ & 8.0 & $10.6\pm0.3$  & 47.8 & 0.229  & \\
$bb\bar{u}\bar{d}$  &$\eta_6$ & 120 & $7.2\sim 8.9$ & 8.0 & $10.4\pm0.2$  & 40.5 & 0.142  & 10.56\\
                    &$\eta_8$ & 120 & $8.2\sim 9.4$ & 8.8 & $10.5\pm0.2$  & 35.9 & 0.492  & \\
                    \hline
                    &$\eta_1$ & 120 & $7.2\sim 8.8$ & 8.0 & $10.5\pm0.2$  & 37.9 & 0.124 & \\
$bb\bar{q}\bar{s}$  &$\eta_6$ & 120 & $7.2\sim 8.9$ & 8.0 & $10.4\pm0.2$  & 40.6 & 0.145 & 10.65\\
                    &$\eta_8$ & 120 & $7.6\sim 9.3$ & 8.4 & $10.5\pm0.2$  & 37.8 & 0.491 & \\
                    \hline
                    & $\eta_1$ & 125 & $6.6\sim 9.6$ & 8.0 & $10.6\pm0.3$  & 47.1 & 0.240 & \\
$bb\bar{s}\bar{s}$  & $\eta_3$ & 120 & $6.7\sim 9.0$ & 8.0 & $10.5\pm0.3$  & 40.1 & 0.490 & 10.73\\
                    & $\eta_4$ & 120 & $6.8\sim 8.9$ & 8.0 & $10.6\pm0.3$  & 38.8 & 0.655 & \\
\hlinewd{.8pt}
\end{tabular}
\tabcaption{The numerical results for the doubly-charmed/bottomed $QQ\bar q\bar q$,
$QQ\bar q\bar s$ and $QQ\bar s\bar s$ systems with $J^P=1^-$.} \label{table3}
\end{center}

\begin{center}
\begin{tabular}{l|c|c|c|c|c|c|c|c}
\hlinewd{.8pt}
  &Current & $s_0$  & [$M^2_{Bmin},M^2_{Bmax}$] & $M^2_B$ & $m_X $ & PC & ~~$f_X$~~ & open charm/bottom \\
                    &  & ($\mbox{GeV}^2$) & ($\mbox{GeV}^2$) &($\mbox{GeV}^2$)& $(\mbox{GeV})$ &(\%)&($\mbox{GeV}^5$) &threshold $(\mbox{GeV})$\\
                    \hline
                    & $\eta_1$ & 28 & $3.0\sim 4.2$ & 3.6 & $4.96\pm0.11$  & 42.1  & 0.0801 &\\
                    & $\eta_2$ & 27 & $3.1\sim 4.0$ & 3.6 & $4.87\pm0.11$  & 38.5  & 0.0726 & \\
$cc\bar{q}\bar{s}$  & $\eta_3$ & 21 & $2.4\sim 3.4$ & 2.8 & $4.12\pm0.17$  & 47.5  & 0.0571 & \\
                    & $\eta_4$ & 21 & $2.5\sim 3.4$ & 2.8 & $4.13\pm0.16$  & 47.9  & 0.0574 & 3.975\\
                    & $\eta_5$ & 21 & $2.8\sim3.7$ & 3.2 & $4.12\pm0.16 $  & 41.7  & 0.0378 & \\
                    & $\eta_6$ & 21 & $3.0\sim3.7$ & 3.2 & $4.17\pm0.12 $  & 41.5  & 0.0718 & \\
                    & $\eta_7$ & 21 & $2.2\sim3.3$ & 2.8 & $4.15\pm0.17 $  & 42.9  & 0.0465 & \\
                    \hline
                    & $\eta_1$ & 29  & $3.2\sim4.5$ & 3.8 & $5.03\pm0.13$  & 42.5  & 0.138  &\\
                    & $\eta_2$ & 30  & $3.2\sim4.6$ & 3.8 & $5.12\pm0.14$  & 45.9  & 0.150  & 4.081\\
$cc\bar{s}\bar{s}$  & $\eta_3$ & 21  & $2.2\sim3.4$ & 2.8 & $4.17\pm0.16$  & 45.4  & 0.0838 & \\
                    & $\eta_4$ & 21  & $2.2\sim3.4$ & 2.8 & $4.19\pm0.16$  & 45.7  & 0.0849 & \\
                    \hline
$bb\bar{q}\bar{q}$  & $\eta_3$ & 115 & $6.5\sim 8.8$ & 7.8 & $10.2\pm0.3$  & 41.4  & 0.459  &\\
                    & $\eta_4$ & 115 & $6.8\sim 8.8$ & 7.8 & $10.2\pm0.3$  & 41.7  & 0.454  & \\
$bb\bar{u}\bar{d}$  & $\eta_5$ & 115 & $7.0\sim 9.0$ & 8.0 & $10.2\pm0.3$  & 42.8  & 0.215  & 10.60\\
                    & $\eta_6$ & 115 & $7.0\sim 9.2$ & 8.0 & $10.2\pm0.3$  & 42.0  & 0.304  & \\
                    & $\eta_7$ & 115 & $6.5\sim 8.6$ & 7.6 & $10.2\pm0.3$  & 43.2  & 0.241  & \\
                    & $\eta_8$ & 115 & $6.8\sim 8.8$ & 7.6 & $10.2\pm0.3$  & 41.7  & 0.343  & \\
                    \hline
                    & $\eta_1$ & 125 & $6.9\sim 8.6$ & 7.6 & $10.7\pm0.3$  & 42.1  & 0.155  &\\
                    & $\eta_2$ & 125 & $6.9\sim 8.8$ & 7.6 & $10.7\pm0.4$  & 44.5  & 0.170  & \\
$bb\bar{q}\bar{s}$  & $\eta_3$ & 120 & $6.2\sim 9.8$ & 8.0 & $10.4\pm0.3$  & 48.9  & 0.452  & \\
                    & $\eta_4$ & 120 & $6.5\sim 9.8$ & 8.0 & $10.4\pm0.3$  & 49.3  & 0.446  & 10.69\\
                    & $\eta_5$ & 120 & $6.6\sim 9.8$ & 8.0 & $10.3\pm0.3$  & 52.3  & 0.298  & \\
                    & $\eta_6$ & 120 & $6.6\sim 9.8$ & 8.0 & $10.3\pm0.4$  & 52.1  & 0.418  & \\
                    & $\eta_7$ & 120 & $6.2\sim 9.6$ & 8.0 & $10.4\pm0.3$  & 48.1  & 0.342  & \\
                    & $\eta_8$ & 120 & $5.8\sim 9.6$ & 8.0 & $10.4\pm0.3$  & 46.3  & 0.491  & \\
                    \hline
                    & $\eta_1$ & 130 & $7.0\sim 9.7$ & 8.5 & $11.0\pm0.3$  & 40.8  & 0.336 &\\
                    & $\eta_2$ & 130 & $7.2\sim 9.9$ & 8.5 & $10.9\pm0.3$  & 42.9  & 0.370 & 10.78\\
$bb\bar{s}\bar{s}$  & $\eta_3$ & 120 & $6.2\sim 9.8$ & 8.0 & $10.4\pm0.3$  & 48.1  & 0.657 & \\
                    & $\eta_4$ & 120 & $6.2\sim 9.8$ & 8.0 & $10.4\pm0.3$  & 48.5  & 0.651 & \\
\hlinewd{.8pt}
\end{tabular}
\tabcaption{The numerical results for the doubly-charmed/bottomed $QQ\bar q\bar q$,
$QQ\bar q\bar s$ and $QQ\bar s\bar s$ systems with $J^P=1^+$.} \label{table4}
\end{center}


\section{Decay Patterns of the $QQ\bar q\bar q$ and $QQ\bar s\bar s$ States}\label{sec:decay}

In this section, we study the decay patterns of the possible
doubly charmed/bottomed states. From the numerical results of
Tables~\ref{table1}$-$\ref{table4}, the extracted masses of the
$cc\bar{q}\bar{q}$, $cc\bar{q}\bar{s}$, and $cc\bar{s}\bar{s}$
doubly charmed states are above the
$D^{(*)(+/0)}_{(0/1)}D^{(*)(+/0)}_{(0/1)}$,
$D^{(*)+}_{(0/1)}D^{(*)+}_{s(0/1)}$, and
$D^{(*)+}_{s(0/1)}D^{(*)+}_{s(0/1)}$ thresholds. The possible
decay modes of the $cc\bar{q}\bar{q}$, $cc\bar{q}\bar{s}$, and
$cc\bar{s}\bar{s}$ states with different quantum numbers are
listed in Table~\ref{decay}. Both the S-wave and P-wave decay
patterns are allowed. These $cc\bar{q}\bar{q}$, $cc\bar{q}\bar{s}$,
and $cc\bar{s}\bar{s}$ tetraquarks will decay easily through
rearrangement or the so-called fall-apart mechanism. They
are very broad resonances. It may be difficult to observe them
experimentally.

The situation is very different for the doubly bottomed states. As
we emphasize in the previous section, the $\eta_1^s$ current with
$J^P=0^+$ and the $\eta_1^s, \eta_2^s$ currents with $J^P=1^+$
explore the excited tetraquark states because of their special
diquark structure. We focus on the doubly bottomed ground states.
The extracted masses of these states as shown in
Tables~\ref{table1}$-$\ref{table4} are lower than the open bottom
thresholds $\bar{B^0}\bar{B^0}=10.56$ GeV and
$\bar{B^0_s}\bar{B^0_s}=10.73$ GeV. These doubly bottomed states
cannot decay into the two $b\bar q$ or $b\bar s$ mesons, which
implies the existence of the doubly bottomed bound states
$bb\bar{q}\bar{q}$ and $bb\bar{s}\bar{s}$. This observation is
consistent with the conclusions of
Refs.~\cite{1988-Carlson-p744-744,1993-Manohar-p17-33,2008-Zhang-p437-440}.

\begin{center}
\renewcommand{\arraystretch}{1.3}
\begin{tabular*}{16cm}{c|c|c}
\hlinewd{.8pt}
  $J^P$ & S-wave & P-wave \\
  \hline
        & $D^0+D_0^*(2400)^0$,$D^++D_0^*(2400)^0$, & $D^0+D^*(2007)^0$,$D^++D^*(2007)^0$, $D^0+D^*(2010)^+$, \\
  $0^-$ & $D^0+D_{s0}^*(2317)^+$,$D^++D_{s0}^*(2317)^+$,&$D^++D^*(2010)^+$, $D^*(2007)^0+D^*(2007)^0$,  \\
        & $D_0^*(2400)^0+D_s^+$,$D^*(2007)^0+D_{s1}^+$, &$D^*(2010)^++D^*(2010)^+$,$D^*(2007)^0+D^*(2010)^+$,\\
        &$D_1(2420)^0+D^*_s(2112)^+$,$D_{s0}^*(2317)^++D_s^+$,&$D^*(2007)^0+D^+_s$,$D^*(2010)^++D^+_s$,$D^++D^*_s(2112)^+$\\
        &\dots&$D^+_s+D^*_s(2112)^+$,$D^*_s(2112)^++D^*_s(2112)^+$,\dots\\
  \hline
        & $D^++D_s^+$,$D^0+D_s^+$,$D_s^++D_s^+$, &    $D_{s0}^*(2317)^++D^*_s(2112)^+$,    \\
  $0^+$ & $D_s^{*+}+D_s^{*+}$,$D_{s0}^{*+}+D_{s0}^{*+}$, &$D_s^++D_{s1}(2460)^+$,$D_s^++D_{s1}(2536)^+$,\\
        & $D_{s1}^{+}(2460)+D_{s1}^{+}(2460)$,\dots   &     \dots  \\
  \hline
        & $D^0+D_1^0$,$D^++D_1^0$,&$D^0+D^0$,$D^++D^+$,$D^0+D^+$,$D^{*0}+D^{*0}$,$D^{*+}+D^{*+}$,\\
  $1^-$ & $D^{*0}+D_{s0}^{*+}$,$D^{*+}+D_{s0}^{*+}$,&$D^0+D^{*0}$,$D^0+D^{*+}$,$D^++D^{*0}$,$D^++D^{*+}$,\\
        &  $D_s^++D_{s1}^+$,     &$D^++D_s^{*+}$,$D^0+D_s^{*+}$,$D^{*0}+D_s^{+}$,$D^{*+}+D_s^{+}$,\\
        & \dots   &$D_s^++D_s^+$,$D_s^{*+}+D_s^{*+}$,\dots \\
  \hline
        & $D^++D^{*+}_s$,$D^0+D^{*+}_s$,$D^{*0}+D_s^+$,&$D^0+D_{s0}^{*+}$,$D^++D_{s0}^{*+}$,$D_0^{*0}+D_s^+$,\\
        & $D^{*+}+D_s^+$,$D_{s}^++D^{*+}_s$,&$D^{*0}+D_{s1}^+$,$D^{*+}+D_{s1}^+$,$D_1^0+D^{*+}_s$,\\
  $1^+$ & $D_{s0}^{*+}+D_{s1}^+$,$D_{s0}^{*+}+D_{s1}^+$,$D_{s1}^++D_{s1}^+$,\dots&$D_s^++D_{s0}^{*+}$, $D_{s0}^{*+}+D^{*+}_s$,$D_s^++D_{s1}^+$,  $D_s^++D_{s1}^+$,\dots\\
\hlinewd{.8pt}
\end{tabular*}
\tabcaption{The possible two-meson decay modes of the
$cc\bar{q}\bar{q}$ and $cc\bar{s}\bar{s}$ states.} \label{decay}
\end{center}

Except for the two-meson decay modes, these
doubly charmed/bottomed tetraquark states may also decay into the
two-baryon final states: a pair of bottom and antibottom baryons
or one doubly bottomed baryon plus one antinucleon. The lightest
bottom baryon is $\Lambda_b$. Its mass is 5.62 GeV. In other
words, these doubly bottomed tetraquark states do not decay into a
pair of bottom and antibottom baryons.

The mass of the doubly charmed baryons discovered by the SELEX
Collaboration is $m_{\Xi_{cc}}=3519\pm1$ MeV. For the other
doubly charmed/bottomed baryons, their masses have been estimated
in the quark model, $m_{\Omega_{cc}}=3.8$ GeV, $m_{\Xi_{bb}}=10.2$
GeV \cite{1994-Bagan-p57-72,1996-Moinester-p349-362}. The possible
two-baryon decay patterns of the doubly charmed tetraquark states
are listed in Table~\ref{decay1}. In contrast, the doubly bottomed
$bb\bar q\bar q$ states cannot decay into $\bar N\Xi_{bb}$
because their masses in Tables~\ref{table1}$-$\ref{table4} are below
$m_{\bar N}+m_{\Xi_{bb}}$. The above analysis also supports the
existence of the doubly bottomed tetraquark states.

\begin{center}
\renewcommand{\arraystretch}{1.5}
\begin{tabular*}{6cm}{ccc}
\hlinewd{.8pt}
  $J^P$ &   ~~~~~~ S-wave   ~~~~~~        &   ~~~~~~    P-wave ~~~~~~\\
  \hline
  $0^-$ &                           &  $\bar N+\Xi_{cc}$ \\
  $0^+$ &   $\bar N+\Omega_{cc}$    &                    \\
  $1^-$ &                           &   -                 \\
  $1^+$ &   $\bar N+\Omega_{cc}$     &                    \\
\hlinewd{.8pt}
\end{tabular*}
\tabcaption{The possible two-baryon decay modes of the
$cc\bar{q}\bar{q}$ and $cc\bar{s}\bar{s}$ states.} \label{decay1}
\end{center}

\section{Summary}\label{sec:summary}

In order to explore the possible $QQ\bar{q}\bar{q}$,
$QQ\bar{q}\bar{s}$, and $QQ\bar{s}\bar{s}$ states with
$J^P=0^-,0^+,1^-$, and $1^+$, we have constructed the possible
tetraquark interpolating operators without derivatives in a
systematic way. Because of Fermi statistics, the wave functions of
$(QQ)$ and $\bar{q}\bar{q}$ should be antisymmetric (color
$\times$ flavor $\times$ orbital $\times$ spin). We obtain 26
color-singlet interpolating currents with these quantum numbers,
in which 16 currents possess symmetric light quark flavor
structure $[\mathbf{ \bar 6_f}]_{\bar{q}\bar q}$ and the other 10
currents belong to $[\mathbf{ 3_f}]_{\bar{q}\bar q}$. The
properties of these currents, such as the isospins, the flavor
structures, and the $J^P$ quantum numbers, are summarized in
Table~\ref{table01}.

Then we make the operator product expansion and extract the
spectral densities for every interpolating current. Because of the
special Lorentz structures of the currents, the quark condensate
$\langle\bar{q}q\rangle$ and $\langle \bar{q}g_s\sigma \cdot G
q\rangle$ vanishes for $QQ\bar{q}\bar{q}$ systems. For the
$QQ\bar{q}\bar{s}$ and $QQ\bar{s}\bar{s}$ systems, we keep the
$m_s$-related terms in the spectral densities. These terms give
important contributions to the correlation functions. Now the most
important corrections come from the quark condensate and the four-quark condensate.

In the working range of the Borel parameter, only $0^-$ and $1^-$
$cc\bar{q}\bar{q}$ systems give a stable mass sum rule. The masses
of the possible $0^-$ and $1^-$ $cc\bar{q}\bar{q}$ states are
$4.45\pm 0.12$ GeV and $4.35\pm 0.14$ GeV. There does not exist a
stable mass sum rule for the $0^+$ and $1^+$ $cc\bar{q}\bar{q}$
systems. The QCD sum rules of the tetraquark systems become more
stable as the quark mass increases. According to our analysis,
stable QCD sum rules exist for the following channels:
$cc\bar{q}\bar{s}$, $cc\bar{s}\bar{s}$, $bb\bar{q}\bar{q}$,
$bb\bar{q}\bar{s}$, and $bb\bar{s}\bar{s}$.

Unfortunately, the doubly charmed tetraquark states are found to
lie above the two-meson threshold. These states will decay very
rapidly through the fall-apart mechanism. Very probably they may
be too broad to be identified as a resonance experimentally. In
contrast, it's very interesting to note that the masses of the
doubly bottomed tetraquark states are below the open bottom
thresholds and the $\bar N+\Omega_{bb}$ threshold. In other words,
the tetraquark states $bb\bar{q}\bar{q}$, $bb\bar{q}\bar{s}$, and
$bb\bar{s}\bar{s}$ are stable. Once produced, they decay via
electromagnetic and weak interactions only. The
$bb\bar{q}\bar{q}$, $bb\bar{q}\bar{s}$, and $bb\bar{s}\bar{s}$
states may be searched for at facilities such as LHCb and RHIC in
the future, where plenty of heavy quarks are produced~\cite{Cho:2010db}.

\section*{Acknowledgments}

This project was supported by the National Natural Science
Foundation of China under Grants No. 11075004 and No. 11021092 and by the Ministry
of Science and Technology of China (Grant No. 2009CB825200). This work is
also supported in part by the DFG and the NSFC through funds
provided to the Sino-Germen CRC 110, ``Symmetries and the Emergence
of Structure in QCD.''





\appendix

\section{The Spectral Densities}\label{sec:rhos}
In this appendix, we list the spectral densities of the tetraquark
interpolating currents with different quantum numbers,
respectively. The spectral densities read
\begin{equation}
\rho^{OPE}(s)=\rho^{pert}(s)+\rho^{\langle\bar{q}q\rangle}(s)+\rho^{\langle
GG\rangle}(s)+\rho^{\langle\bar{q}Gq\rangle}(s)+\rho^{\langle
\bar{q}q\rangle^2}(s).
\end{equation}

The Borel transformation of the correlation functions reads
\begin{equation}
\Pi(M_B^2)=\int_{4(m_Q+m_q)^2}^{\infty}ds\rho^{OPE}(s)e^{-s/M_B^2}+\Pi^{\langle\bar{q}q\rangle\langle\bar{q}g_s\sigma\cdot
Gq\rangle}(M_B^2).
\end{equation}

For the interpolating currents with symmetric light quark flavor
structure $[\mathbf{ \bar 6_f}]_{\bar{q}\bar q}$, there are two types
of Wick contraction when we calculate the two-point correlations
for the $QQ\bar q\bar q$ and $QQ\bar q\bar s$ systems, as
mentioned in Sec.~\ref{sec:current} and Sec.~\ref{sec:num}. We
list the spectral densities for both of them. For the currents
with antisymmetric light quark flavor structure $[\mathbf{
3_f}]_{\bar{q}\bar q}$, we just list the spectral densities for
the $QQ\bar q\bar s$ systems while keeping the $m_s$-related
terms. The spectral densities for the $QQ\bar s\bar s$ and $QQ\bar
u\bar d$ systems can be obtained from the expressions of the
$QQ\bar q\bar q$ and $QQ\bar q\bar s$ systems, respectively, by
replacing the corresponding parameters.

\subsection{The spectral densities for the currents in the $QQ\bar q\bar q$ systems}
For the interpolating currents with $J^P=0^-$,
\begin{equation}
\begin{split}
\rho^{pert}_1(s)=&\int_{\alpha_{min}}^{\alpha_{max}}d\alpha\int_{\beta_{min}}^{\beta_{max}}d\beta
\frac{(1-\alpha -\beta ) \left(m^2 (\alpha +\beta -2)+\alpha  \beta  s\right) \left(\alpha  \beta  s-m^2 (\alpha +\beta )\right)^3}{64 \pi ^6 \alpha ^3 \beta ^3},\\
\rho^{\langle\bar{q} q\rangle}_1(s)=&-m_q\langle\bar{q} q\rangle
\int_{\alpha_{min}}^{\alpha_{max}}d\alpha\int_{\beta_{min}}^{\beta_{max}}d\beta
\frac{\left(m^2 (\alpha +\beta )-\alpha  \beta  s\right) \left(m^2 (\alpha +\beta +1)-2 \alpha  \beta  s\right)}{2 \pi ^4 \alpha  \beta },\\
\rho^{\langle GG\rangle}_1(s)=&\langle g_s^2
GG\rangle\int_{\alpha_{min}}^{\alpha_{max}}d\alpha\int_{\beta_{min}}^{\beta_{max}}d\beta
\Big{ \{ } \frac{m^2(1-\alpha-\beta)^2\big( 2m^2(\alpha+\beta)+m^2-3\alpha\beta s\big)}{192\pi^6 \alpha^3}+\\
&(m^2(\alpha+\beta)-\alpha\beta
s)\Big[\frac{(1-\alpha-\beta)^2\big(2\alpha^2\beta
s-m^2(\alpha(\alpha+\beta+2)-8\beta)\big)}{512
\pi^6 \alpha^3\beta^2}-\frac{m^2(\alpha+\beta+1)-2\alpha \beta s}{256\pi^6\alpha \beta}\Big] \Big{\}},\\
\rho^{\langle \bar{q}Gq\rangle}_1(s)=&-\frac{m_q\langle\bar{q}\sigma\cdot G q\rangle (s-4m^2)}{8\pi^4}\sqrt{1-\frac{4m^2}{s}},\\
\rho^{\langle\bar{q}q\rangle^2}_1(s)=&\frac{\langle\bar{q}q\rangle^2 (s-4m^2)}{3\pi^2}\sqrt{1-\frac{4m^2}{s}},\\
\Pi_1^{\langle\bar{q}q\rangle\langle
\bar{q}Gq\rangle}(M_B^2)=&\frac{\langle\bar{q}q\rangle\langle
\bar{q}\sigma \cdot Gq\rangle}{3\pi^2} \int^1_0dx\Big{\{ }
\frac{m^4(2x-1)}{M_B^2(1-x)x^2}+\frac{m^2}{x(1-x)}+M_B^2\Big{\}}e^{-\frac{m^2}{M^2_B(1-x)x}}.
\end{split}
\end{equation}

\begin{equation}
\begin{split}
\rho^{pert}_2(s)=&\int_{\alpha_{min}}^{\alpha_{max}}d\alpha\int_{\beta_{min}}^{\beta_{max}}d\beta
\frac{(1-\alpha-\beta)\big( m^2(\alpha+\beta)-\alpha \beta s\big)^3\big(m^2(3\alpha+3\beta-2)-\alpha\beta s\big)}{64\pi^6\alpha^3\beta^3},\\
\rho^{\langle\bar{q} q\rangle}_2(s)=&m_q\langle\bar{q} q\rangle
\int_{\alpha_{min}}^{\alpha_{max}}d\alpha\int_{\beta_{min}}^{\beta_{max}}d\beta
\frac{3\big( m^2(\alpha+\beta-1)-2\alpha \beta s\big)\big( m^2(\alpha+\beta)-\alpha \beta s\big)}{2\pi^4\alpha\beta},\\
\rho^{\langle GG\rangle}_2(s)=&\langle g_s^2
GG\rangle\int_{\alpha_{min}}^{\alpha_{max}}d\alpha\int_{\beta_{min}}^{\beta_{max}}d\beta
\Big{ \{ } \frac{m^2(1-\alpha-\beta)^2\big( m^2(2\alpha+2\beta-1)+m^2-3\alpha\beta s\big)}{192\pi^6 \alpha^3}+\\
&(m^2(\alpha+\beta)-\alpha\beta
s)\Big[\frac{(1-\alpha-\beta)^2\big(2\alpha^2\beta
s-m^2(\alpha(\alpha+\beta-2)+8\beta)\big)}{512
\pi^6 \alpha^3\beta^2}-\frac{m^2(\alpha+\beta-1)-2\alpha \beta s}{256\pi^6\alpha \beta}\Big] \Big{\}},\\
\rho^{\langle \bar{q}Gq\rangle}_2(s)=&\frac{m_q\langle\bar{q}\sigma\cdot G q\rangle s}{8\pi^4}\sqrt{1-\frac{4m^2}{s}},\\
\rho^{\langle\bar{q}q\rangle^2}_2(s)=&-\frac{\langle\bar{q}q\rangle^2 s}{3\pi^2}\sqrt{1-\frac{4m^2}{s}},\\
\Pi_2^{\langle\bar{q}q\rangle\langle
\bar{q}Gq\rangle}(M_B^2)=&-\frac{\langle\bar{q}q\rangle\langle
\bar{q}\sigma \cdot Gq\rangle}{3\pi^2} \int^1_0dx\Big{\{ }
\frac{m^4}{M_B^2(1-x)x^2}+\frac{m^2}{x(1-x)}+M_B^2\Big{\}}e^{-\frac{m^2}{M^2_B(1-x)x}}.
\end{split}
\end{equation}

\begin{equation}
\begin{split}
\rho^{pert}_3(s)=&\int_{\alpha_{min}}^{\alpha_{max}}d\alpha\int_{\beta_{min}}^{\beta_{max}}d\beta
\frac{3(1-\alpha-\beta)\big( m^2(\alpha+\beta)-\alpha \beta s\big)^4}{16\pi^6\alpha^3\beta^3},\\
\rho^{\langle\bar{q} q\rangle}_3(s)=&m_q\langle\bar{q} q\rangle
\int_{\alpha_{min}}^{\alpha_{max}}d\alpha\int_{\beta_{min}}^{\beta_{max}}d\beta
\frac{6\big( m^2(\alpha+\beta-2)-2\alpha \beta s\big)\big( m^2(\alpha+\beta)-\alpha \beta s\big)}{\pi^4\alpha\beta},\\
\rho^{\langle GG\rangle}_3(s)=&\langle g_s^2
GG\rangle\int_{\alpha_{min}}^{\alpha_{max}}d\alpha\int_{\beta_{min}}^{\beta_{max}}d\beta
\Big{\{} \frac{m^2(1-\alpha-\beta)^2\big( 2m^2(\alpha+\beta)-3\alpha\beta s\big)}{16\pi^6\alpha^3}\\
&-\frac{\big(\alpha(\alpha+12\beta-2)+\beta(9\beta-10)+1\big)\big(m^2(\alpha+\beta)-2\alpha \beta s\big)\big(m^2(\alpha+\beta)-\alpha \beta s\big)}{64\pi^6\alpha^2\beta^2}\Big{\}},\\
\rho^{\langle
\bar{q}Gq\rangle}_3(s)=&\frac{m_q\langle\bar{q}\sigma\cdot G
q\rangle}{\pi^4} \Big{\{}3m^2\sqrt{1-\frac{4m^2}{s}}
-\int_{\alpha_{min}}^{\alpha_{max}}d\alpha\int_{\beta_{min}}^{\beta_{max}}d\beta
\frac{2m^2}{ \alpha}\Big{\}},\\
\rho^{\langle\bar{q}q\rangle^2}_3(s)=&-\frac{8\langle\bar{q}q\rangle^2 m^2}{\pi^2}\sqrt{1-\frac{4m^2}{s}},\\
\Pi_3^{\langle\bar{q}q\rangle\langle
\bar{q}Gq\rangle}(M_B^2)=&-\frac{4m^2\langle\bar{q}q\rangle\langle
\bar{q}\sigma \cdot Gq\rangle}{3\pi^2} \int^1_0dx\Big{\{ }
\frac{3m^2-2xM_B^2}{x^2M_B^2 }\Big{\}
}e^{-\frac{m^2}{M^2_B(1-x)x}}.
\end{split}
\end{equation}

For the interpolating currents with $J^P=0^+$,
\begin{equation}
\begin{split}
\rho^{pert}_1(s)=&\int_{\alpha_{min}}^{\alpha_{max}}d\alpha\int_{\beta_{min}}^{\beta_{max}}d\beta
\frac{(1-\alpha-\beta)\big( m^2(\alpha+\beta-2)+\alpha\beta s\big)\big( \alpha \beta s-m^2(\alpha+\beta)\big)^3}{64\pi^6\alpha^3\beta^3},\\
\rho^{\langle\bar{q} q\rangle}_1(s)=&3m_q\langle\bar{q} q\rangle
\int_{\alpha_{min}}^{\alpha_{max}}d\alpha\int_{\beta_{min}}^{\beta_{max}}d\beta
\frac{\big( m^2(\alpha+\beta)-\alpha\beta s\big)\big( m^2(\alpha+\beta+1)-2\alpha\beta s\big)}{2\pi^4\alpha\beta},\\
\rho^{\langle GG\rangle}_1(s)=&\langle g_s^2
GG\rangle\int_{\alpha_{min}}^{\alpha_{max}}d\alpha\int_{\beta_{min}}^{\beta_{max}}d\beta
\Big{ \{ } \frac{m^2(1-\alpha-\beta)^2\big( 2m^2(\alpha+\beta)+m^2-3\alpha\beta s\big)}{192\pi^6 \alpha^3}+\\
&(m^2(\alpha+\beta)-\alpha\beta
s)\Big[\frac{(1-\alpha-\beta)^2\big(2\alpha^2\beta
s-m^2(\alpha(\alpha+\beta+2)-8\beta)\big)}{512
\pi^6 \alpha^3\beta^2}-\frac{m^2(\alpha+\beta+1)-2\alpha \beta s}{256\pi^6\alpha \beta}\Big] \Big{\}},\\
\rho^{\langle \bar{q}Gq\rangle}_1(s)=&\frac{m_q\langle\bar{q}\sigma\cdot G q\rangle (s-4m^2)}{8\pi^4}\sqrt{1-\frac{4m^2}{s}},\\
\rho^{\langle\bar{q}q\rangle^2}_1(s)=&\frac{\langle\bar{q}q\rangle^2 (s-4m^2)}{3\pi^2}\sqrt{1-\frac{4m^2}{s}},\\
\Pi_1^{\langle\bar{q}q\rangle\langle
\bar{q}Gq\rangle}(M_B^2)=&-\frac{\langle\bar{q}q\rangle\langle
\bar{q}\sigma \cdot Gq\rangle}{3\pi^2} \int^1_0dx\Big{\{ }
\frac{m^4(2x-1)}{M_B^2(1-x)x^2}+\frac{m^2}{x(1-x)}+M_B^2\Big{\}}e^{-\frac{m^2}{M^2_B(1-x)x}}.
\end{split}
\end{equation}

\begin{equation}
\begin{split}
\rho^{pert}_2(s)=&\int_{\alpha_{min}}^{\alpha_{max}}d\alpha\int_{\beta_{min}}^{\beta_{max}}d\beta
\frac{(1-\alpha-\beta)\big( m^2(\alpha+\beta)-\alpha \beta s\big)^3\big(m^2(3\alpha+3\beta-2)-\alpha\beta s\big)}{64\pi^6\alpha^3\beta^3},\\
\rho^{\langle\bar{q} q\rangle}_2(s)=&-m_q\langle\bar{q} q\rangle
\int_{\alpha_{min}}^{\alpha_{max}}d\alpha\int_{\beta_{min}}^{\beta_{max}}d\beta
\frac{\big( m^2(\alpha+\beta-1)-2\alpha \beta s\big)\big( m^2(\alpha+\beta)-\alpha \beta s\big)}{2\pi^4\alpha\beta},\\
\rho^{\langle GG\rangle}_2(s)=&\langle g_s^2
GG\rangle\int_{\alpha_{min}}^{\alpha_{max}}d\alpha\int_{\beta_{min}}^{\beta_{max}}d\beta
\Big{ \{ } \frac{m^2(1-\alpha-\beta)^2\big( m^2(2\alpha+2\beta-1)+m^2-3\alpha\beta s\big)}{192\pi^6 \alpha^3}+\\
&(m^2(\alpha+\beta)-\alpha\beta
s)\Big[\frac{(1-\alpha-\beta)^2\big(2\alpha^2\beta
s-m^2(\alpha(\alpha+\beta-2)+8\beta)\big)}{512
\pi^6 \alpha^3\beta^2}-\frac{m^2(\alpha+\beta-1)-2\alpha \beta s}{256\pi^6\alpha \beta}\Big] \Big{\}},\\
\rho^{\langle \bar{q}Gq\rangle}_2(s)=&-\frac{m_q\langle\bar{q}\sigma\cdot G q\rangle s}{8\pi^4}\sqrt{1-\frac{4m^2}{s}},\\
\rho^{\langle\bar{q}q\rangle^2}_2(s)=&\frac{\langle\bar{q}q\rangle^2 s}{3\pi^2}\sqrt{1-\frac{4m^2}{s}},\\
\Pi_2^{\langle\bar{q}q\rangle\langle
\bar{q}Gq\rangle}(M_B^2)=&\frac{\langle\bar{q}q\rangle\langle
\bar{q}\sigma \cdot Gq\rangle}{3\pi^2} \int^1_0dx\Big{\{ }
\frac{m^4}{M_B^2(1-x)x^2}+\frac{m^2}{x(1-x)}+M_B^2\Big{\}}e^{-\frac{m^2}{M^2_B(1-x)x}}.
\end{split}
\end{equation}

\begin{equation}
\begin{split}
\rho^{pert}_3(s)=&\int_{\alpha_{min}}^{\alpha_{max}}d\alpha\int_{\beta_{min}}^{\beta_{max}}d\beta
\frac{(1-\alpha-\beta)\big(m^2(2\alpha+2\beta-1)-\alpha\beta s\big)\big(m^2(\alpha+\beta)-\alpha\beta s\big)}{32\pi^6\alpha^3\beta^3},\\
\rho^{\langle\bar{q} q\rangle}_3(s)=&m_q\langle\bar{q} q\rangle
\int_{\alpha_{min}}^{\alpha_{max}}d\alpha\int_{\beta_{min}}^{\beta_{max}}d\beta
\frac{3m^2\big(m^2(\alpha+\beta)-\alpha\beta s\big)}{2\pi^4\alpha\beta},\\
\rho^{\langle GG\rangle}_3(s)=&\langle g_s^2
GG\rangle\int_{\alpha_{min}}^{\alpha_{max}}d\alpha\int_{\beta_{min}}^{\beta_{max}}d\beta
\Big{ \{ }
\frac{m^2(1-\alpha-\beta)^2\big(m^2(4\alpha\beta-3\alpha+4\beta^2-4\beta)+3\alpha(1-2\beta)\beta
s\big)}
{192\pi^6\alpha^3\beta}\\
&+\big(m^2(\alpha+\beta)-\alpha\beta s\big)\Big[ \frac{(1-\alpha-\beta)\big(m^2(\alpha+\beta-1)-2\alpha\beta s\big)}{64\pi^6\alpha^2\beta}+\frac{m^2\big(m^2(\alpha+\beta)-\alpha\beta s\big)}{128\pi^6\alpha \beta}\Big] \Big{\}},\\
\rho^{\langle
\bar{q}Gq\rangle}_3(s)=&-\frac{m_q\langle\bar{q}\sigma\cdot G
q\rangle}{8\pi^4}\Big{\{}\big(2m^2+s\big)\sqrt{1-
\frac{4m^2}{s}}-\int_{\alpha_{min}}^{\alpha_{max}}d\alpha\int_{\beta_{min}}^{\beta_{\max}}d\beta
\frac{2m^2(\alpha+\beta-1)-3\alpha
\beta s}{\alpha}\Big{\}},\\
\rho^{\langle\bar{q}q\rangle^2}_3(s)=&\frac{\langle\bar{q}q\rangle^2 \big(2m^2+s\big)}{3\pi^2}\sqrt{1-\frac{4m^2}{s}},\\
\Pi_3^{\langle\bar{q}q\rangle\langle
\bar{q}Gq\rangle}(M_B^2)=&\frac{\langle\bar{q}q\rangle\langle
\bar{q}\sigma \cdot Gq\rangle}{6\pi^2} \int^1_0dx\Big{\{ }
\frac{2m^4(2-x)}{M_B^2(1-x)x^2}+\frac{m^2}{1-x}+2M_B^2(1-x)\Big{\}
}e^{-\frac{m^2}{M^2_B(1-x)x}}.
\end{split}
\end{equation}

\begin{equation}
\begin{split}
\rho^{pert}_4(s)=&\int_{\alpha_{min}}^{\alpha_{max}}d\alpha\int_{\beta_{min}}^{\beta_{max}}d\beta
\frac{(1-\alpha-\beta)(m^2-\alpha\beta s)\big( m^2(\alpha+\beta)-\alpha\beta s\big)^3}{16\pi^6\alpha^3\beta^3},\\
\rho^{\langle\bar{q} q\rangle}_4(s)=&m_q\langle\bar{q} q\rangle
\int_{\alpha_{min}}^{\alpha_{max}}d\alpha\int_{\beta_{min}}^{\beta_{max}}d\beta
\frac{\big(m^2(\alpha+\beta)-\alpha\beta s\big)\big(m^2(4\alpha+4\beta+5)-8\alpha\beta s\big)}{\pi^4\alpha\beta},\\
\rho^{\langle GG\rangle}_4(s)=&\langle g_s^2
GG\rangle\int_{\alpha_{min}}^{\alpha_{max}}d\alpha\int_{\beta_{min}}^{\beta_{max}}d\beta
\Big{ \{ } \frac{m^2(1-\alpha-\beta)^2\big(m^2(4\alpha\beta+3\alpha+4\beta^2+4\beta)-3\alpha\beta(2\beta+1)s\big)}{96\pi^6\alpha^2\beta}\\
&+\big(m^2(\alpha+\beta)-\alpha\beta s\big)\Big[ \frac{5(1-\alpha-\beta)\big(m^2(\alpha+\beta+1)-2\alpha\beta s\big)}{64\pi^6\alpha^2\beta}+\frac{m^2}{128\pi^6\alpha\beta}\Big] \Big{\} },\\
\rho^{\langle
\bar{q}Gq\rangle}_4(s)=&\frac{m_q\langle\bar{q}\sigma\cdot G
q\rangle}{4\pi^4}\Big{\{}\big(s-6m^2\big)\sqrt{1-
\frac{4m^2}{s}}+\int_{\alpha_{min}}^{\alpha_{max}}d\alpha\int_{\beta_{min}}^{\beta_{max}}d\beta
\frac{10m^2(\alpha+\beta+1)-15\alpha
\beta s}{2\alpha}\Big{\}},\\
\rho^{\langle\bar{q}q\rangle^2}_4(s)=&-\frac{2\langle\bar{q}q\rangle^2\big(s-6m^2\big)}{3\pi^2} \sqrt{1-\frac{4m^2}{s}},\\
\Pi_4^{\langle\bar{q}q\rangle\langle
\bar{q}Gq\rangle}(M_B^2)=&\frac{\langle\bar{q}q\rangle\langle
\bar{q}\sigma \cdot Gq\rangle}{6\pi^2} \int^1_0dx\Big{\{ }
\frac{4m^4(2-3x)}{M_B^2(1-x)x^2}+\frac{m^2(15x-14)}{x(1-x)}-2M_B^2\frac{5x^2-7x+2}{1-x}\Big{\}
}e^{-\frac{m^2}{M^2_B(1-x)x}}.
\end{split}
\end{equation}

\begin{equation}
\begin{split}
\rho^{pert}_5(s)=&\int_{\alpha_{min}}^{\alpha_{max}}d\alpha\int_{\beta_{min}}^{\beta_{max}}d\beta
\frac{3(1-\alpha-\beta)\big( m^2(\alpha+\beta)-\alpha \beta s\big)^4}{16\pi^6\alpha^3\beta^3},\\
\rho^{\langle\bar{q} q\rangle}_5(s)=&m_q\langle\bar{q} q\rangle
\int_{\alpha_{min}}^{\alpha_{max}}d\alpha\int_{\beta_{min}}^{\beta_{max}}d\beta
\frac{6\big( m^2(\alpha+\beta+2)-2\alpha \beta s\big)\big( m^2(\alpha+\beta)-\alpha \beta s\big)}{\pi^4\alpha\beta},\\
\rho^{\langle GG\rangle}_5(s)=&\langle g_s^2
GG\rangle\int_{\alpha_{min}}^{\alpha_{max}}d\alpha\int_{\beta_{min}}^{\beta_{max}}d\beta
\Big{\{} \frac{m^2(1-\alpha-\beta)^2\big( 2m^2(\alpha+\beta)-3\alpha\beta s\big)}{16\pi^6\alpha^3}\\
&-\frac{\big(\alpha(\alpha+12\beta-2)+\beta(9\beta-10)+1\big)\big(m^2(\alpha+\beta)-2\alpha \beta s\big)\big(m^2(\alpha+\beta)-\alpha \beta s\big)}{64\pi^6\alpha^2\beta^2}\Big{\}},\\
\rho^{\langle
\bar{q}Gq\rangle}_5(s)=&-\frac{m_q\langle\bar{q}\sigma\cdot G
q\rangle}{\pi^4} \Big{\{}3m^2\sqrt{1-\frac{4m^2}{s}}
-\int_{\alpha_{min}}^{\alpha_{max}}d\alpha\int_{\beta_{min}}^{\beta_{max}}d\beta
\frac{2m^2}{ \alpha}\Big{\}},\\
\rho^{\langle\bar{q}q\rangle^2}_5(s)=&\frac{8\langle\bar{q}q\rangle^2 m^2}{\pi^2}\sqrt{1-\frac{4m^2}{s}},\\
\Pi_5^{\langle\bar{q}q\rangle\langle
\bar{q}Gq\rangle}(M_B^2)=&\frac{4m^2\langle\bar{q}q\rangle\langle
\bar{q}\sigma \cdot Gq\rangle}{3\pi^2} \int^1_0dx\Big{\{ }
\frac{3m^2-2xM_B^2}{x^2M_B^2 }\Big{\}
}e^{-\frac{m^2}{M^2_B(1-x)x}}.
\end{split}
\end{equation}

For the interpolating currents with $J^P=1^-$,
\begin{equation}
\begin{split}
\rho^{pert}_1(s)=&\int_{\alpha_{min}}^{\alpha_{max}}d\alpha\int_{\beta_{min}}^{\beta_{max}}d\beta
\Big{\{} \frac{(1-\alpha-\beta)^2\big(\alpha\beta s-m^2(\alpha+\beta-4)\big)\big(m^2(\alpha+\beta)-\alpha\beta s\big)^3}{128\pi^6\alpha^3\beta^3}\\
&+\frac{(1-\alpha-\beta)\big(m^2(\alpha+\beta)-\alpha\beta s\big)^4}{64\pi^6\alpha^3\beta^3}\Big{\}},\\
\rho^{\langle\bar{q} q\rangle}_1(s)=&-m_q\langle\bar{q} q\rangle
\int_{\alpha_{min}}^{\alpha_{max}}d\alpha\int_{\beta_{min}}^{\beta_{max}}d\beta
\frac{\big(m^2(\alpha+\beta)-\alpha\beta s\big)\big(m^2(\alpha+\beta+2)-3\alpha\beta s\big)}{4\pi^4\alpha\beta},\\
\rho^{\langle GG\rangle}_1(s)=&\langle g_s^2
GG\rangle\int_{\alpha_{min}}^{\alpha_{max}}d\alpha\int_{\beta_{min}}^{\beta_{max}}d\beta
\Big{\{} \frac{\big(m^2(\alpha+\beta+2)-\alpha\beta s\big)\big(\alpha\beta s-m^2(\alpha+\beta)\big)}{512\pi^6\alpha\beta}\\
&+(1-\alpha-\beta)^2\Big[ \frac{\big(m^2(\alpha+\beta)-\alpha\beta s\big)\big(m^2(3\alpha^2+3\alpha\beta-16\beta^3+48\beta)-5\alpha^2\beta s\big)}{3072\pi^6\alpha^3\beta^2}\\
&+\frac{m^2\big(2m^2(\alpha+\beta)+m^2-3\alpha\beta s\big)}{192\pi^6\alpha^3}\Big]\Big\},\\
\rho^{\langle
\bar{q}Gq\rangle}_1(s)=&-\frac{m_q\langle\bar{q}\sigma\cdot G
q\rangle\big(s-4m^2)}{12\pi^4}
\sqrt{1-\frac{4m^2}{s}},\\
\rho^{\langle\bar{q}q\rangle^2}_1(s)=&\frac{2\langle\bar{q}q\rangle^2 \big(s-4m^2\big)}{9\pi^2}\sqrt{1-\frac{4m^2}{s}},\\
\Pi_1^{\langle\bar{q}q\rangle\langle
\bar{q}Gq\rangle}(M_B^2)=&\frac{\langle\bar{q}q\rangle\langle
\bar{q}\sigma \cdot Gq\rangle}{3\pi^2} \int^1_0dx\Big{\{ }
\frac{m^4(2x-1)}{M_B^2 x^2(1-x)}+\frac{m^2(2-x)}{1-x}+M_B^2
x\Big{\} }e^{-\frac{m^2}{M^2_B(1-x)x}}.
\end{split}
\end{equation}

\begin{equation}
\begin{split}
\rho^{pert}_2(s)=&\int_{\alpha_{min}}^{\alpha_{max}}d\alpha\int_{\beta_{min}}^{\beta_{max}}d\beta
\\ &
\frac{(1-\alpha-\beta)\big(m^2(\alpha+\beta)-\alpha\beta
s\big)^3\big(m^2(7(\alpha+\beta)(\alpha+\beta+1)-8)-3
\alpha\beta s(\alpha+\beta+1)\big)}{384\pi^6\alpha^3\beta^3},\\
\rho^{\langle\bar{q} q\rangle}_2(s)=&m_q\langle\bar{q} q\rangle
\int_{\alpha_{min}}^{\alpha_{max}}d\alpha\int_{\beta_{min}}^{\beta_{max}}d\beta
\frac{3\big(m^2(\alpha+\beta)-\alpha\beta s\big)\big(m^2(\alpha+\beta-1)-2\alpha\beta s\big)}{2\pi^4\alpha\beta}\\
&+\frac{(1-\alpha-\beta)\big(m^2(\alpha+\beta)-\alpha\beta s\big)\big(5\alpha\beta s-m^2(3\alpha+3\beta-2)\big)}{4\pi^4\alpha\beta},\\
\rho^{\langle GG\rangle}_2(s)=&\langle g_s^2
GG\rangle\int_{\alpha_{min}}^{\alpha_{max}}d\alpha\int_{\beta_{min}}^{\beta_{max}}d\beta
\\&\Big{\{}
\frac{(1-\alpha-\beta)^3\big(m^2(\alpha+\beta)-\alpha\beta
s\big)\big( 5\alpha^2\beta s
-m^2(3\alpha^2+3\alpha\beta-4\alpha+16\beta)\big)}{3072\pi^6\alpha^3\beta^2}\\
&-\frac{(1-\alpha-\beta)^2\big(m^2(\alpha+\beta)-\alpha\beta
s\big)\big(m^2(\alpha^2+\alpha\beta-2\alpha+8\beta)\big)}
{512\pi^6\alpha^3\beta^2}\\
&+\frac{m^2(1-\alpha-\beta)^2\big(m^2(3\alpha^2+6\alpha\beta+2\alpha+3\beta^2+2\beta-2)-\alpha\beta
s(4\alpha+4\beta+5)\big)}
{576\pi^6\alpha^3}\\
&+\frac{(1-\alpha-\beta)\big(m^2(\alpha+\beta)-\alpha\beta s\big)\big(m^2(3\alpha+3\beta-2)-5\alpha\beta s\big)}{1536\pi^6\alpha\beta}\\
&+\frac{\big(m^2(\alpha+\beta)-\alpha\beta s\big)\big(m^2(\alpha+\beta-1)-2\alpha\beta s\big)}{768\pi^6\alpha\beta}\Big{\}},\\
\rho^{\langle
\bar{q}Gq\rangle}_2(s)=&\frac{m_q\langle\bar{q}\sigma\cdot G
q\rangle s}{8\pi^4}
\sqrt{1-\frac{4m^2}{s}},\\
\rho^{\langle\bar{q}q\rangle^2}_2(s)=&-\frac{2\langle\bar{q}q\rangle^2 s}{3\pi^2}\sqrt{1-\frac{4m^2}{s}},\\
\Pi_2^{\langle\bar{q}q\rangle\langle
\bar{q}Gq\rangle}(M_B^2)=&-\frac{\langle\bar{q}q\rangle\langle
\bar{q}\sigma \cdot Gq\rangle}{3\pi^2} \int^1_0dx\Big{\{ }
\frac{m^4}{M_B^2 x^2(1-x)}+\frac{m^2}{(1-x)x}+M_B^2 \Big{\}
}e^{-\frac{m^2}{M^2_B(1-x)x}}.
\end{split}
\end{equation}

\begin{equation}
\begin{split}
\rho^{pert}_3(s)=&\int_{\alpha_{min}}^{\alpha_{max}}d\alpha\int_{\beta_{min}}^{\beta_{max}}d\beta
\Big{\{}
\frac{(1-\alpha-\beta)^3\big(m^2(\alpha+\beta)-\alpha\beta
s\big)\big( m^2(3\alpha+3\beta+1)-7\alpha\beta s\big)}
{192\pi^6\alpha^3\beta^3}\\
&+\frac{(1-\alpha-\beta)\big(m^2(\alpha+\beta)-\alpha\beta
s\big)\big(\alpha\beta s(\alpha+\beta-7)-m^2(\alpha^2+2\alpha\beta
-3\alpha+\beta^3-3\beta-4)\big)}{256\pi^6\alpha^3\beta^3}\Big{\}},\\
\rho^{\langle\bar{q} q\rangle}_3(s)=&m_q\langle\bar{q} q\rangle
\int_{\alpha_{min}}^{\alpha_{max}}d\alpha\int_{\beta_{min}}^{\beta_{max}}d\beta
\Big{\{} \frac{(1-\alpha-\beta)\big(m^2(\alpha+\beta)-\alpha\beta
s\big)\big(m^2(15\alpha+15\beta+2)-25\alpha\beta s\big)}
{8\pi^4\alpha\beta}\\
&+\frac{\big( m^2(\alpha+\beta)-\alpha\beta s\big)\big(m^2(3\alpha+3\beta-5)-4\alpha\beta s\big)}{4\pi^4\alpha\beta}\Big{\}},\\
\rho^{\langle GG\rangle}_3(s)=&\langle g_s^2
GG\rangle\int_{\alpha_{min}}^{\alpha_{max}}d\alpha\int_{\beta_{min}}^{\beta_{max}}d\beta
\Big{\{} \frac{m^2(1-\alpha-\beta)^3\big(m^2(15\alpha+15\beta+1)-20\alpha\beta s\big)}{1152\pi^6\alpha^3}\\
&+\frac{(1-\alpha-\beta)^3\big(m^2(\alpha+\beta)-\alpha\beta
s\big)\big(25\alpha^2\beta s-m^2(15\alpha^2+15\alpha\beta
+4\alpha-24\beta)\big)}{9126\pi^6\alpha^3\beta^2}\\
&+\frac{(1-\alpha-\beta)^2\big(m^2(\alpha+\beta)^2-\alpha\beta
s\big)\big( m^2(15\alpha^2\beta-3\alpha^2+15\alpha\beta^2+
\alpha\beta-2\alpha+12\beta)+\alpha^2(6-25\beta)\beta s\big)}{1536\pi^6\alpha^3\beta^2}\\
&+\frac{ (\alpha +\beta -1) \left(m^2 (\alpha +\beta )-\alpha  \beta  s\right) \left(\alpha  (25 \alpha +14) \beta  s-m^2 \left(15 \alpha ^2+\alpha  (15 \beta +8)+6 \beta +8\right)\right)}{1536 \pi ^6 \alpha ^2 \beta }\\
&+\frac{m^2 (\alpha +\beta -1)^2 \left(m^2 (6 \alpha +6 \beta
+1)-9 \alpha  \beta  s\right)}{384 \pi ^6 \alpha ^3}
+\frac{ \left(m^2 (\alpha +\beta )-\alpha  \beta  s\right) \left(m^2 (3 \alpha +3 \beta -5)-4 \alpha  \beta  s\right)}{768 \pi ^6 \alpha  \beta }\Big{\}},\\
\rho^{\langle
\bar{q}Gq\rangle}_3(s)=&\frac{m_q\langle\bar{q}\sigma\cdot G
q\rangle s}{8\pi^4}\Big{\{}
\frac{16m^2-s}{6}\sqrt{1-\frac{4m^2}{s}}+\int_{\alpha_{min}}^{\alpha_{max}}d\alpha\int_{\beta_{min}}^{\beta_{max}}d\beta
\frac{\alpha\beta s-3m^2}{3\alpha} \Big{\}},\\
\rho^{\langle\bar{q}q\rangle^2}_3(s)=&\frac{\langle\bar{q}q\rangle^2 \big(s-16m^2\big)}{18\pi^2}\sqrt{1-\frac{4m^2}{s}},\\
\Pi_3^{\langle\bar{q}q\rangle\langle
\bar{q}Gq\rangle}(M_B^2)=&\frac{\langle\bar{q}q\rangle\langle
\bar{q}\sigma \cdot Gq\rangle}{18\pi^2} \int^1_0dx\Big{\{ }
\frac{m^4(12x-9)}{M_B^2x^2(1-x)}-\frac{2m^2(3x-4)}{(1-x)}-\frac{3M_B^2(2x^2-3x+1)}{1-x}\Big{\}}e^{-\frac{m^2}{M^2_B(1-x)x}}.
\end{split}
\end{equation}

\begin{equation}
\begin{split}
\rho^{pert}_4(s)=&\int_{\alpha_{min}}^{\alpha_{max}}d\alpha\int_{\beta_{min}}^{\beta_{max}}d\beta
\Big{\{}
\frac{(1-\alpha-\beta)^3\big(m^2(\alpha+\beta)-\alpha\beta s
\big)^3\big(m^2(3\alpha+3\beta+2)-7\alpha\beta s\big)}
{192\pi^6\alpha^3\beta^3},\\
\rho^{\langle\bar{q} q\rangle}_4(s)=&m_q\langle\bar{q} q\rangle
\int_{\alpha_{min}}^{\alpha_{max}}d\alpha\int_{\beta_{min}}^{\beta_{max}}d\beta
\Big{\{} \frac{(1-\alpha-\beta)\big(m^2(\alpha+\beta)-\alpha\beta s\big)\big(5\alpha\beta s-m^2(3\alpha+3\beta+1)\big)}{2\pi^4\alpha\beta}\\
&+\frac{\big(m^2(\alpha+\beta)-\alpha\beta s\big)\big(m^2(13\alpha+13\beta-14)-23\alpha\beta s\big)}{8\pi^4\alpha\beta}\Big{\}},\\
\rho^{\langle GG\rangle}_4(s)=&\langle g_s^2
GG\rangle\int_{\alpha_{min}}^{\alpha_{max}}d\alpha\int_{\beta_{min}}^{\beta_{max}}d\beta
\Big{\{} \frac{(1-\alpha-\beta)^3\big(m^2(\alpha+\beta)-\alpha\beta s\big)\big(5\alpha^2\beta s-3m^2(\alpha^2+\alpha\beta+2\beta)\big)}{1152\pi^6\alpha^3\beta^2}\\
&+\frac{(1-\alpha-\beta)^2\big(m^2(\alpha+\beta)-\alpha\beta s\big)\big(5\alpha^2\beta(6\beta+1)s-m^2(18 \alpha ^2 \beta +3 \alpha ^2+18 \alpha  \beta ^2+19 \alpha  \beta +8 \beta ^2-24 \beta)\big)}{3072\pi^6\alpha^3\beta^2}\\
&-\frac{(1-\alpha-\beta)\big(m^2(\alpha+\beta)-\alpha\beta s\big)\big(m^2(-6\alpha^2+3\alpha-6\alpha\beta+5\beta-4)+\alpha(10\alpha-9)\beta s\big)}{768\pi^6\alpha^2\beta}\\
&+\frac{\big(m^2\alpha+\beta)-\alpha\beta s\big)\big(m^2(7\alpha+7\beta-2)-13\alpha\beta s\big)}{1536\pi^6\alpha\beta}\\
&-\frac{(1-\alpha-\beta)^2m^2}{1152\pi^6\alpha^3}\Big[
2(1-\alpha-\beta)\big(m^2(6\alpha+6\beta+1)-8\alpha\beta s\big)+
3\big(m^2(6\alpha+6\beta-1)-9\alpha\beta s\big)\Big]\Big{\}},\\
\rho^{\langle
\bar{q}Gq\rangle}_4(s)=&\frac{m_q\langle\bar{q}\sigma\cdot G
q\rangle s}{12\pi^4}\Big{\{}
(2m^2+s)\sqrt{1-\frac{4m^2}{s}}+\int_{\alpha_{min}}^{\alpha_{max}}d\alpha\int_{\beta_{min}}^{\beta_{max}}d\beta
\frac{3m^2(\alpha+\beta-1)-4\alpha\beta s}{2\alpha}\Big{\}},\\
\rho^{\langle\bar{q}q\rangle^2}_4(s)=&\frac{2\langle\bar{q}q\rangle^2 \big(2m^2+s\big)}{9\pi^2}\sqrt{1-\frac{4m^2}{s}},\\
\Pi_4^{\langle\bar{q}q\rangle\langle
\bar{q}Gq\rangle}(M_B^2)=&\frac{\langle\bar{q}q\rangle\langle
\bar{q}\sigma \cdot Gq\rangle}{18\pi^2} \int^1_0dx\Big{\{ }
\frac{m^4(6x-9)}{M_B^2x^2(1-x)}-\frac{m^2(3x^2-4x+3)}{x(1-x)}-6M_B^2(1-x)\Big{\}}e^{-\frac{m^2}{M^2_B(1-x)x}}.
\end{split}
\end{equation}

For the interpolating currents with $J^P=1^+$,
\begin{equation}
\begin{split}
\rho^{pert}_1(s)=&\int_{\alpha_{min}}^{\alpha_{max}}d\alpha\int_{\beta_{min}}^{\beta_{max}}d\beta
\\&
\Big{\{} \frac{(1-\alpha-\beta)^2\big(\alpha\beta s-m^2(\alpha+\beta-4)\big)\big(m^2(\alpha+\beta)-\alpha\beta s\big)^3}{128\pi^6\alpha^3\beta^3}\\
&+\frac{(1-\alpha-\beta)\big(m^2(\alpha+\beta)-\alpha\beta s\big)^4}{64\pi^6\alpha^3\beta^3}\Big{\}},\\
\rho^{\langle\bar{q} q\rangle}_1(s)=&3m_q\langle\bar{q} q\rangle
\int_{\alpha_{min}}^{\alpha_{max}}d\alpha\int_{\beta_{min}}^{\beta_{max}}d\beta
\frac{\big(m^2(\alpha+\beta)-\alpha\beta s\big)\big(m^2(\alpha+\beta+2)-3\alpha\beta s\big)}{4\pi^4\alpha\beta},\\
\rho^{\langle GG\rangle}_1(s)=&\langle g_s^2
GG\rangle\int_{\alpha_{min}}^{\alpha_{max}}d\alpha\int_{\beta_{min}}^{\beta_{max}}d\beta
\\&
\Big{\{} \frac{\big(m^2(\alpha+\beta+2)-\alpha\beta s\big)\big(\alpha\beta s-m^2(\alpha+\beta)\big)}{512\pi^6\alpha\beta}\\
&+(1-\alpha-\beta)^2\Big[ \frac{\big(m^2(\alpha+\beta)-\alpha\beta s\big)\big(m^2(3\alpha^2+3\alpha\beta-16\beta^3+48\beta)-5\alpha^2\beta s\big)}{3072\pi^6\alpha^3\beta^2}\\
&+\frac{m^2\big(2m^2(\alpha+\beta)+m^2-3\alpha\beta s\big)}{192\pi^6\alpha^3},\\
\rho^{\langle
\bar{q}Gq\rangle}_1(s)=&\frac{m_q\langle\bar{q}\sigma\cdot G
q\rangle\big(s-4m^2)}{12\pi^4}
\sqrt{1-\frac{4m^2}{s}},\\
\rho^{\langle\bar{q}q\rangle^2}_1(s)=&-\frac{2\langle\bar{q}q\rangle^2 \big(s-4m^2\big)}{9\pi^2}\sqrt{1-\frac{4m^2}{s}},\\
\Pi_1^{\langle\bar{q}q\rangle\langle
\bar{q}Gq\rangle}(M_B^2)=&-\frac{\langle\bar{q}q\rangle\langle
\bar{q}\sigma \cdot Gq\rangle}{3\pi^2} \int^1_0dx\Big{\{ }
\frac{m^4(2x-1)}{M_B^2 x^2(1-x)}+\frac{m^2(2-x)}{1-x}+M_B^2
x\Big{\} }e^{-\frac{m^2}{M^2_B(1-x)x}}.
\end{split}
\end{equation}

\begin{equation}
\begin{split}
\rho^{pert}_2(s)=&\int_{\alpha_{min}}^{\alpha_{max}}d\alpha\int_{\beta_{min}}^{\beta_{max}}d\beta
\frac{(1-\alpha-\beta)\big(\alpha\beta s-m^2(\alpha+\beta)\big)\big(m^2(\alpha+\beta+1)(\alpha+\beta)-8)+3\alpha\beta s(\alpha+\beta+1)\big)}{384\pi^6\alpha^3\beta^3},\\
\rho^{\langle\bar{q} q\rangle}_2(s)=&m_q\langle\bar{q} q\rangle
\int_{\alpha_{min}}^{\alpha_{max}}d\alpha\int_{\beta_{min}}^{\beta_{max}}d\beta
\Big{\{} \frac{(1-\alpha\beta s)\big(m^2(\alpha+\beta)-\alpha\beta s\big)\big(5\alpha\beta s-m^2(3\alpha+3\beta+2)\big)}{4\pi^4\alpha\beta}\\
&+\frac{3\big(m^2(\alpha+\beta)-\alpha\beta s\big)\big(m^2(\alpha+\beta+1)-2\alpha\beta s\big)}{2\pi^4\alpha\beta},\\
\rho^{\langle GG\rangle}_2(s)=&\langle g_s^2
GG\rangle\int_{\alpha_{min}}^{\alpha_{max}}d\alpha\int_{\beta_{min}}^{\beta_{max}}d\beta
\\&
\Big{\{} \frac{m^2 (\alpha +\beta -1)^2 \left(m^2 (3 \alpha ^2+6 \alpha  \beta +4 \alpha +3 \beta ^2+4 \beta -1)-\alpha  \beta  s (4 \alpha +4 \beta +5)\right)}{576 \pi ^6 \alpha ^3}\\
&+\big(m^2(\alpha+\beta)-\alpha\beta s\big)\Big[ \frac{(1-\alpha-\beta)^3(\alpha-4\beta)}{768\pi^6\alpha^3\beta^2}-\frac{(1-\alpha-\beta)^2\big(m^2(\alpha^2+\alpha\beta+2\alpha-8\beta)-2\alpha\beta s\big)}{512\pi^6\alpha^3\beta^2}\\
&+\frac{(1-\alpha-\beta)\big(m^2(3\alpha+3\beta+2)-5\alpha\beta s\big)}{1536\pi^6\alpha\beta}-\frac{m^2(\alpha+\beta+1)-2\alpha\beta s}{768\pi^6\alpha\beta}\Big]\Big{\}},\\
\rho^{\langle
\bar{q}Gq\rangle}_2(s)=&\frac{m_q\langle\bar{q}\sigma\cdot G
q\rangle\big(s-4m^2)}{8\pi^4}
\sqrt{1-\frac{4m^2}{s}},\\
\rho^{\langle\bar{q}q\rangle^2}_2(s)=&-\frac{\langle\bar{q}q\rangle^2 \big(s-4m^2\big)}{3\pi^2}\sqrt{1-\frac{4m^2}{s}},\\
\Pi_2^{\langle\bar{q}q\rangle\langle
\bar{q}Gq\rangle}(M_B^2)=&\frac{\langle\bar{q}q\rangle\langle
\bar{q}\sigma \cdot Gq\rangle}{3\pi^2} \int^1_0dx\Big{\{ }
\frac{m^4(1-2x)}{M_B^2 x^2(1-x)}-\frac{m^2}{x(1-x)}-M_B^2 \Big{\}
}e^{-\frac{m^2}{M^2_B(1-x)x}}.
\end{split}
\end{equation}

\begin{equation}
\begin{split}
\rho^{pert}_3(s)=&\int_{\alpha_{min}}^{\alpha_{max}}d\alpha\int_{\beta_{min}}^{\beta_{max}}d\beta
(1-\alpha-\beta)\big(m^2(\alpha+\beta)-\alpha\beta s\big)\Big[ \frac{(7-\alpha-\beta)\big(m^2(\alpha+\beta)-\alpha\beta s\big)}{256\pi^6\alpha^3\beta^3} \\
&+\frac{(1-\alpha-\beta)\big(m^2(3\alpha^2+6\alpha\beta-4\alpha+3\beta^2-4\beta-4)-7(\alpha+\beta-1)\alpha\beta s\big)}{192\pi^6\alpha^3\beta^3}\Big],\\
\rho^{\langle\bar{q} q\rangle}_3(s)=&m_q\langle\bar{q} q\rangle
\int_{\alpha_{min}}^{\alpha_{max}}d\alpha\int_{\beta_{min}}^{\beta_{max}}d\beta
\\&
\frac{m^2(\alpha+\beta)-\alpha\beta s}{8\pi^4\alpha\beta}\Big[ 2\big( m^2 (3 \alpha +3 \beta +5)-4 \alpha  \beta  s\big)\\
&-(1-\alpha-\beta)\big(m^2(-15\alpha-15\beta+2)+25\alpha\beta s\big)\Big],\\
\rho^{\langle GG\rangle}_3(s)=&\langle g_s^2
GG\rangle\int_{\alpha_{min}}^{\alpha_{max}}d\alpha\int_{\beta_{min}}^{\beta_{max}}d\beta
\Big{\{} \big(m^2(\alpha+\beta)-\alpha\beta s\big) \Big[
\frac{(1-\alpha-\beta)^3\big(25\alpha^2\beta
s-m^2(15\alpha^2+15\alpha\beta-4\alpha
+24\beta)\big)}{9216\pi^6\alpha^3\beta^2}\\
&+\frac{(1-\alpha-\beta)\big(m^2(15\alpha^2\beta-3\alpha^2+15\alpha\beta^2-7\alpha\beta+2\alpha-12\beta)+\alpha^2(6-25\beta)\beta s\big)}{1536\pi^6\alpha^3\beta^2}\\
&+\frac{(1-\alpha-\beta)\big(m^2(3\alpha^2+3\alpha\beta+4\alpha+6\beta-8)-\alpha^2(5\alpha+14\beta)\beta s\big)}{1536\pi^6\alpha^2\beta}+\frac{m^2(3\alpha+3\beta+5)-4\alpha\beta s}{768\pi^6\alpha\beta}\Big]\\
&+\frac{ m^2 (\alpha +\beta -1)^2 \left(\alpha  \beta  s (20 \alpha +20 \beta -47)-m^2 \left(15 \alpha ^2+\alpha  (30 \beta -34)+15 \beta ^2-34 \beta +4\right)\right)}{1152 \pi ^6 \alpha ^3}\Big{\}},\\
\rho^{\langle
\bar{q}Gq\rangle}_3(s)=&\frac{m_q\langle\bar{q}\sigma\cdot G
q\rangle s}{48\pi^4}\Big{\{}
(16m^2-s)\sqrt{1-\frac{4m^2}{s}}+\int_{\alpha_{min}}^{\alpha_{max}}d\alpha\int_{\beta_{min}}^{\beta_{max}}d\beta
\frac{6m^2+2\alpha\beta s}{\alpha} \Big{\}},\\
\rho^{\langle\bar{q}q\rangle^2}_3(s)=&\frac{\langle\bar{q}q\rangle^2 \big(s+20m^2\big)}{18\pi^2}\sqrt{1-\frac{4m^2}{s}},\\
\Pi_3^{\langle\bar{q}q\rangle\langle
\bar{q}Gq\rangle}(M_B^2)=&\frac{\langle\bar{q}q\rangle\langle
\bar{q}\sigma \cdot Gq\rangle}{18\pi^2} \int^1_0dx\Big{\{ }
\frac{m^4(9-6x)}{M_B^2
x^2(1-x)}-\frac{2m^2(3x^2-7x+3)}{x(1-x)}-\frac{3M_B^2(2x^2-3x+1)}{1-x}
\Big{\} }e^{-\frac{m^2}{M^2_B(1-x)x}}.
\end{split}
\end{equation}

\begin{equation}
\begin{split}
\rho^{pert}_4(s)=&\int_{\alpha_{min}}^{\alpha_{max}}d\alpha\int_{\beta_{min}}^{\beta_{max}}d\beta
\Big{\{}
\frac{(1-\alpha-\beta)^3\big(m^2(\alpha+\beta)-\alpha\beta s
\big)^3\big(m^2(3\alpha+3\beta+2)-7\alpha\beta s\big)}
{192\pi^6\alpha^3\beta^3},\\
\rho^{\langle\bar{q} q\rangle}_4(s)=-&m_q\langle\bar{q} q\rangle
\int_{\alpha_{min}}^{\alpha_{max}}d\alpha\int_{\beta_{min}}^{\beta_{max}}d\beta
\Big{\{} \frac{(1-\alpha-\beta)\big(m^2(\alpha+\beta)-\alpha\beta s\big)\big(5\alpha\beta s-m^2(3\alpha+3\beta+1)\big)}{2\pi^4\alpha\beta}\\
&-\frac{\big(m^2(\alpha+\beta)-\alpha\beta s\big)\big(m^2(\alpha+\beta+10)-3\alpha\beta s\big)}{8\pi^4\alpha\beta}\Big{\}},\\
\rho^{\langle GG\rangle}_4(s)=&\langle g_s^2
GG\rangle\int_{\alpha_{min}}^{\alpha_{max}}d\alpha\int_{\beta_{min}}^{\beta_{max}}d\beta
\Big{\{} \frac{(1-\alpha-\beta)^3\big(m^2(\alpha+\beta)-\alpha\beta s\big)\big(5\alpha^2\beta s-3m^2(\alpha^2+\alpha\beta+2\beta)\big)}{1152\pi^6\alpha^3\beta^2}\\
&+\frac{(1-\alpha-\beta)^2\big(m^2(\alpha+\beta)-\alpha\beta s\big)\big(5\alpha^2\beta(6\beta+1)s-m^2(18 \alpha ^2 \beta +3 \alpha ^2+18 \alpha  \beta ^2+19 \alpha  \beta +8 \beta ^2-24 \beta)\big)}{3072\pi^6\alpha^3\beta^2}\\
&-\frac{(1-\alpha-\beta)\big(m^2(\alpha+\beta)-\alpha\beta s\big)\big(m^2(-6\alpha^2+3\alpha-6\alpha\beta+5\beta-4)+\alpha(10\alpha-9)\beta s\big)}{768\pi^6\alpha^2\beta}\\
&+\frac{\big(m^2\alpha+\beta)-\alpha\beta s\big)\big(m^2(7\alpha+7\beta-2)-13\alpha\beta s\big)}{1536\pi^6\alpha\beta}\\
&-\frac{(1-\alpha-\beta)^2m^2}{1152\pi^6\alpha^3}\Big[
2(1-\alpha-\beta)\big(m^2(6\alpha+6\beta+1)-8\alpha\beta s\big)+
3\big(m^2(6\alpha+6\beta-1)-9\alpha\beta s\big)\Big]\Big{\}},\\
\rho^{\langle
\bar{q}Gq\rangle}_4(s)=&-\frac{m_q\langle\bar{q}\sigma\cdot G
q\rangle s}{12\pi^4}\Big{\{}
(2m^2+s)\sqrt{1-\frac{4m^2}{s}}+\int_{\alpha_{min}}^{\alpha_{max}}d\alpha\int_{\beta_{min}}^{\beta_{max}}d\beta
\frac{3m^2(\alpha+\beta-1)-4\alpha\beta s)}{2\alpha}\Big{\}},\\
\rho^{\langle\bar{q}q\rangle^2}_4(s)=&\frac{2\langle\bar{q}q\rangle^2 \big(2m^2+s\big)}{9\pi^2}\sqrt{1-\frac{4m^2}{s}},\\
\Pi_4^{\langle\bar{q}q\rangle\langle
\bar{q}Gq\rangle}(M_B^2)=&\frac{\langle\bar{q}q\rangle\langle
\bar{q}\sigma \cdot Gq\rangle}{18\pi^2} \int^1_0dx\Big{\{ }
\frac{m^4(9-6x)}{M_B^2x^2(1-x)}+\frac{m^2(3x^2-4x+3)}{x(1-x)}+6M_B^2(1-x)\Big{\}}e^{-\frac{m^2}{M^2_B(1-x)x}}.
\end{split}
\end{equation}

\subsection{The spectral densities for the currents in the $QQ\bar q\bar s$ systems}

For the interpolating currents with $J^P=0^-$,
\begin{eqnarray}
\nonumber
\rho^{pert}_1(s)&=&\frac{1}{2^7\pi^6}\dab\frac{(1-\alpha-\beta)[(2-\alpha-\beta)m^2-\alpha\beta s]\f(s)^3}{\alpha^3\beta^3}
\, , \non
\rho^{\qq}_1(s)&=&-\frac{m_s(2\qq-\ss)}{8\pi^4}\dab\frac{[(1+\alpha+\beta)m^2-2\alpha\beta s]\f(s)}{\alpha\beta} \, , \non
\rho^{\GGa}_1(s)&=&\frac{\GGb}{3\times2^{10}\pi^6}\dab \non
&&\bigg\{\frac{8(1-\alpha-\beta)^2m^2}{\alpha^3}\bigg[\frac{[(3\alpha+4\beta)m^2-3\alpha\beta s]}{\beta}+[(2\alpha+2\beta)m^2-3\alpha\beta s]\bigg]-\non
&&\frac{3\f(s)}{\alpha\beta}\bigg[\frac{(1-\alpha-\beta)^2[(2+\alpha+\beta)m^2-2\alpha\beta s]}{\alpha\beta}+2[(1+\alpha+\beta)m^2-2\alpha\beta s]\bigg]
\bigg\}
\, , \non
\rho^{\qGqa}_1(s)&=&\frac{m_s\qGqb(4m^2-s)}{32\pi^4}\sqrt{1-4m^2/s}\, , \\
\rho^{\qq^2}_1(s)&=&-\frac{\qq\ss(4m^2-s)}{6\pi^2}\sqrt{1-4m_c^2/s}
\, ,
\end{eqnarray}
\begin{eqnarray}
\nonumber
\Pi^{\qGqa\qq}_1(M_B^2)&=&\frac{\qGqb\ss+\sGs\qq}{12\pi^2}\int_0^1d\alpha\bigg
\{\frac{m^4(2\alpha-1)}{M_B^2\alpha^2(1-\alpha)}+\frac{m^2}{\alpha(1-\alpha)}+M_B^2\bigg\}\efun
\, .
\end{eqnarray}

\begin{eqnarray}
\nonumber
\rho^{pert}_2(s)&=&\frac{1}{2^7\pi^6}\dab\frac{(1-\alpha-\beta)[(3\alpha+3\beta-2)m^2-\alpha\beta s]\f(s)^3}{\alpha^3\beta^3}
\, , \non
\rho^{\qq}_2(s)&=&\frac{m_s(2\qq+\ss)}{8\pi^4}\dab\frac{[(\alpha+\beta-1)m^2-2\alpha\beta s]\f(s)}{\alpha\beta} \, , \non
\rho^{\GGa}_2(s)&=&-\frac{\GGb}{3\times2^{10}\pi^6}\dab \non
&&\bigg\{\frac{8(1-\alpha-\beta)^2m^2}{\alpha^3}\bigg[\frac{[(3\alpha+4\beta)m^2-3\alpha\beta s]}{\beta}-[(2\alpha+2\beta)m^2-3\alpha\beta s]\bigg]+\non
&&\frac{3\f(s)}{\alpha\beta}\bigg[\frac{(1-\alpha-\beta)^2[(\alpha+\beta-2)m^2-2\alpha\beta s]}{\alpha\beta}+2[(\alpha+\beta-1)m^2-2\alpha\beta s]\bigg]
\bigg\}
\, , \non
\rho^{\qGqa}_2(s)&=&\frac{m_s\qGqb s}{32\pi^4}\sqrt{1-4m^2/s}\, , \\
\rho^{\qq^2}_2(s)&=&-\frac{\qq\ss s}{6\pi^2}\sqrt{1-4m_c^2/s}
\, ,
\end{eqnarray}
\begin{eqnarray}
\nonumber
\Pi^{\qGqa\qq}_2(M_B^2)&=&-\frac{\qGqb\ss+\sGs\qq}{12\pi^2}\int_0^1d\alpha\bigg
\{\frac{m^4}{M_B^2\alpha^2(1-\alpha)}+\frac{m^2}{\alpha(1-\alpha)}+M_B^2\bigg\}\efun
\, .
\end{eqnarray}

\begin{eqnarray}
\nonumber
\rho^{pert}_3(s)&=&\frac{3}{2^6\pi^6}\dab\frac{(1-\alpha-\beta)\f(s)^4}{\alpha^3\beta^3}
\, , \non
\rho^{\qq}_3(s)&=&\frac{3m_s}{4\pi^4}\dab\frac{\f(s)}{\alpha\beta}\Big[[(\alpha+\beta)m^2-2\alpha\beta s]\ss
-2m^2\qq\Big] \, , \non
\rho^{\GGa}_3(s)&=&\frac{\GGb}{2^{7}\pi^6}\dab
\bigg\{\frac{4(1-\alpha-\beta)^2[2(\alpha+\beta)m^2-3\alpha\beta s]m^2}{\alpha^3}+\non
&&\frac{[(\alpha+\beta)m^2-2\alpha\beta s]\f(s)}{\alpha\beta}\bigg[\frac{(1-\alpha-\beta)(\alpha+9\beta-1)}{\alpha\beta}-2\bigg]\bigg\}
\, , \non
\rho^{\qGqa}_3(s)&=&\frac{3m_s\qGqb m^2}{8\pi^4}\sqrt{1-4m^2/s}+
\frac{m_s\qGqb m^2}{2\pi^4}\dab\frac{1}{\alpha}\, , \\
\rho^{\qq^2}_3(s)&=&-\frac{2\qq\ss m^2}{\pi^2}\sqrt{1-4m_c^2/s}
\, ,
\end{eqnarray}
\begin{eqnarray}
\nonumber
\Pi^{\qGqa\qq}_3(M_B^2)&=&\frac{\qGqb\ss+\sGs\qq}{6\pi^2}\int_0^1d\alpha\bigg
\{\frac{2m^2}{3\alpha}-\frac{m^4}{2M_B^2\alpha^2}\bigg\}\efun
\, .
\end{eqnarray}

\begin{eqnarray}
\nonumber
\rho^{pert}_4(s)&=&\frac{1}{2^6\pi^6}\dab\frac{(1-\alpha-\beta)[(2\alpha+2\beta-1)m^2-\alpha\beta s]\f(s)^3}{\alpha^3\beta^3}
\, , \non
\rho^{\qq}_4(s)&=&\frac{m_s}{8\pi^4}\dab\frac{\f(s)}{\alpha\beta}\bigg\{2(\qq+\ss)[(\alpha+\beta)m^2-2\alpha\beta s]-(4\qq+\ss)m^2\bigg\} \, , \non
\rho^{\GGa}_4(s)&=&-\frac{\GGb}{3\times2^{7}\pi^6}\dab \non
&&\bigg\{\frac{(1-\alpha-\beta)^2m^2}{\alpha^3}\bigg[\frac{3\f(s)}{\beta}-[(4\alpha+4\beta-1)m^2-6\alpha\beta s]\bigg]-\non
&&\frac{\f(s)}{\alpha\beta}\bigg[\frac{3(1-\alpha-\beta)[(\alpha+\beta-1)m^2-2\alpha\beta s]}{\alpha}+\frac{3m^2}{2}\bigg]
\bigg\}
\, , \non
\rho^{\qGqa}_4(s)&=&\frac{m_s\qGqb (2m^2+s)}{32\pi^4}\sqrt{1-4m^2/s}-
\frac{m_s\qGqb}{32\pi^4}\dab\frac{[2(\alpha+\beta-1)m^2-3\alpha\beta s]}{\alpha}\, , \\
\rho^{\qq^2}_4(s)&=&-\frac{\qq\ss (2m^2+s)}{6\pi^2}\sqrt{1-4m_c^2/s}
\, ,
\end{eqnarray}
\begin{eqnarray}
\nonumber\Pi^{\qGqa\qq}_4(M_B^2)&=&-\frac{\qGqb\ss+\sGs\qq}{12\pi^2}\int_0^1d\alpha\bigg
\{\frac{m^4(2-\alpha)}{M_B^2\alpha^2(1-\alpha)}-\frac{m^2(1-3\alpha)}{2\alpha(1-\alpha)}+(1-\alpha)M_B^2\bigg\}\efun
\, .
\end{eqnarray}

\begin{eqnarray}
\nonumber
\rho^{pert}_5(s)&=&\frac{1}{2^5\pi^6}\dab\frac{(1-\alpha-\beta)(m^2-\alpha\beta s)\f(s)^3}{\alpha^3\beta^3}
\, , \non
\rho^{\qq}_5(s)&=&-\frac{m_s}{4\pi^4}\dab\frac{\f(s)}{\alpha\beta}\bigg\{2(\qq-\ss)[(\alpha+\beta)m^2-2\alpha\beta s]+(4\qq-\ss)m^2\bigg\} \, , \non
\rho^{\GGa}_5(s)&=&\frac{\GGb}{3\times2^{8}\pi^6}\dab \non
&&\bigg\{\frac{4(1-\alpha-\beta)^2m^2}{\alpha^3}\bigg[\frac{3\f(s)}{\beta}+[(4\alpha+4\beta+1)m^2-6\alpha\beta s]\bigg]+\non
&&\frac{3\f(s)}{\alpha\beta}\bigg[\frac{10(1-\alpha-\beta)[(\alpha+\beta+1)m^2-2\alpha\beta s]}{\alpha}+m^2\bigg]
\bigg\}
\, , \non
\rho^{\qGqa}_5(s)&=&\frac{m_s\qGqb (6m^2-s)}{16\pi^4}\sqrt{1-4m^2/s}+
\frac{5m_s\qGqb}{32\pi^4}\dab\frac{[2(\alpha+\beta+1)m^2-3\alpha\beta s]}{\alpha}\, , \\
\rho^{\qq^2}_5(s)&=&-\frac{\qq\ss (6m^2-s)}{3\pi^2}\sqrt{1-4m_c^2/s}
\, ,
\end{eqnarray}
\begin{eqnarray}
\nonumber\Pi^{\qGqa\qq}_5(M_B^2)&=&\frac{\qGqb\ss+\sGs\qq}{24\pi^2}\int_0^1d\alpha\bigg
\{\frac{m^4(12\alpha-8)}{M_B^2\alpha^2(1-\alpha)}+\frac{m^2(9-5\alpha)}{\alpha(1-\alpha)}+(4-10\alpha)M_B^2\bigg\}\efun
\, .
\end{eqnarray}

For the interpolating currents with $J^P=0^+$,

\begin{equation}
\begin{split}
\rho^{pert}_1(s)=&\int_{\alpha_{min}}^{\alpha_{max}}d\alpha\int_{\beta_{min}}^{\beta_{max}}d\beta
\frac{(1-\alpha-\beta)\big( m^2(\alpha+\beta-2)+\alpha\beta s\big)\big( \alpha \beta s-m^2(\alpha+\beta)\big)^3}{128\pi^6\alpha^3\beta^3},\\
\rho^{\langle\bar{q} q\rangle}_1(s)=&m_q(2\qq+\ss) \int_{\alpha_{min}}^{\alpha_{max}}d\alpha\int_{\beta_{min}}^{\beta_{max}}d\beta
\frac{\big( m^2(\alpha+\beta)-\alpha\beta s\big)\big( m^2(\alpha+\beta+1)-2\alpha\beta s\big)}{8\pi^4\alpha\beta},\\
\rho^{\langle GG\rangle}_1(s)=&\frac{\langle g_s^2 GG\rangle}{2}\int_{\alpha_{min}}^{\alpha_{max}}d\alpha\int_{\beta_{min}}^{\beta_{max}}d\beta
\Big{ \{ } \frac{m^2(1-\alpha-\beta)^2\big( 2m^2(\alpha+\beta)+m^2-3\alpha\beta s\big)}{192\pi^6 \alpha^3}+\\
&(m^2(\alpha+\beta)-\alpha\beta s)\Big[\frac{(1-\alpha-\beta)^2\big(2\alpha^2\beta  s-m^2(\alpha(\alpha+\beta+2)-8\beta)\big)}{512
\pi^6 \alpha^3\beta^2}-\frac{m^2(\alpha+\beta+1)-2\alpha \beta s}{256\pi^6\alpha \beta}\Big] \Big{\}},\\
\rho^{\langle \bar{q}Gq\rangle}_1(s)=&\frac{m_q(4\qGqb+\sGs) (s-4m^2)}{128\pi^4}\sqrt{1-\frac{4m^2}{s}},\\
\rho^{\langle\bar{q}q\rangle^2}_1(s)=&\frac{\qq \ss (s-4m^2)}{6\pi^2}\sqrt{1-\frac{4m^2}{s}},\\
\Pi_1^{\langle\bar{q}q\rangle\langle \bar{q}Gq\rangle}(M_B^2)=&-\frac{\qq\sGs+\ss\qGqb}{12\pi^2}
\int^1_0dx\Big{\{ } \frac{m^4(2x-1)}{M_B^2(1-x)x^2}+\frac{m^2}{x(1-x)}+M_B^2\Big{\}}e^{-\frac{m^2}{M^2_B(1-x)x}}.
\end{split}
\end{equation}

\begin{equation}
\begin{split}
\rho^{pert}_2(s)=&\int_{\alpha_{min}}^{\alpha_{max}}d\alpha\int_{\beta_{min}}^{\beta_{max}}d\beta
\frac{(1-\alpha-\beta)\big( m^2(\alpha+\beta)-\alpha \beta s\big)^3\big(m^2(3\alpha+3\beta-2)-\alpha\beta s\big)}{128\pi^6\alpha^3\beta^3},\\
\rho^{\langle\bar{q} q\rangle}_2(s)=&-m_q(2\qq-\ss) \int_{\alpha_{min}}^{\alpha_{max}}d\alpha\int_{\beta_{min}}^{\beta_{max}}d\beta
\frac{\big( m^2(\alpha+\beta-1)-2\alpha \beta s\big)\big( m^2(\alpha+\beta)-\alpha \beta s\big)}{8\pi^4\alpha\beta},\\
\rho^{\langle GG\rangle}_2(s)=&\frac{\langle g_s^2 GG\rangle}{2}\int_{\alpha_{min}}^{\alpha_{max}}d\alpha\int_{\beta_{min}}^{\beta_{max}}d\beta
\Big{ \{ } \frac{m^2(1-\alpha-\beta)^2\big( m^2(2\alpha+2\beta-1)+m^2-3\alpha\beta s\big)}{192\pi^6 \alpha^3}+\\
&(m^2(\alpha+\beta)-\alpha\beta s)\Big[\frac{(1-\alpha-\beta)^2\big(2\alpha^2\beta  s-m^2(\alpha(\alpha+\beta-2)+8\beta)\big)}{512
\pi^6 \alpha^3\beta^2}-\frac{m^2(\alpha+\beta-1)-2\alpha \beta s}{256\pi^6\alpha \beta}\Big] \Big{\}},\\
\rho^{\langle \bar{q}Gq\rangle}_2(s)=&-\frac{m_q(4\qGqb-\sGs) s}{128\pi^4}\sqrt{1-\frac{4m^2}{s}},\\
\rho^{\langle\bar{q}q\rangle^2}_2(s)=&\frac{\qq\ss s}{6\pi^2}\sqrt{1-\frac{4m^2}{s}},\\
\Pi_2^{\langle\bar{q}q\rangle\langle \bar{q}Gq\rangle}(M_B^2)=&\frac{\qq\sGs+\ss\qGqb}{12\pi^2}
\int^1_0dx\Big{\{ } \frac{m^4}{M_B^2(1-x)x^2}+\frac{m^2}{x(1-x)}+M_B^2\Big{\}}e^{-\frac{m^2}{M^2_B(1-x)x}}.
\end{split}
\end{equation}

\begin{equation}
\begin{split}
\rho^{pert}_3(s)=&\int_{\alpha_{min}}^{\alpha_{max}}d\alpha\int_{\beta_{min}}^{\beta_{max}}d\beta
\frac{(1-\alpha-\beta)\big(m^2(2\alpha+2\beta-1)-\alpha\beta s\big)\big(m^2(\alpha+\beta)-\alpha\beta s\big)}{64\pi^6\alpha^3\beta^3},\\
\rho^{\langle\bar{q} q\rangle}_3(s)=& \int_{\alpha_{min}}^{\alpha_{max}}d\alpha\int_{\beta_{min}}^{\beta_{max}}d\beta
\frac{m_q m^2\big(m^2(\alpha+\beta)-\alpha\beta s\big)}{8\pi^4\alpha\beta}\\
&\qquad \times \big(\langle \bar{q_2}q_2\rangle(m^2(2\alpha+2\beta-1)-4\alpha\beta s)-2\langle \bar{q_1}q_1\rangle
(m^2(\alpha+\beta-2)-2\alpha\beta s\big),\\
\rho^{\langle GG\rangle}_3(s)=&\frac{\langle g_s^2 GG\rangle}{2}\int_{\alpha_{min}}^{\alpha_{max}}d\alpha\int_{\beta_{min}}^{\beta_{max}}d\beta
\Big{ \{ } \frac{m^2(1-\alpha-\beta)^2\big(m^2(4\alpha\beta-3\alpha+4\beta^2-4\beta)+3\alpha(1-2\beta)\beta s\big)}
{192\pi^6\alpha^3\beta}\\
&+\big(m^2(\alpha+\beta)-\alpha\beta s\big)\Big[ \frac{(1-\alpha-\beta)\big(m^2(\alpha+\beta-1)-2\alpha\beta s\big)}{64\pi^6\alpha^2\beta}+\frac{m^2\big(m^2(\alpha+\beta)-\alpha\beta s\big)}{128\pi^6\alpha \beta}\Big] \Big{\}},\\
\rho^{\langle \bar{q}Gq\rangle}_3(s)=&\int_{\alpha_{min}}^{\alpha_{max}}d\alpha\int_{\beta_{min}}^{\beta_{max}}d\beta
\\&
\frac{m_q\big(\sGs\big(m^2(2\alpha+2\beta-1)-3\alpha\beta s\big)+\qGqb
\big(3\alpha\beta s-2m^2(\alpha+\beta-1)\big)\big)}{32\pi^4\alpha}\\
&+\frac{m_q\big(m^2 \sGs-\qGqb(2m^2+s)\big)}{32\pi^4},\\
\rho^{\langle\bar{q}q\rangle^2}_3(s)=&\frac{\qq \ss \big(2m^2+s\big)}{6\pi^2}\sqrt{1-\frac{4m^2}{s}},\\
\Pi_3^{\langle\bar{q}q\rangle\langle \bar{q}Gq\rangle}(M_B^2)=&\frac{\qq\sGs+\ss\qGqb}{24\pi^2}
\int^1_0dx\Big{\{ } \frac{2m^4(2-x)}{M_B^2(1-x)x^2}+\frac{m^2}{1-x}+2M_B^2(1-x)\Big{\} }e^{-\frac{m^2}{M^2_B(1-x)x}}.
\end{split}
\end{equation}

\begin{equation}
\begin{split}
\rho^{pert}_4(s)=&\int_{\alpha_{min}}^{\alpha_{max}}d\alpha\int_{\beta_{min}}^{\beta_{max}}d\beta
\frac{(1-\alpha-\beta)(m^2-\alpha\beta s)\big( m^2(\alpha+\beta)-\alpha\beta s\big)^3}{32\pi^6\alpha^3\beta^3},\\
\rho^{\langle\bar{q} q\rangle}_4(s)=& \int_{\alpha_{min}}^{\alpha_{max}}d\alpha\int_{\beta_{min}}^{\beta_{max}}d\beta
\frac{m_q m^2\big(m^2(\alpha+\beta)-\alpha\beta s\big)}{4\pi^4\alpha\beta}\\
&\qquad \times \big(\langle \bar{q_2}q_2\rangle(m^2(2\alpha+2\beta+1)-4\alpha\beta s)+2\langle \bar{q_1}q_1\rangle
(m^2(\alpha+\beta+2)-2\alpha\beta s\big),\\
\rho^{\langle GG\rangle}_4(s)=&\frac{\langle g_s^2 GG\rangle}{2}\int_{\alpha_{min}}^{\alpha_{max}}d\alpha\int_{\beta_{min}}^{\beta_{max}}d\beta
\Big{ \{ } \frac{m^2(1-\alpha-\beta)^2\big(m^2(4\alpha\beta+3\alpha+4\beta^2+4\beta)-3\alpha\beta(2\beta+1)s\big)}{96\pi^6\alpha^2\beta}\\
&+\big(m^2(\alpha+\beta)-\alpha\beta s\big)\Big[ \frac{5(1-\alpha-\beta)\big(m^2(\alpha+\beta+1)-2\alpha\beta s\big)}{64\pi^6\alpha^2\beta}+\frac{m^2}{128\pi^6\alpha\beta}\Big] \Big{\} },\\
\rho^{\langle \bar{q}Gq\rangle}_4(s)=&\int_{\alpha_{min}}^{\alpha_{max}}d\alpha\int_{\beta_{min}}^{\beta_{max}}d\beta
\\&
\frac{5 m_q\big(\sGs\big(m^2(2\alpha+2\beta+1)-3\alpha\beta s\big)+\qGqb
\big(2m^2(\alpha+\beta+1)-3\alpha\beta s\big)\big)}{32\pi^4\alpha}\\
&+\frac{m_q\big(m^2 \sGs+\qGqb(s-6m^2)\big)}{32\pi^4}\sqrt{1-\frac{4m^2}{s}},\\
\rho^{\langle\bar{q}q\rangle^2}_4(s)=&-\frac{\qq\ss\big(s-6m^2\big)}{3\pi^2} \sqrt{1-\frac{4m^2}{s}},\\
\Pi_4^{\langle\bar{q}q\rangle\langle \bar{q}Gq\rangle}(M_B^2)=&\frac{\qq\sGs+\ss\qGqb}{24\pi^2}\\
&\qquad \times
\int^1_0dx\Big{\{ } \frac{4m^4(2-3x)}{M_B^2(1-x)x^2}+\frac{m^2(15x-14)}{x(1-x)}-2M_B^2\frac{5x^2-7x+2}{1-x}\Big{\} }e^{-\frac{m^2}{M^2_B(1-x)x}}.
\end{split}
\end{equation}

\begin{equation}
\begin{split}
\rho^{pert}_5(s)=&\int_{\alpha_{min}}^{\alpha_{max}}d\alpha\int_{\beta_{min}}^{\beta_{max}}d\beta
\frac{3(1-\alpha-\beta)\big( m^2(\alpha+\beta)-\alpha \beta s\big)^4}{32\pi^6\alpha^3\beta^3},\\
\rho^{\langle\bar{q} q\rangle}_5(s)=&\int_{\alpha_{min}}^{\alpha_{max}}d\alpha\int_{\beta_{min}}^{\beta_{max}}d\beta
\frac{3m_q\big(m^2(\alpha+\beta)-\alpha\beta s\big)\big( (m^2(\alpha+\beta)-2\alpha\beta s)\ss +2m^2\qq \big)}{2\pi^4\alpha\beta},\\
\rho^{\langle GG\rangle}_5(s)=&\frac{\langle g_s^2 GG\rangle}{2}\int_{\alpha_{min}}^{\alpha_{max}}d\alpha\int_{\beta_{min}}^{\beta_{max}}d\beta
\Big{\{} \frac{m^2(1-\alpha-\beta)^2\big( 2m^2(\alpha+\beta)-3\alpha\beta s\big)}{16\pi^6\alpha^3}\\
&-\frac{\big(\alpha(\alpha+12\beta-2)+\beta(9\beta-10)+1\big)\big(m^2(\alpha+\beta)-2\alpha \beta s\big)\big(m^2(\alpha+\beta)-\alpha \beta s\big)}{64\pi^6\alpha^2\beta^2}\Big{\}},\\
\rho^{\langle \bar{q}Gq\rangle}_5(s)=&\int_{\alpha_{min}}^{\alpha_{max}}d\alpha\int_{\beta_{min}}^{\beta_{max}}d\beta
\frac{m_q\big(2m^2(\alpha+\beta)-3\alpha\beta s\big)\sGs+2m^2\qGqb}{ 4\pi^4\alpha}\\
&\frac{m_q \sGs(s-2m^2)-12m_q m^2\qGqb}{16\pi^4}\sqrt{1-\frac{4m^2}{s}},\\
\rho^{\langle\bar{q}q\rangle^2}_5(s)=&\frac{4\qq \ss m^2}{\pi^2}\sqrt{1-\frac{4m^2}{s}},\\
\Pi_5^{\langle\bar{q}q\rangle\langle \bar{q}Gq\rangle}(M_B^2)=&\frac{m^2(\qq\sGs+\ss\qGqb)}{3\pi^2}
\int^1_0dx\Big{\{ } \frac{3m^2-2xM_B^2}{x^2M_B^2 }\Big{\} }e^{-\frac{m^2}{M^2_B(1-x)x}}.
\end{split}
\end{equation}

For the interpolating currents with $J^P=1^-$,

\begin{eqnarray}
\nonumber
\rho^{pert}_1(s)&=&\frac{1}{2^8\pi^6}\dab
\non&&
\frac{(1-\alpha-\beta)\f(s)^3}{\alpha^3\beta^3}
\Big\{(1+\alpha+\beta)\f(s)+4(1-\alpha-\beta)m^2\Big\}
\, , \non
\rho^{\qq}_1(s)&=&-\frac{m_s(2\qq-\ss)}{16\pi^4}\dab\frac{[(2+\alpha+\beta)m^2-3\alpha\beta s]\f(s)}{\alpha\beta} \, , \non
\rho^{\GGa}_1(s)&=&\frac{\GGb}{3\times2^{11}\pi^6}\dab \non
&&\bigg\{\frac{16(1-\alpha-\beta)^2m^2}{\alpha^3}\bigg[\frac{[(3\alpha+4\beta)m^2-3\alpha\beta s]}{\beta}+[(\alpha+\beta)m^2-2\alpha\beta s]\bigg]+\non
&&\frac{\f(s)}{\alpha\beta}\bigg[\frac{(1-\alpha-\beta)^2[3(\alpha+\beta)m^2-5\alpha\beta s]}{\alpha\beta}-6[(2+\alpha+\beta)m^2-3\alpha\beta s]\bigg]
\bigg\}
\, , \non
\rho^{\qGqa}_1(s)&=&\frac{m_s\qGqb(4m^2-s)}{48\pi^4}\sqrt{1-4m^2/s}\, , \\
\rho^{\qq^2}_1(s)&=&-\frac{\qq\ss(4m^2-s)}{9\pi^2}\sqrt{1-4m_c^2/s}
\, ,
\end{eqnarray}
\begin{eqnarray}
\nonumber\Pi^{\qGqa\qq}_1(M_B^2)&=&\frac{\qGqb\ss+\sGs\qq}{12\pi^2}\int_0^1d\alpha\bigg
\{\frac{m^4(2\alpha-1)}{M_B^2\alpha^2(1-\alpha)}+\frac{m^2(2-\alpha)}{(1-\alpha)}+M_B^2 \alpha\bigg\}\efun
\, .
\end{eqnarray}

\begin{eqnarray}
\nonumber
\rho^{pert}_2(s)&=&\frac{1}{3\times2^8\pi^6}\dab\frac{(1-\alpha-\beta)\f(s)^3}{\alpha^3\beta^3}
\non&&
\Big\{3(1+\alpha+\beta)\f(s)-4(1-\alpha-\beta)(2+\alpha+\beta)m^2\Big\}
\, , \non
\rho^{\qq}_2(s)&=&-\frac{m_s}{16\pi^4}\dab\frac{\f(s)}{\alpha\beta}
\non &&
\Big\{[(\alpha+\beta)m^2+3\alpha\beta s]\ss-4[(\alpha+\beta-1)m^2-2\alpha\beta s]\qq\Big\} \, , \non
\rho^{\GGa}_2(s)&=&-\frac{\GGb}{3^2\times2^{11}\pi^6}\dab \non
&&\bigg\{\frac{16(1-\alpha-\beta)^2m^2}{\alpha^3}\bigg[\frac{[(3\alpha+4\beta)m^2-3\alpha\beta s](2+\alpha+\beta)}{\beta}-3[(\alpha+\beta)m^2-2\alpha\beta s]\bigg]-\non
&&\frac{3\f(s)}{\alpha\beta}\bigg[\frac{(1-\alpha-\beta)^2[(\alpha+\beta+8)m^2+9\alpha\beta s]}{\alpha\beta}+2[(5\alpha+5\beta-4)m^2-5\alpha\beta s]\bigg]
\bigg\}
\, , \non
\rho^{\qGqa}_2(s)&=&\frac{m_s\qGqb s}{32\pi^4}\sqrt{1-4m^2/s}\, , \\
\rho^{\qq^2}_2(s)&=&-\frac{\qq\ss s}{6\pi^2}\sqrt{1-4m_c^2/s}
\, ,
\end{eqnarray}
\begin{eqnarray}
\nonumber\Pi^{\qGqa\qq}_2(M_B^2)&=&-\frac{\qGqb\ss+\sGs\qq}{12\pi^2}\int_0^1d\alpha\bigg
\{\frac{m^4}{M_B^2\alpha^2(1-\alpha)}+\frac{m^2}{(1-\alpha)}+M_B^2\bigg\}\efun
\, .
\end{eqnarray}

\begin{eqnarray}
\nonumber
\rho^{pert}_3(s)&=&-\frac{1}{3\times2^9\pi^6}\dab\frac{(1-\alpha-\beta)\f(s)^3}{\alpha^3\beta^3}
\non&&
\Big\{9(1+\alpha+\beta)\f(s)+4(1-\alpha-\beta)(4-\alpha-\beta)m^2\Big\}
\, , \non
\rho^{\qq}_3(s)&=&-\frac{m_s}{32\pi^4}\dab\frac{\f(s)}{\alpha\beta}
\Big\{[(\alpha+\beta+4)m^2-9\alpha\beta s]\ss-4(3m^2-\alpha\beta s)\qq\Big\} \, , \non
\rho^{\GGa}_3(s)&=&\frac{\GGb}{3^2\times2^{11}\pi^6}\dab \non
&&\bigg\{\frac{8(1-\alpha-\beta)^2m^2}{\alpha^3}\bigg[\frac{[(3\alpha+4\beta)m^2-3\alpha\beta s](\alpha+\beta-4)}{\beta}-9[(\alpha+\beta)m^2-2\alpha\beta s]\bigg]+\non
&&\frac{\f(s)}{\alpha\beta}\bigg[\frac{(1-\alpha-\beta)^2[(5\alpha+5\beta+16)m^2-27\alpha\beta s]}{\alpha\beta}
\non&&
-\frac{72(1-\alpha-\beta)[(\alpha+\beta+1)m^2-2\alpha\beta s]}{\alpha}-6[(\alpha+\beta-8)m^2-5\alpha\beta s]\bigg]
\bigg\}
\, , \non
\rho^{\qGqa}_3(s)&=&\frac{m_s\qGqb (16m^2-s)}{192\pi^4}\sqrt{1-4m^2/s}+
\frac{m_s\qGqb}{96\pi^4}\dab\frac{(3m^2-\alpha\beta s)}{\alpha}\, , \\
\rho^{\qq^2}_3(s)&=&-\frac{\qq\ss (16m^2-s)}{36\pi^2}\sqrt{1-4m_c^2/s}
\, ,
\end{eqnarray}
\begin{eqnarray}
\nonumber\Pi^{\qGqa\qq}_3(M_B^2)&=&\frac{\qGqb\ss+\sGs\qq}{72\pi^2}\int_0^1d\alpha
\non&&
\bigg\{\frac{3(3-4\alpha)m^4}{M_B^2\alpha^2(1-\alpha)}-\frac{3m^2}{\alpha(1-\alpha)}+\frac{m^2(4-6\alpha)}{(1-\alpha)}
+(3-6\alpha)M_B^2\bigg\}\efun
\, .
\end{eqnarray}

\begin{eqnarray}
\nonumber
\rho^{pert}_4(s)&=&-\frac{1}{3\times2^9\pi^6}\dab\frac{(1-\alpha-\beta)\f(s)^3}{\alpha^3\beta^3}
\non&&
\Big\{9(1+\alpha+\beta)\f(s)-4(1-\alpha-\beta)(1+2\alpha+2\beta)m^2\Big\}
\, , \non
\rho^{\qq}_4(s)&=&\frac{m_s}{16\pi^4}\dab
\non&&
\frac{\f(s)}{\alpha\beta}
\Big\{[(2\alpha+2\beta-1)m^2+3\alpha\beta s]\ss-[3(\alpha+\beta-2)m^2-5\alpha\beta s]\qq\Big\} \, , \non
\rho^{\GGa}_4(s)&=&\frac{\GGb}{3^2\times2^{11}\pi^6}\dab \non
&&\bigg\{\frac{8(1-\alpha-\beta)^2m^2}{\alpha^3}\bigg[\frac{3[(3\alpha+4\beta)m^2-3\alpha\beta s]}{\beta}+2(1-\alpha-\beta)[3(\alpha+\beta)m^2-4\alpha\beta s]
\non&&
-9[(\alpha+\beta)m^2-2\alpha\beta s]\bigg]-\frac{3\f(s)}{\alpha\beta}\bigg[\frac{(1-\alpha-\beta)^2[7(\alpha+\beta)m^2-13\alpha\beta s]}{\alpha\beta}
\non&&
+\frac{16(1-\alpha-\beta)[(\alpha+\beta-3)m^2-3\alpha\beta s]}{\alpha}+2[(19\alpha+19\beta-2)m^2-33\alpha\beta s]\bigg]
\bigg\}
\, , \non
\rho^{\qGqa}_4(s)&=&\frac{m_s\qGqb (2m^2+s)}{48\pi^4}\sqrt{1-4m^2/s}-
\frac{m_s\qGqb}{96\pi^4}\dab\frac{[3(\alpha+\beta-1)m^2-4\alpha\beta s]}{\alpha}\, , \\
\rho^{\qq^2}_4(s)&=&-\frac{\qq\ss (2m^2+s)}{9\pi^2}\sqrt{1-4m_c^2/s}
\, ,
\end{eqnarray}
\begin{eqnarray}
\nonumber\Pi^{\qGqa\qq}_4(M_B^2)&=&\frac{\qGqb\ss+\sGs\qq}{72\pi^2}\int_0^1d\alpha
\non&&
\bigg\{\frac{3(3-2\alpha)m^4}{M_B^2\alpha^2(1-\alpha)}+\frac{(2+\alpha)m^2}{\alpha(1-\alpha)}
-3m^2+6(1-\alpha)M_B^2\bigg\}\efun
\, .
\end{eqnarray}

\begin{eqnarray}
\nonumber
\rho^{pert}_5(s)&=&\frac{1}{2^9\pi^6}\dab
\non&&
\frac{(1-\alpha-\beta)\f(s)^3}{\alpha^3\beta^3}
\Big\{(1+\alpha+\beta)\f(s)-4(1-\alpha-\beta)m^2\Big\}
\, , \non
\rho^{\qq}_5(s)&=&\frac{m_s(2\qq+\ss)}{32\pi^4}\dab\frac{[(\alpha+\beta-2)m^2-3\alpha\beta s]\f(s)}{\alpha\beta} \, , \non
\rho^{\GGa}_5(s)&=&-\frac{\GGb}{3\times2^{11}\pi^6}\dab \non
&&\bigg\{\frac{8(1-\alpha-\beta)^2m^2}{\alpha^3}\bigg[\frac{3\f(s)}{\beta}-[(\alpha+\beta-1)m^2-2\alpha\beta s]\bigg]+\non
&&\frac{\f(s)}{\alpha\beta}\bigg[\frac{(1-\alpha-\beta)^2[3(\alpha+\beta)m^2-5\alpha\beta s]}{\alpha\beta}-6[(\alpha+\beta-2)m^2-3\alpha\beta s]\bigg]
\bigg\}
\, , \non
\rho^{\qGqa}_5(s)&=&\frac{m_s\qGqb(2m^2+s)}{96\pi^4}\sqrt{1-4m^2/s}\, , \\
\rho^{\qq^2}_5(s)&=&-\frac{\qq\ss(2m^2+s)}{18\pi^2}\sqrt{1-4m_c^2/s}
\, ,
\end{eqnarray}
\begin{eqnarray}
\nonumber\Pi^{\qGqa\qq}_5(M_B^2)&=&-\frac{\qGqb\ss+\sGs\qq}{24\pi^2}\int_0^1d\alpha\bigg
\{\frac{m^4}{M_B^2\alpha^2(1-\alpha)}+\frac{m^2(2-\alpha)}{(1-\alpha)}+M_B^2 \alpha\bigg\}\efun
\, .
\end{eqnarray}

\begin{eqnarray}
\nonumber
\rho^{pert}_6(s)&=&-\frac{1}{3\times2^8\pi^6}\dab\frac{(1-\alpha-\beta)\f(s)^3}{\alpha^3\beta^3}
\non&&
\Big\{3(1+\alpha+\beta)\f(s)+4(1-\alpha-\beta)(2+\alpha+\beta)m^2\Big\}
\, , \non
\rho^{\qq}_6(s)&=&-\frac{m_s}{16\pi^4}\dab\frac{\f(s)}{\alpha\beta}
\non &&
\Big\{3\f(s)\ss-4[(\alpha+\beta+1)m^2-2\alpha\beta s]\qq\Big\} \, , \non
\rho^{\GGa}_6(s)&=&-\frac{\GGb}{3^2\times2^{11}\pi^6}\dab \non
&&\bigg\{\frac{16(1-\alpha-\beta)^2m^2}{\alpha^3}\bigg[\frac{3(2+\alpha+\beta)\f(s)}{\beta}
+2[(2\alpha+2\beta+1)m^2-3\alpha\beta s]\bigg]-\non
&&\frac{3\f(s)}{\alpha\beta}\bigg[\frac{(1-\alpha-\beta)^2[(7\alpha+7\beta+8)m^2-9\alpha\beta s]}{\alpha\beta}-2[(\alpha+\beta+4)m^2-5\alpha\beta s]\bigg]
\bigg\}
\, , \non
\rho^{\qGqa}_6(s)&=&\frac{m_s\qGqb (4m^2-s)}{32\pi^4}\sqrt{1-4m^2/s}\, , \\
\rho^{\qq^2}_6(s)&=&-\frac{\qq\ss (4m^2-s)}{6\pi^2}\sqrt{1-4m_c^2/s}
\, ,
\end{eqnarray}
\begin{eqnarray}
\nonumber\Pi^{\qGqa\qq}_6(M_B^2)&=&-\frac{\qGqb\ss+\sGs\qq}{12\pi^2}\int_0^1d\alpha\bigg
\{\frac{m^4(1-2\alpha)}{M_B^2\alpha^2(1-\alpha)}-\frac{m^2}{\alpha(1-\alpha)}-M_B^2\bigg\}\efun
\, .
\end{eqnarray}

\begin{eqnarray}
\nonumber
\rho^{pert}_7(s)&=&-\frac{1}{3\times2^9\pi^6}\dab\frac{(1-\alpha-\beta)\f(s)^3}{\alpha^3\beta^3}
\non&&
\Big\{9(1+\alpha+\beta)\f(s)-4(1-\alpha-\beta)(4-\alpha-\beta)m^2\Big\}
\, , \non
\rho^{\qq}_7(s)&=&-\frac{m_s}{32\pi^4}\dab\frac{\f(s)}{\alpha\beta}
\Big\{[(5\alpha+5\beta-4)m^2-9\alpha\beta s]\ss-4(3m^2+\alpha\beta s)\qq\Big\} \, , \non
\rho^{\GGa}_7(s)&=&\frac{\GGb}{3^2\times2^{11}\pi^6}\dab \non
&&\bigg\{\frac{8(1-\alpha-\beta)^2m^2}{\alpha^3}\bigg[\frac{3(4-\alpha-\beta)\f(s)}{\beta}-
[(10\alpha+10\beta-4)m^2-18\alpha\beta s]\bigg]+\non
&&\frac{\f(s)}{\alpha\beta}\bigg[\frac{(1-\alpha-\beta)^2[(13\alpha+13\beta-16)m^2-27\alpha\beta s]}{\alpha\beta}
\non&&
-\frac{24(1-\alpha-\beta)[(5\alpha+5\beta-3)m^2-6\alpha\beta s]}{\alpha}-6[(5\alpha+5\beta+8)m^2-5\alpha\beta s]\bigg]
\bigg\}
\, , \non
\rho^{\qGqa}_7(s)&=&\frac{m_s\qGqb (20m^2+s)}{192\pi^4}\sqrt{1-4m^2/s}+
\frac{m_s\qGqb}{96\pi^4}\dab\frac{(3m^2+\alpha\beta s)}{\alpha}\, , \\
\rho^{\qq^2}_7(s)&=&-\frac{\qq\ss (20m^2+s)}{36\pi^2}\sqrt{1-4m_c^2/s}
\, ,
\end{eqnarray}
\begin{eqnarray}
\nonumber\Pi^{\qGqa\qq}_7(M_B^2)&=&-\frac{\qGqb\ss+\sGs\qq}{72\pi^2}\int_0^1d\alpha
\non&&
\bigg\{\frac{3(3-2\alpha)m^4}{M_B^2\alpha^2(1-\alpha)}-\frac{(7-10\alpha)m^2}{\alpha(1-\alpha)}+6m^2
+(3-6\alpha)M_B^2\bigg\}\efun
\, .
\end{eqnarray}

\begin{eqnarray}
\nonumber
\rho^{pert}_8(s)&=&-\frac{1}{3\times2^8\pi^6}\dab\frac{(1-\alpha-\beta)\f(s)^3}{\alpha^3\beta^3}
\non&&
\Big\{9(1+\alpha+\beta)\f(s)+4(1-\alpha-\beta)(1+2\alpha+2\beta)m^2\Big\}
\, , \non
\rho^{\qq}_8(s)&=&-\frac{m_s}{16\pi^4}\dab\frac{\f(s)}{\alpha\beta}
\non&&
\Big\{[(7\alpha+7\beta-2)m^2-9\alpha\beta s]\ss-2[3(\alpha+\beta+2)m^2-5\alpha\beta s]\qq\Big\} \, , \non
\rho^{\GGa}_8(s)&=&-\frac{\GGb}{3^2\times2^{11}\pi^6}\dab \non
&&\bigg\{\frac{16(1-\alpha-\beta)^2m^2}{\alpha^3}\bigg[\frac{3(1+2\alpha+2\beta)\f(s)}{\beta}+
[(11\alpha+11\beta+1)m^2-18\alpha\beta s]\bigg]-\non
&&\frac{3\f(s)}{\alpha\beta}\bigg[\frac{(1-\alpha-\beta)^2[3(\alpha+\beta)m^2-5\alpha\beta s]}{\alpha\beta}
\non&&
-\frac{240(1-\alpha-\beta)\f(s)}{\alpha}-2[(7\alpha+7\beta-2)m^2-9\alpha\beta s]\bigg]
\bigg\}
\, , \non
\rho^{\qGqa}_8(s)&=&\frac{m_s\qGqb (7m^2-s)}{24\pi^4}\sqrt{1-4m^2/s}+
\frac{5m_s\qGqb}{96\pi^4}\dab\frac{[3(\alpha+\beta+1)m^2-4\alpha\beta s]}{\alpha}\, , \\
\rho^{\qq^2}_8(s)&=&-\frac{2\qq\ss (7m^2-s)}{9\pi^2}\sqrt{1-4m_c^2/s}
\, ,
\end{eqnarray}
\begin{eqnarray}
\nonumber\Pi^{\qGqa\qq}_8(M_B^2)&=&-\frac{\qGqb\ss+\sGs\qq}{72\pi^2}\int_0^1d\alpha
\non&&
\bigg\{\frac{6(3-4\alpha)m^4}{M_B^2\alpha^2(1-\alpha)}+\frac{2(2\alpha-11)m^2}{\alpha(1-\alpha)}
-\frac{6(\alpha-3)m^2}{(1-\alpha)}
+(21\alpha-12)M_B^2\bigg\}\efun
\, .
\end{eqnarray}

For the interpolating currents with $J^P=1^+$,

\begin{equation}
\begin{split}
\rho^{pert}_1(s)=&\int_{\alpha_{min}}^{\alpha_{max}}d\alpha\int_{\beta_{min}}^{\beta_{max}}d\beta
\Big{\{} \frac{(1-\alpha-\beta)^2\big(\alpha\beta s-m^2(\alpha+\beta-4)\big)\big(m^2(\alpha+\beta)-\alpha\beta s\big)^3}{256\pi^6\alpha^3\beta^3}\\
&+\frac{(1-\alpha-\beta)\big(m^2(\alpha+\beta)-\alpha\beta s\big)^4}{128\pi^6\alpha^3\beta^3}\Big{\}},\\
\rho^{\langle\bar{q} q\rangle}_1(s)=&m_q(\qq+\ss) \int_{\alpha_{min}}^{\alpha_{max}}d\alpha\int_{\beta_{min}}^{\beta_{max}}d\beta
\frac{\big(m^2(\alpha+\beta)-\alpha\beta s\big)\big(m^2(\alpha+\beta+2)-3\alpha\beta s\big)}{16\pi^4\alpha\beta},\\
\rho^{\langle GG\rangle}_1(s)=&\frac{\langle g_s^2 GG\rangle}{2}\int_{\alpha_{min}}^{\alpha_{max}}d\alpha\int_{\beta_{min}}^{\beta_{max}}d\beta
\Big{\{} \frac{\big(m^2(\alpha+\beta+2)-\alpha\beta s\big)\big(\alpha\beta s-m^2(\alpha+\beta)\big)}{512\pi^6\alpha\beta}\\
&+(1-\alpha-\beta)^2\Big[ \frac{\big(m^2(\alpha+\beta)-\alpha\beta s\big)\big(m^2(3\alpha^2+3\alpha\beta-16\beta^3+48\beta)-5\alpha^2\beta s\big)}{3072\pi^6\alpha^3\beta^2}\\
&+\frac{m^2\big(2m^2(\alpha+\beta)+m^2-3\alpha\beta s\big)}{192\pi^6\alpha^3},\\
\rho^{\langle \bar{q}Gq\rangle}_1(s)=&\frac{m_q(4\qGqb+\sGs)\big(s-4m^2)}{192\pi^4}
\sqrt{1-\frac{4m^2}{s}},\\
\rho^{\langle\bar{q}q\rangle^2}_1(s)=&-\frac{\qq\ss \big(s-4m^2\big)}{9\pi^2}\sqrt{1-\frac{4m^2}{s}},\\
\Pi_1^{\langle\bar{q}q\rangle\langle \bar{q}Gq\rangle}(M_B^2)=&-\frac{\qq\sGs+\ss\qGqb}{12\pi^2}
\int^1_0dx\Big{\{ } \frac{m^4(2x-1)}{M_B^2 x^2(1-x)}+\frac{m^2(2-x)}{1-x}+M_B^2 x\Big{\} }e^{-\frac{m^2}{M^2_B(1-x)x}}.
\end{split}
\end{equation}

\begin{equation}
\begin{split}
\rho^{pert}_2(s)=&\int_{\alpha_{min}}^{\alpha_{max}}d\alpha\int_{\beta_{min}}^{\beta_{max}}d\beta
\\&
\frac{(1-\alpha-\beta)\big(\alpha\beta s-m^2(\alpha+\beta)\big)\big(m^2(\alpha+\beta+1)(\alpha+\beta)-8)+3\alpha\beta s(\alpha+\beta+1)\big)}{768\pi^6\alpha^3\beta^3},\\
\rho^{\langle\bar{q} q\rangle}_1(s)=& \int_{\alpha_{min}}^{\alpha_{max}}d\alpha\int_{\beta_{min}}^{\beta_{max}}d\beta
\Big{\{} \frac{m_q\qq\big(m^2(\alpha+\beta)-\alpha\beta s\big)\big(m^2(\alpha+\beta+1)-2\alpha\beta s\big)}{4\pi^4\alpha\beta}\\
&+\frac{m_q\ss\big(m^2(\alpha+\beta)(3\alpha+3\beta+1)-(5\alpha+5\beta-1)\alpha\beta s\big)}{16\pi^4\alpha\beta},\\
\rho^{\langle GG\rangle}_2(s)=&\frac{\langle g_s^2 GG\rangle}{2}\int_{\alpha_{min}}^{\alpha_{max}}d\alpha\int_{\beta_{min}}^{\beta_{max}}d\beta
\\&
\Big{\{} \frac{m^2 (\alpha +\beta -1)^2 \left(m^2 (3 \alpha ^2+6 \alpha  \beta +4 \alpha +3 \beta ^2+4 \beta -1)-\alpha  \beta  s (4 \alpha +4 \beta +5)\right)}{576 \pi ^6 \alpha ^3}\\
&+\big(m^2(\alpha+\beta)-\alpha\beta s\big)\Big[ \frac{(1-\alpha-\beta)^3(\alpha-4\beta)}{768\pi^6\alpha^3\beta^2}-\frac{(1-\alpha-\beta)^2\big(m^2(\alpha^2+\alpha\beta+2\alpha-8\beta)-2\alpha\beta s\big)}{512\pi^6\alpha^3\beta^2}\\
&+\frac{(1-\alpha-\beta)\big(m^2(3\alpha+3\beta+2)-5\alpha\beta s\big)}{1536\pi^6\alpha\beta}-\frac{m^2(\alpha+\beta+1)-2\alpha\beta s}{768\pi^6\alpha\beta}\Big]\Big{\}},\\
\rho^{\langle \bar{q}Gq\rangle}_2(s)=&\frac{m_q\qGqb\big(s-4m^2)}{32\pi^4}
\sqrt{1-\frac{4m^2}{s}},\\
\rho^{\langle\bar{q}q\rangle^2}_2(s)=&-\frac{\qq\ss \big(s-4m^2\big)}{6\pi^2}\sqrt{1-\frac{4m^2}{s}},\\
\Pi_2^{\langle\bar{q}q\rangle\langle \bar{q}Gq\rangle}(M_B^2)=&\frac{\qq\sGs+\ss\qGqb}{12\pi^2}
\int^1_0dx\Big{\{ } \frac{m^4(1-2x)}{M_B^2 x^2(1-x)}-\frac{m^2}{x(1-x)}-M_B^2 \Big{\} }e^{-\frac{m^2}{M^2_B(1-x)x}}.
\end{split}
\end{equation}

\begin{equation}
\begin{split}
\rho^{pert}_3(s)=&\int_{\alpha_{min}}^{\alpha_{max}}d\alpha\int_{\beta_{min}}^{\beta_{max}}d\beta
(1-\alpha-\beta)\big(m^2(\alpha+\beta)-\alpha\beta s\big)\Big[ \frac{(7-\alpha-\beta)\big(m^2(\alpha+\beta)-\alpha\beta s\big)}{512\pi^6\alpha^3\beta^3} \\
&+\frac{(1-\alpha-\beta)\big(m^2(3\alpha^2+6\alpha\beta-4\alpha+3\beta^2-4\beta-4)-7(\alpha+\beta-1)\alpha\beta s\big)}{384\pi^6\alpha^3\beta^3}\Big],\\
\rho^{\langle\bar{q} q\rangle}_3(s)=&\int_{\alpha_{min}}^{\alpha_{max}}d\alpha\int_{\beta_{min}}^{\beta_{max}}d\beta
\Big{\{} \frac{m_q\qq (3m^2+\alpha\beta s)\big(m^2(\alpha+\beta)-\alpha\beta\big)}{8\pi^4\alpha\beta}\\
&+\frac{m_q\ss \big(m^2(\alpha+\beta)-\alpha\beta s\big)\big(\alpha\beta s(25\alpha+25\beta-37)-m^2((\alpha+\beta)(15(\alpha+\beta)-23)+4)\big)}{32\pi^4\alpha\beta}\Big{\}},\\
\rho^{\langle GG\rangle}_3(s)=&\frac{\langle g_s^2 GG\rangle}{2}\int_{\alpha_{min}}^{\alpha_{max}}d\alpha\int_{\beta_{min}}^{\beta_{max}}d\beta
\\&
\Big{\{} \big(m^2(\alpha+\beta)-\alpha\beta s\big) \Big[ \frac{(1-\alpha-\beta)^3\big(25\alpha^2\beta s-m^2(15\alpha^2+15\alpha\beta-4\alpha
+24\beta)\big)}{9216\pi^6\alpha^3\beta^2}\\
&+\frac{(1-\alpha-\beta)\big(m^2(15\alpha^2\beta-3\alpha^2+15\alpha\beta^2-7\alpha\beta+2\alpha-12\beta)+\alpha^2(6-25\beta)\beta s\big)}{1536\pi^6\alpha^3\beta^2}\\
&+\frac{(1-\alpha-\beta)\big(m^2(3\alpha^2+3\alpha\beta+4\alpha+6\beta-8)-\alpha^2(5\alpha+14\beta)\beta s\big)}{1536\pi^6\alpha^2\beta}+\frac{m^2(3\alpha+3\beta+5)-4\alpha\beta s}{768\pi^6\alpha\beta}\Big]\\
&+\frac{ m^2 (\alpha +\beta -1)^2 \left(\alpha  \beta  s (20 \alpha +20 \beta -47)-m^2 \left(15 \alpha ^2+\alpha  (30 \beta -34)+15 \beta ^2-34 \beta +4\right)\right)}{1152 \pi ^6 \alpha ^3}\Big{\}},\\
\rho^{\langle \bar{q}Gq\rangle}_3(s)=&\Big{\{} \int_{\alpha_{min}}^{\alpha_{max}}d\alpha\int_{\beta_{min}}^{\beta_{max}}d\beta
\frac{m_q\qGqb (3m^2+\alpha\beta s)}{96\pi^4\alpha}\\
&-\frac{m_q\sGs \big(m^2(72\alpha^2+117\alpha\beta+45\beta^2-56\alpha-56\beta+8)-3\alpha\beta s(32\alpha+20\beta-25)\big)}{576\pi^4\alpha}\Big{\}}\\
&-\frac{6m_q\qGqb(20m^2+s)+\sGs(7s-52m^2)}{1152\pi^4}\sqrt{1-\frac{4m^2}{2}},\\
\rho^{\langle\bar{q}q\rangle^2}_3(s)=&\frac{\qq\ss \big(s+20m^2\big)}{36\pi^2}\sqrt{1-\frac{4m^2}{s}},\\
\Pi_3^{\langle\bar{q}q\rangle\langle \bar{q}Gq\rangle}(s)=&\frac{\qq\sGs+\ss\qGqb}{72\pi^2}\\
&\times \int^1_0dx\Big{\{ } \frac{m^4(9-6x)}{M_B^2 x^2(1-x)}-\frac{2m^2(3x^2-7x+3)}{x(1-x)}-\frac{3M_B^2(2x^2-3x+1)}{1-x} \Big{\} }e^{-\frac{m^2}{M^2_B(1-x)x}}.
\end{split}
\end{equation}

\begin{equation}
\begin{split}
\rho^{pert}_4(s)=&\int_{\alpha_{min}}^{\alpha_{max}}d\alpha\int_{\beta_{min}}^{\beta_{max}}d\beta
\Big{\{} \frac{(1-\alpha-\beta)^3\big(m^2(\alpha+\beta)-\alpha\beta s \big)^3\big(m^2(3\alpha+3\beta+2)-7\alpha\beta s\big)}
{384\pi^6\alpha^3\beta^3},\\
\rho^{\langle\bar{q}q\rangle}_4(s)=&\int_{\alpha_{min}}^{\alpha_{max}}d\alpha\int_{\beta_{min}}^{\beta_{max}}d\beta
\Big{\{} \frac{m_q\langle \bar{q_1}q_1\rangle \big(m^2(\alpha+\beta)-\alpha\beta s\big)\big(5\alpha\beta s-3m^2(\alpha+\beta-2)\big)}{16\pi^4\alpha\beta}\\
&+\frac{m_q\ss \big(m^2(\alpha+\beta)-\alpha\beta s\big)\big(m^2(-12\alpha^2-24\alpha\beta-12\beta^2+15\alpha+15\beta+2)+\alpha\beta s(20\alpha+20\beta-33)\big)}{32\pi^4\alpha\beta}\Big{\}},\\
\rho^{\langle GG\rangle}_3(s)=&\frac{\langle g_s^2 GG\rangle}{2}\int_{\alpha_{min}}^{\alpha_{max}}d\alpha\int_{\beta_{min}}^{\beta_{max}}d\beta
\Big{\{} \frac{(1-\alpha-\beta)^3\big(m^2(\alpha+\beta)-\alpha\beta s\big)\big(5\alpha^2\beta s-3m^2(\alpha^2+\alpha\beta+2\beta)\big)}{1152\pi^6\alpha^3\beta^2}\\
&+\frac{(1-\alpha-\beta)^2\big(m^2(\alpha+\beta)-\alpha\beta s\big)\big(5\alpha^2\beta(6\beta+1)s-m^2(18 \alpha ^2 \beta +3 \alpha ^2+18 \alpha  \beta ^2+19 \alpha  \beta +8 \beta ^2-24 \beta)\big)}{3072\pi^6\alpha^3\beta^2}\\
&-\frac{(1-\alpha-\beta)\big(m^2(\alpha+\beta)-\alpha\beta s\big)\big(m^2(-6\alpha^2+3\alpha-6\alpha\beta+5\beta-4)+\alpha(10\alpha-9)\beta s\big)}{768\pi^6\alpha^2\beta}\\
&+\frac{\big(m^2\alpha+\beta)-\alpha\beta s\big)\big(m^2(7\alpha+7\beta-2)-13\alpha\beta s\big)}{1536\pi^6\alpha\beta}\\
&-\frac{(1-\alpha-\beta)^2m^2}{1152\pi^6\alpha^3}\Big[ 2(1-\alpha-\beta)\big(m^2(6\alpha+6\beta+1)-8\alpha\beta s\big)+
3\big(m^2(6\alpha+6\beta-1)-9\alpha\beta s\big)\Big]\Big{\}},\\
\rho^{\langle \bar{q}Gq\rangle}_4(s)=&\int_{\alpha_{min}}^{\alpha_{max}}d\alpha\int_{\beta_{min}}^{\beta_{max}}d\beta
\Big{\{} \frac{m_q\qGqb\big(4\alpha\beta s-3m^2(\alpha+\beta-1)\big)}{96\pi^4\alpha}\\
&\frac{m_q\sGs\big(m^2(9\alpha^2-18\alpha\beta-27\beta^2+38\alpha+38\beta-2)-3\alpha\beta s(4\alpha-12\beta+19)\big)}{576\pi^4\alpha}\Big{\}}\\
&+\frac{m_q\big(12\qGqb(2m^2+s)-\sGs(5s-8m^2)\big)}{576\pi^4}\sqrt{1-\frac{4m^2}{s}},\\
\rho^{\langle\bar{q}q\rangle^2}_4(s)=&\frac{\qq\ss \big(2m^2+s\big)}{9\pi^2}\sqrt{1-\frac{4m^2}{s}},\\
\Pi_4^{\langle\bar{q}q\rangle\langle \bar{q}Gq\rangle}(s)=&\frac{\qq\sGs+\ss\qGqb}{72\pi^2}\\
&\times
\int^1_0dx\Big{\{ } \frac{m^4(9-6x)}{M_B^2x^2(1-x)}+\frac{m^2(3x^2-4x+3)}{x(1-x)}+6M_B^2(1-x)\Big{\}}e^{-\frac{m^2}{M^2_B(1-x)x}}.
\end{split}
\end{equation}

\begin{equation}
\begin{split}
\rho^{pert}_5(s)=&\int_{\alpha_{min}}^{\alpha_{max}}d\alpha\int_{\beta_{min}}^{\beta_{max}}d\beta
\\&
 \frac{(1-\alpha-\beta)\big(m^2(\alpha+\beta)-\alpha\beta s\big)^3\big(m^2((\alpha+\beta)(\alpha+\beta+5)-4)-\alpha\beta s(\alpha+\beta+1)\big)}{512\pi^6\alpha^3\beta^3},\\
\rho^{\langle\bar{q} q\rangle}_5(s)=&3m_q(\ss-2\qq) \int_{\alpha_{min}}^{\alpha_{max}}d\alpha\int_{\beta_{min}}^{\beta_{max}}d\beta
\frac{\big(m^2(\alpha+\beta)-\alpha\beta s\big)\big(m^2(\alpha+\beta-2)-3\alpha\beta s\big)}{32\pi^4\alpha\beta},\\
\rho^{\langle GG\rangle}_5(s)=&\langle g_s^2 GG\rangle\int_{\alpha_{min}}^{\alpha_{max}}d\alpha\int_{\beta_{min}}^{\beta_{max}}d\beta
\Big{\{} \big(m^2(\alpha+\beta)-\alpha\beta s\big)\Big[-\frac{\big(\alpha^2+\alpha(8\beta-2)+(\beta-1)^2\big)}{6144\pi^6\alpha^2\beta^2}\\
&+\frac{6\alpha^2\beta\big(m^2(\alpha+\beta-1)-2\alpha\beta s\big)-(\alpha-\beta-1)\big(m^2(\alpha^2+\alpha\beta+4\beta(\beta+3))-2\alpha^2\beta s\big)}{3072\pi^6\alpha^3}\Big]\\
&+\frac{m^2(1-\alpha-\beta)^2\big(m^2(2\alpha+2\beta-1)-3\alpha\beta s\big)}{768\pi^6\alpha^3}\Big{\}},\\
\rho^{\langle \bar{q}Gq\rangle}_5(s)=&-\frac{m_q(\qGqb+\sGs)\big(s+m^2)}{192\pi^4}
\sqrt{1-\frac{4m^2}{s}},\\
\rho^{\langle\bar{q}q\rangle^2}_5(s)=&\frac{\qq\ss \big(s+2m^2\big)}{18\pi^2}\sqrt{1-\frac{4m^2}{s}},\\
\Pi_5^{\langle\bar{q}q\rangle\langle \bar{q}Gq\rangle}(M_B^2)=&\frac{\qq\sGs+\ss\qGqb}{24\pi^2}\\
&\times \int^1_0dx\Big{\{ } \frac{m^4}{M_B^2 x^2(1-x)}+\frac{m^2(2-x)}{1-x}+M_B^2 x\Big{\} }e^{-\frac{m^2}{M^2_B(1-x)x}}.
\end{split}
\end{equation}

\begin{equation}
\begin{split}
\rho^{pert}_6(s)=&\int_{\alpha_{min}}^{\alpha_{max}}d\alpha\int_{\beta_{min}}^{\beta_{max}}d\beta
\\&
\frac{(1-\alpha-\beta)\big(m^2(\alpha+\beta)-\alpha\beta s\big)^3\big(m^2(7(\alpha+\beta)(\alpha+\beta+1)-8)-3
\alpha\beta s(\alpha+\beta+1)\big)}{768\pi^6\alpha^3\beta^3},\\
\rho^{\langle\bar{q} q\rangle}_6(s)=&\int_{\alpha_{min}}^{\alpha_{max}}d\alpha\int_{\beta_{min}}^{\beta_{max}}d\beta
\big(m^2(\alpha+\beta)-\alpha\beta s\big)\Big{\{} \frac{m_q\qq\big(2\alpha\beta s-
m^2(\alpha+\beta-1)\big)}{4\pi^4\alpha\beta}\\
&+\frac{m_q\ss \big(3m^2(\alpha^2+\alpha(2\beta-1)+\beta(\beta01))+\alpha\beta s(-5\alpha-5\beta+1)\big)}{16\pi^4\alpha\beta}\Big{\}},\\
\rho^{\langle GG\rangle}_6(s)=&\frac{\langle g_s^2 GG\rangle}{2}\int_{\alpha_{min}}^{\alpha_{max}}d\alpha\int_{\beta_{min}}^{\beta_{max}}d\beta
\\&
\Big{\{} \frac{(1-\alpha-\beta)^3\big(m^2(\alpha+\beta)-\alpha\beta s\big)\big( 5\alpha^2\beta s
-m^2(3\alpha^2+3\alpha\beta-4\alpha+16\beta)\big)}{3072\pi^6\alpha^3\beta^2}\\
&-\frac{(1-\alpha-\beta)^2\big(m^2(\alpha+\beta)-\alpha\beta s\big)\big(m^2(\alpha^2+\alpha\beta-2\alpha+8\beta)\big)}
{512\pi^6\alpha^3\beta^2}\\
&+\frac{m^2(1-\alpha-\beta)^2\big(m^2(3\alpha^2+6\alpha\beta+2\alpha+3\beta^2+2\beta-2)-\alpha\beta s(4\alpha+4\beta+5)\big)}
{576\pi^6\alpha^3}\\
&+\frac{(1-\alpha-\beta)\big(m^2(\alpha+\beta)-\alpha\beta s\big)\big(m^2(3\alpha+3\beta-2)-5\alpha\beta s\big)}{1536\pi^6\alpha\beta}\\
&+\frac{\big(m^2(\alpha+\beta)-\alpha\beta s\big)\big(m^2(\alpha+\beta-1)-2\alpha\beta s\big)}{768\pi^6\alpha\beta}\Big{\}},\\
\rho^{\langle \bar{q}Gq\rangle}_6(s)=&-\frac{m_q\qGqb s}{32\pi^4}
\sqrt{1-\frac{4m^2}{s}},\\
\rho^{\langle\bar{q}q\rangle^2}_6(s)=&\frac{\qq\ss s}{6\pi^2}\sqrt{1-\frac{4m^2}{s}},\\
\Pi_6^{\langle\bar{q}q\rangle\langle \bar{q}Gq\rangle}(M_B^2)=&-\frac{\qq\sGs+\ss\qGqb}{12\pi^2}\int^1_0dx\Big{\{ } \frac{m^4}{M_B^2 x^2(1-x)}+\frac{m^2}{(1-x)x}+M_B^2 \Big{\} }e^{-\frac{m^2}{M^2_B(1-x)x}}.
\end{split}
\end{equation}

\begin{equation}
\begin{split}
\rho^{pert}_7(s)=&\int_{\alpha_{min}}^{\alpha_{max}}d\alpha\int_{\beta_{min}}^{\beta_{max}}d\beta
\Big{\{} \frac{(1-\alpha-\beta)^3\big(m^2(\alpha+\beta)-\alpha\beta s\big)\big( m^2(3\alpha+3\beta+1)-7\alpha\beta s\big)}
{384\pi^6\alpha^3\beta^3}\\
&+\frac{(1-\alpha-\beta)\big(m^2(\alpha+\beta)-\alpha\beta s\big)\big(\alpha\beta s(\alpha+\beta-7)-m^2(\alpha^2+2\alpha\beta
-3\alpha+\beta^3-3\beta-4)\big)}{512\pi^6\alpha^3\beta^3}\Big{\}},\\
\rho^{\langle\bar{q} q\rangle}_7(s)=&\int_{\alpha_{min}}^{\alpha_{max}}d\alpha\int_{\beta_{min}}^{\beta_{max}}d\beta
\big(m^2(\alpha+\beta)-\alpha\beta s\big)\Big{\{} \frac{m_q\qq\big(3m^2-\alpha\beta s\big)}{8\pi^4\alpha\beta}\\
&\frac{m_q\ss\big(m^2(-15\alpha^2+\alpha(19-30\beta)-15\beta^2+19\beta+4)+
\alpha\beta s(25\alpha+25\beta-37)\big)}{32\pi^4\alpha\beta}
\Big{\}},\\
\rho^{\langle GG\rangle}_7(s)=&\frac{\langle g_s^2 GG\rangle}{2}\int_{\alpha_{min}}^{\alpha_{max}}d\alpha\int_{\beta_{min}}^{\beta_{max}}d\beta
\Big{\{} \frac{m^2(1-\alpha-\beta)^3\big(m^2(15\alpha+15\beta+1)-20\alpha\beta s\big)}{1152\pi^6\alpha^3}\\
&+\frac{(1-\alpha-\beta)^3\big(m^2(\alpha+\beta)-\alpha\beta s\big)\big(25\alpha^2\beta s-m^2(15\alpha^2+15\alpha\beta
+4\alpha-24\beta)\big)}{9126\pi^6\alpha^3\beta^2}\\
&+\frac{(1-\alpha-\beta)^2\big(m^2(\alpha+\beta)^2-\alpha\beta s\big)\big( m^2(15\alpha^2\beta-3\alpha^2+15\alpha\beta^2+
\alpha\beta-2\alpha+12\beta)+\alpha^2(6-25\beta)\beta s\big)}{1536\pi^6\alpha^3\beta^2}\\
&+\frac{ (\alpha +\beta -1) \left(m^2 (\alpha +\beta )-\alpha  \beta  s\right) \left(\alpha  (25 \alpha +14) \beta  s-m^2 \left(15 \alpha ^2+\alpha  (15 \beta +8)+6 \beta +8\right)\right)}{1536 \pi ^6 \alpha ^2 \beta }\\
&+\frac{m^2 (\alpha +\beta -1)^2 \left(m^2 (6 \alpha +6 \beta +1)-9 \alpha  \beta  s\right)}{384 \pi ^6 \alpha ^3}
+\frac{ \left(m^2 (\alpha +\beta )-\alpha  \beta  s\right) \left(m^2 (3 \alpha +3 \beta -5)-4 \alpha  \beta  s\right)}{768 \pi ^6 \alpha  \beta }\Big{\}},\\
\rho^{\langle \bar{q}Gq\rangle}_7(s)=&\int_{\alpha_{min}}^{\alpha_{max}}d\alpha\int_{\beta_{min}}^{\beta_{max}}d\beta
\Big{\{} \frac{m_q\qGqb\big(3m^2-\alpha\beta s\big)}{96\pi^4\alpha}\\
&+\frac{m_q\sGs m^2 \left(-72 \alpha ^2-13 \alpha  (9 \beta -4)-45 \beta ^2+52 \beta +8\right)+3 \alpha  \beta  s (32 \alpha +20 \beta -25)}{576 \pi ^4 \alpha }\Big{\}}\\
&-\frac{m_q\qGqb\big(96m^2-6s\big)+m_q\sGs\big(8m^2+8s\big)}{1152\pi^4}\sqrt{1-\frac{4m^2}{s}}
,\\
\rho^{\langle\bar{q}q\rangle^2}_7(s)=&-\frac{\qq\ss \big(s-16m^2\big)}{36\pi^2}\sqrt{1-\frac{4m^2}{s}},\\
\Pi_7^{\langle\bar{q}q\rangle\langle \bar{q}Gq\rangle}(M_B^2)=&\frac{\qq\sGs+\ss\qGqb}{72\pi^2}\\
&\times
\int^1_0dx\Big{\{ } \frac{m^4(12x-9)}{M_B^2x^2(1-x)}-\frac{2m^2(3x-4)}{(1-x)}-\frac{3M_B^2(2x^2-3x+1)}{1-x}\Big{\}}e^{-\frac{m^2}{M^2_B(1-x)x}}.
\end{split}
\end{equation}

\begin{equation}
\begin{split}
\rho^{pert}_8(s)=&\int_{\alpha_{min}}^{\alpha_{max}}d\alpha\int_{\beta_{min}}^{\beta_{max}}d\beta
\Big{\{}\frac{(1-\alpha-\beta)(7-\alpha-\beta)\big(m^2(\alpha+\beta)-\alpha\beta s\big)}{256\pi^6\alpha^3\beta^3}\\
&+\frac{m^2 \left(-3 \alpha ^2+\alpha  (5-6 \beta )-3 \beta ^2+5 \beta +1\right)+7 \alpha  \beta  s (\alpha +\beta -1)}{192 \pi ^6 \alpha ^3 \beta ^3}(1-\alpha-\beta)^2\big(m^2(\alpha+\beta)-\alpha\beta s\big)\Big{\}},\\
\rho^{\langle\bar{q} q\rangle}_8(s)=&\int_{\alpha_{min}}^{\alpha_{max}}d\alpha\int_{\beta_{min}}^{\beta_{max}}d\beta
\big(m^2(\alpha+\beta)-\alpha\beta s\big)\Big{\{} \frac{m_q\qq \big(3m^2(\alpha+\beta+2)-5\alpha\beta s\big)}{8\pi^4\alpha\beta}\\
&+\frac{m_q\ss\big(\alpha  \beta  s (20 \alpha +20 \beta -33)-m^2 (12 \alpha ^2+\alpha  (24 \beta -23)+12 \beta ^2-23 \beta +2)\big)}{16 \pi ^4 \alpha  \beta }\Big{\}},\\
\rho^{\langle GG\rangle}_8(s)=&\langle g_s^2 GG\rangle\int_{\alpha_{min}}^{\alpha_{max}}d\alpha\int_{\beta_{min}}^{\beta_{max}}d\beta
\Big{\{}-\frac{\left(\alpha ^2+\alpha  (20 \beta -2)+21 \beta ^2-22 \beta +1\right) \left(\alpha  \beta  s-m^2 (\alpha +\beta )\right)^2}{6144 \pi ^6 \alpha ^2 \beta ^2}\\
&+\big(m^2(\alpha+\beta)-\alpha\beta s\big)\Big[ \frac{(1-\alpha-\beta)^3\big(5\alpha\beta s-3m^2(\alpha+\beta)\big)}{2304\pi^6\alpha^2\beta^2}\\
&-\frac{(\alpha +\beta -1)^2 \left(m^2 \left(\alpha ^2 (1-45 \beta )-5 \alpha  \beta  (9 \beta -5)-8 \beta  (3 \beta +1)\right)+\alpha ^2 \beta  (75 \beta -2) s\right)}{3072 \pi ^6 \alpha ^3 \beta ^2}\\
&-\frac{m^2 \left(6 \alpha ^3+3 \alpha ^2 (4 \beta +3)+\alpha  \left(6 \beta ^2+29 \beta +1\right)+20 \left(\beta ^2-1\right)\right)-2 \alpha  \beta  s \left(5 \alpha ^2+\alpha  (5 \beta +12)+20 (\beta -1)\right)}{1536 \pi ^6 \alpha ^2 \beta }\Big]\\
&+\frac{m^2 (\alpha +\beta -1)^2 \left(m^2 \left(-12 \alpha ^2-8 \alpha  (3 \beta -4)-12 \beta ^2+32 \beta +1\right)+\alpha  \beta  s (16 \alpha +16 \beta -43)\right)}{1152 \pi ^6 \alpha ^3}\Big{\}},\\
\rho^{\langle \bar{q}Gq\rangle}_8(s)=&\int_{\alpha_{min}}^{\alpha_{max}}d\alpha\int_{\beta_{min}}^{\beta_{max}}d\beta
\Big{\{} \frac{5m_q\qGqb\big(3m^2(\alpha+\beta-1)-4\alpha\beta s\big)}{96\pi^4\alpha}\\
&+\frac{m_q\sGs \big(m^2 \left(-171 \alpha ^2+\alpha  (230-306 \beta )+5 (-27 \beta ^2+46 \beta +2)\right)+3 \alpha  \beta  s (76 \alpha +60 \beta -95)\big)}{576 \pi ^4 \alpha }\Big{\}}\\
&-\frac{24m_q\qGqb\big(7m^2-s\big)-m_q\sGs\big(14m^2-5s\big)}{576\pi^4}\sqrt{1-\frac{4m^2}{s}},\\
\rho^{\langle\bar{q}q\rangle^2}_8(s)=&\frac{2\qq\ss \big(7m^2-s\big)}{9\pi^2}\sqrt{1-\frac{4m^2}{s}},\\
\Pi_8^{\langle\bar{q}q\rangle\langle \bar{q}Gq\rangle}(M_B^2)=&\frac{\qq\sGs+\ss\qGqb}{72\pi^2}\\
&\times
\int^1_0dx\Big{\{ } \frac{6m^4(3-4x)}{M_B^2 x^2(1-x)}-\frac{m^2(6x^2-32x+27)}{x(1-x)}-\frac{3M_B^2(7x^2-11x+4)}{1-x}
\Big{\}}e^{-\frac{m^2}{M^2_B(1-x)x}}.
\end{split}
\end{equation}
where $m$ is the heavy quark $Q$ mass $m_Q$. Some other
notations are
\begin{align}
\nonumber\qquad \alpha_{max}= & \frac{1+\sqrt{1-4m^2/s}}{2} & \alpha_{min}= & \frac{1-\sqrt{1-4m^2/s}}{2}\qquad \\
\qquad\beta_{max}= & 1-\alpha & \beta_{min}= & \frac{\alpha
m^2}{\alpha s-m^2}.\qquad
\end{align}

\end{document}